\documentclass[webpdf,modern,medium,namedate]{oup-authoring-template}
\onecolumn 
\usepackage[mathlines, right]{lineno}
\usepackage[fontsize=12pt]{fontsize}

\usepackage{amsgen,amsmath,amstext,amsbsy,amsopn,amssymb,latexsym}
\usepackage{booktabs, makecell, multirow, tabularx,epsfig}
\usepackage{graphicx, pdflscape, verbatim, setspace, bm}
\usepackage[algo2e]{algorithm2e}
\usepackage{array}
\usepackage{natbib}
\usepackage{setspace}
\usepackage{enumerate}

\usepackage{enumitem}
 \usepackage{algorithm, algpseudocode}

\usepackage[utf8]{inputenc}
\usepackage{grffile}
\numberwithin{equation}{section}
\usepackage[normalem]{ulem}

\newcommand{\red}{\color{red}}

\newcommand{\green}{\color{teal}}

\newcommand{\R}{\mathbb{R}}

\newcommand{\1}{\mathbf{1}}

\newcommand{\I}{\operatorname{I}}

\newcommand{\E}{\mathbf{E}}

\newcommand{\cI}{\mathcal{I} }
\newcommand{\cA}{\mathcal{A} }

\newcommand{\eps}{\epsilon}

\newtheorem{Corollary}{Corollary}

\newtheorem{Assumption}{Assumption}
\newtheorem{Lemma}{Lemma}

\newtheorem{Theorem}{Theorem}

\newcommand{\cov}{\rm Cov}

\newcommand{\tr}{^{ \mathrm{\scriptscriptstyle T} }}

\newcommand{\ignore}[1]{}

\def\Var{{\rm Var}\,}
\def\var{{\rm Var}\,}
\def\Cov{{\rm Cov}\,}
\def\E{{\rm E}\,}
\DeclareMathOperator{\Trace}{tr}
\def\lam{ \lambda }

\newcommand{\tf }{ t^{(1)} }
\newcommand{\ts }{ t^{(2)} }

\def\dagger{ -1}

\def\Qbeta{{\theta}^2 }
\def\hs{ \widehat \theta^2_{\text{Block-Sum}} }

\def\P{\mathrm P}
\def\beq{\begin{equation}}
\def\eeq{\end{equation}}

\newcommand{\Proposed}{SMILE}
\DeclareMathOperator{\rk}{r}
\def\rX{ \rk(X) }
\newcommand{\Smat}{ S }
\newcommand{\PhiS}{ {\Phi_S} }
\newcommand{\xiS}{ {\xi} }
\newcommand{\sg}{ {\eta} }

\hypersetup{hidelinks}

\graphicspath{{./figure/}}
\makeatletter
\newcommand{\arxivexternallabel}[2]{%
  \expandafter\gdef\csname r@#1\endcsname{{#2}{0}{}{}{}}}
\arxivexternallabel{append:likelihood}{D.1}
\arxivexternallabel{appd-compute}{D.2}
\arxivexternallabel{fig:W}{S1}
\arxivexternallabel{fig:eigenvar}{S2}
\arxivexternallabel{supp_sec:setup}{E.4}
\arxivexternallabel{tab:supp-cover-1}{S1}
\arxivexternallabel{tab:supp-cover-2}{S2}
\arxivexternallabel{sec:stability}{F}
\makeatother

\usepackage{etoolbox}

\makeatletter

\def\secsize{%
  \sffamilyfontbold\fontsize{13}{15}\selectfont}
\def\subsecsize{%
  \sffamilyfont\fontsize{12}{14}\selectfont}
\def\subsubsecsize{%
  \normalfont\mathversion{bold}\fontsize{11}{13}\selectfont\bfseries}

\def\section{%
  \@startsection{section}{1}{\z@}
  {-1\p@ plus -.5\p@}{2.5\p@}
  {\reset@font\raggedright\secsize}}
\def\subsection{%
  \@startsection{subsection}{2}{\z@}
  {-4\p@ plus -1\p@}{1.5\p@}
  {\reset@font\raggedright\subsecsize}}
\def\subsubsection{%
  \@startsection{subsubsection}{3}{\z@}
  {-3\p@ plus -1\p@}{0.2em}
  {\reset@font\raggedright\subsubsecsize}}

\patchcmd{\@oddhead}
  {\vspace{5\p@}\rule{\textwidth}{1\p@}}
  {\vspace{5\p@}\raisebox{3\p@}[0pt][0pt]{\rule{\textwidth}{1\p@}}}
  {}{}
\patchcmd{\@evenhead}
  {\vspace{5\p@}\rule{\textwidth}{1\p@}}
  {\vspace{5\p@}\raisebox{3\p@}[0pt][0pt]{\rule{\textwidth}{1\p@}}}
  {}{}

\AtBeginDocument{%
	\abovedisplayskip=4pt plus 1pt minus 1pt
	\belowdisplayskip=\abovedisplayskip
	\abovedisplayshortskip=2pt plus 1pt minus 1pt
	\belowdisplayshortskip=\abovedisplayskip}

\patchcmd{\@maketitle}
  {\if@twocolumn\else\vspace*{-10pt}\fi\if@modern}
  {\if@twocolumn\else\vspace*{-10pt}\fi\iffalse}
  {}{}
\makeatother

\begin{document}
\journaltitle{Journals of the Royal Statistical Society}
\DOI{DOI HERE}
\copyrightyear{XXXX}
\pubyear{XXXX}
\access{Advance Access Publication Date: Day Month Year}
\appnotes{Original article}

\firstpage{1}

\DOI{}

\title[Heritability estimation using genetic similarity representation]{ Heritability estimation using genetic similarity representation
}
\author[1]{Jianqiao Wang}
\author[2,3, $\ast$]{Xihong Lin}
\address[1]{Department of Statistics and Data Science, Tsinghua University,  Beijing, China} 
\address[2]{Department of Statistics, Harvard University, Cambridge, USA} 
\address[3]{Department of Biostatistics, Harvard T.H. Chan School of Public Health,  Boston, USA}

\authormark{Wang and Lin}
\corresp[$\ast$]{Corresponding author.\ \href{email:xlin@hsph.harvard.edu}{xlin@hsph.harvard.edu}}

	
\abstract{
	We introduce a similarity representation framework for robust heritability estimation in Genome-Wide Association Studies (GWAS). 
	This problem parallels the signal-to-noise ratio estimation problem in linear models with a large number of predictors.  
	Traditional fixed- and random-effects methods for heritability estimation often impose restrictive assumptions on regression coefficients or the design (genotype) matrix. 
 These assumptions are usually violated by the heterogeneous effects of genetic variants (regression coefficients) that depend on the genotype distribution and the correlation among genotypes due to linkage disequilibrium. This leads to the non-robust estimation of heritability in practice.
To overcome these limitations, we propose a SiMILarity rEpresentation method (SMILE) which models the relationship between the outcome similarity and genetic similarity through Gram matrices. 
  SMILE represents genetic similarity using a weighted Gram matrix of genotypes, where a data-dependent weight matrix is used to disentangle the heterogeneous variant effects from the genotype distribution.  
	SMILE includes the classical random-effects model as a special case and improves the fixed-effects model by not requiring accurate estimation of the precision matrix or the regression coefficients.   
	We develop a scalable implementation for efficient analysis of large biobank GWAS data. Extensive simulations and 
	the analysis of the UK Biobank data 
	demonstrate the robustness of the proposed method over the existing methods across a range of genetic architectures,  and show that SMILE provides a versatile approach for heritability estimation. }
        
\maketitle
\thispagestyle{plain}
\setstretch{1.45} 
\section{Introduction}
\subsection{Background and motivation}
		
  The problem of estimating the signal variance, the residual variance and the	signal-to-noise ratio in regression with 
		 high-dimensional predictors is of substantial practical interest.
		Our work is motivated by estimating genetic heritability in Genome-Wide Association Studies (GWAS), which collect data of a phenotype ($Y$)  and the genotypes of a large number of genetic variants across the genome  ($X$) from a large number of individuals.
		The goal is to estimate the proportion of the phenotypic variance attributable to genetic effects \citep{yang2017concepts}. 	
	
		Many methods have been proposed for estimating heritability using the linear regression model $y=X\beta+\epsilon$, which can be broadly categorized into random-effects models and fixed-effects models based on whether regression coefficients $\beta$ are assumed to be random or fixed  \citep{
  yang2010common,dicker2014variance,janson2017eigenprism}.
  These traditional approaches face two challenges.
		First, the distribution of genetic effect sizes (regression coefficients $\beta$)  can be heterogeneous and depend on the genotype distribution under natural selection.
		For example,  individuals with disadvantageous genetic factors are eliminated through negative selection, 
		which suggests a larger effect size may correspond to a smaller deleterious allele frequency \citep{pritchard2001rare}. The effect sizes also depend on the linkage disequilibrium (LD), the correlation structure among genetic variants \citep{gazal2017linkage,zeng2018signatures}. Failing to account for the heterogeneous genetic effect sizes can lead to  biased estimation of heritability \citep{finucane2015partitioning}.

		Second, the high correlation among physically nearby genotypes, especially in high-density genotype arrays or using imputed genotypes,  poses a statistical challenge to traditional regression models that assume a non-singular genotype covariance matrix. 
		This intrinsic singularity can cause unstable estimation of regression coefficients.
  Removing highly correlated variants, though simplifying analysis, leads to information loss and cancels out the advantage of increased genotyping density.  
		
			These challenges motivate us to develop a robust heritability estimation method that avoids restrictive assumptions on regression coefficients $\beta$ or the design matrix $X$.
			We propose to directly model the Gram matrix of the outcome vector (outcome similarity) and the Gram matrix of the underlying genetic signal vector (genetic similarity) and develop an estimation framework by representing the genetic similarity as the weighted Gram matrix of the observed genotypes.
			The proposed framework is robust to different modeling assumptions on $X$ and $\beta$,  and provides a versatile approach for heritability estimation.

	 \vspace{-0.1in}

		\subsection{Problem Formulation and Definition}
		Suppose  $y = (y_1, \ldots, y_n)\tr \in \R^n $ is an outcome vector  and $X=(x_1,\ldots, x_n)^T \in \R^{n \times p}$ is a genotype matrix of $n$ unrelated individuals. The row vector $x_i = (x_{i1}, \ldots, x_{ip})\tr \in \R^p$ is drawn i.i.d. from a population distribution, representing the observed genotype vector of the $i$th subject in GWAS. Thus, $X$ is a random matrix. The outcome vector  and genotype matrix are  centered  with  
		  $\E[y_i] = 0$, $\E[x_{ij}] = 0$ and 
	$\Cov(x_i)
	= \Sigma$. The genotype data usually have large $p$ and $n$. 
 
		 We formulate the heritability estimation  problem using a signal-noise variance decomposition
		 \begin{equation}
		 	\label{eq:var-model}
		 	y_{n\times1} =  g_{n\times1} + \varepsilon _{n\times1} \text{ and } \var(y_i) = \theta^2 + \sigma^2_{\eps},
		 \end{equation}
	 where the genetic signal vector $g(X)=\left(g_{1}, \ldots, g_{n} \right)\tr$  is a function of the genotype matrix $X$, i.e. $g_i=g(x_i)$ with  $\E[g_i] = 0$ and  the signal variance $\E[g_i^2] = \theta^2$.  The noise vector \(\varepsilon=\left(\varepsilon_{1}, \ldots, \varepsilon_{n}\right)\tr\) is independent of $g(X)$  with $\E[\varepsilon_i] = 0$ and  $\E[\varepsilon_i^2] = \sigma^2_{\eps}.$  To focus on the key idea, model (\ref{eq:var-model}) does not include covariates. It can be easily extended to accommodate covariates. 

	 This paper considers a linear model for the genetic signal $g(X) = X \beta$, in which  the genetic effects shaped by natural selection are considered as 
\begin{equation}
\label{eq:effect}
\beta(\Sigma) \text{ with support $\cA(\Sigma)$ and effect size $\beta_{\cA}(\Sigma)$ }.  
\end{equation}
Specifically, $\cA(\Sigma)$ and $\beta_{\cA} (\Sigma) $ represent how the causal variant locations and non-zero effect sizes are related to the allele frequency and the correlation structure of the genotype $X$ due to LD. 
As $\beta$ represents the population-level joint genetic effects, we treat $\beta$ as a fixed population parameter.  

The signal variance $\theta^2$ and the heritability $h^2$ are defined as   
		\begin{equation}
  \label{def: quadratic term}
   \theta^2 = \Var( x_i\tr \beta)  = \beta\tr \Sigma \beta 
   \quad  \text{and} \quad	h^2 = \frac{ \theta^2}{\var(y_i) }. 
		\end{equation}
        Correspondingly, $\gamma=\theta^2/ \sigma^2_{\eps}$ is the signal-noise ratio.
 We measure heritability using $\beta\tr \Sigma \beta$ instead of $\| \beta\|^2=\beta^T\beta$ in \eqref{def: quadratic term} for two reasons.
 First, $\|\beta\|^2$ is not identifiable with singular  $\Sigma$ while $\beta\tr \Sigma \beta$ is well-defined regardless of the covariance structure $\Sigma$. Second, by accounting for the effects of $\Sigma$,  \eqref{def: quadratic term} is able to distinguish heritability in settings where the effect sizes $\beta$ are identical but the genotype covariance structures $\Sigma$ are different.   

To robustly estimate ($\theta^2,\sigma^2_\eps)$ and the heritability $h^2$ under the possible heterogeneous effects \eqref{eq:effect},     
we formulate an estimation framework  by modeling the relationship  between the Gram matrix of the outcome vector $yy\tr$, 
and the Gram matrix  of the  genetic signal $X \beta \beta\tr X\tr$,
\begin{align}
\label{eq:lm}
\E[yy\tr | X \beta \beta\tr X\tr ] = X \beta \beta\tr X\tr + \sigma^2_{\eps} \I_n = \theta^2 ( \sg \sg\tr)_{n \times n} + \sigma^2_{\eps} \I_n, 
\end{align}
where $\sg(X)= \theta^{-1}X\beta$ 
represents the standardized signal vectors.  
From a genetic perspective  \citep{haseman1972investigation}, $yy\tr$ represents the observed trait similarity between individuals and  
$\eta \eta\tr$  represents a standardized genetic similarity.  

This paper proposes to use an observed genotype-based similarity matrix $S(X)$ to represent $\eta\eta\tr$ for robust heritability estimation.  Through a suitable genetic similarity representation, our framework avoids making restrictive modeling assumptions on the dependence of regression coefficients $\beta$ on the design matrix $X$ in fitting the high-dimensional regression.

  
	\subsection{Related literature and challenges}
	\label{sec:lit}
	Existing methods for estimating heritability focus on the linear regression model 
	(\ref{eq:lm}), and can be grouped into two strategies: {\it random-effects modeling},  which treats regression coefficients $\beta$ as random and estimates the variance components specified in the distribution of $\beta$ using REML \citep{yang2010common, loh2015efficient};
	and {\it  fixed-effects modeling,} which treats regression coefficients $\beta$ as fixed parameters to estimate.

{\it  Random-effects modeling:}  
The random-effects approach  is one of the most popular methods for  heritability estimation in statistical genetics
 \citep{
 yang2010common,
 loh2015efficient}.
 In the linear model (\ref{eq:lm}),  the elements of the
 regression coefficient vector $\beta$ are assumed to be i.i.d. and follow a normal distribution with mean 0 and variance $p^{-1}\sigma_{\beta}^2$. It gives that
\beq
\label{random0}
\beta \sim N(0, p^{-1}\sigma^2_\beta \I) \text{ and } y|X \sim N(0,  \frac{X X\tr}{p} \sigma^2_\beta   + \sigma^2_{\eps} \I) .
\eeq
 This formulation connects the variance parameter to the total genetic effects via $\sigma^2_\beta =  \E[\beta\tr \beta]$, which can be estimated directly using either MLE or REML. 

While the homogeneous random-effects assumption is popular, its validity is challenged by the heterogeneity of genetic effects in different regions in \eqref{eq:effect}, which can result in biased heritability estimation.  
To reduce the bias,
 \cite{speed2012improved}  introduce a parametric model that models the dependence of the genetic effect sizes on  LD and allele frequency.   \cite{
 yang2015genetic} consider the stratified random-effects model 
 by stratifying  genetic variants 
   into a small number of LD and allele frequency categories and assigning separate effect distribution parameters to each variant group. 
However, these methods often rely on specific parametric partitioning strategies and are subject to the potential partitioning bias.
Besides, as the number of partitions increases, computation on large-scale genotype data becomes difficult.

{\it Fixed-effects modeling:}  
 The fixed-effects modeling aims to quantify $\beta\tr \Sigma \beta$ by treating the coefficients  $\beta$ as fixed parameters. For example, in the high-dimensional setting, 
genetic variance or covariance estimators are proposed with $\beta$ estimated from sparse penalized regression  \citep{reid2016study,cai2018semi,wang2024regression}. 
Nevertheless, challenges arise for selecting and estimating weak or non-sparse effects,  which are common for GWAS studies of many traits, 
such as human height \citep{yengo2022saturated} and body mass index \citep{smit2025polyg}, especially when $p>>n$. 

Another line of fixed-effects model approaches relies on knowing or estimating the precision (inverse covariance) matrix $\Sigma^{-1}$ 
\citep{dicker2014variance,janson2017eigenprism}.  
The estimation accuracy of $\Sigma^{-1}$ for the high-dimensional genotype data $X$  determines the applicability of these methods. 
When the population inverse matrix is known or an accurate sparse inverse estimation is available,  these methods are valid in high-dimensional regimes with $p/n \to c \in (0, \infty)$ \citep{ dicker2014variance, verzelen2018adaptive, wang2022estimation}. In contrast, approaches based on the generalized inverse of the sample covariance matrix are  limited to settings with $p < n$ \citep{hou2019accurate}. 


In summary, the existing random-effects and fixed-effects frameworks for heritability estimation suffer from several drawbacks. The random-effects approach is sensitive to the misspecification of the heterogeneous genetic variant effect distributions, i.e., the distribution of $\beta$.  The fixed-effects approach requires accurate estimation of regression coefficients $\beta$ and the precision matrix $\Sigma^{-1}$,  complicated by the presence of a large number of correlated genetic variants. 
 The discrepancy between the two modeling formulations and real data analyses motivates us to investigate the necessity of these restrictive assumptions and develop methods to reconcile different modeling strategies.


\subsection{Overview of the framework and contributions}
\label{sec:contribution}

    We propose a similarity representation framework that robustly estimates the signal and noise variance parameters ($\theta^2,\sigma_{\eps}^2)$ and the heritability $h^2$,  which allows the distribution of genetic effects $\beta$ to depend on $\Sigma$.
    We term the proposed method as the SiMILarity rEpresentation method (SMILE).  Encompassing both the classical random-effects and fixed-effects models as special cases, the proposed SMILE method has several features. First,
unlike traditional random-effects models, it robustly estimates $(\theta^2,\sigma_{\eps}^2)$  without requiring a parametric model for the dependence of the effects $\beta$ on $\Sigma$ in \eqref{eq:effect}.  
Second, it improves the fixed-effects model approach by not requiring accurate estimation of the precision matrix or the regression coefficients.   
Finally, the framework reconciles different modeling assumptions and makes distinguishing random-effects or fixed-effects unnecessary for heritability estimation.  


The SMILE framework first uses the observed genotype to construct a robust genetic similarity representation $S(X)$  for the underlying standardized similarity $\sg(X)\sg(X)^T$ in \eqref{eq:lm}.  
 Motivated by the similarity $X (\beta \beta\tr) X\tr$ in the linear model, we specify  $S(X)$ as the weighted genetic Gram matrix, 
\beq
\label{ll-model}
S = XWX\tr \text{ with }  \Trace(S) = n,
\eeq 
 where $W=W(X, \Sigma, \beta)$  could be a function of $X$, $\Sigma$ or $\beta$.  The oracle $W$ for  recovering the standardized similarity is $W = \beta\beta^T/\| \beta \|^2$ (up to a constant). However, as $\beta$ is unknown, we will construct  $W$ that depends on the data $X$ and makes the information contained in data-dependent $S=XWX^T$  similar to the information contained in the pairwise similarity $X\beta\beta^TX^T$. 
 In addition,  to satisfy the property of a similarity matrix, we construct data-dependent $W=W(X)$ to be positive semi-definite, symmetric, and sign-invariant with $W(X) = W(-X)$. 

We propose estimation of $(\theta^2 , \sigma^2_{\eps}$) based on \eqref{eq:lm} and construct the estimating equations 
\beq
\label{s-eq}
 \Psi_n(\theta^2 , \sigma^2_{\eps}; S, y) = 0
\eeq
to obtain $\widehat \theta^2(S)$ and $\widehat \sigma_{\eps}^2(S)$.  The heritability estimator of $h^2$ follows accordingly.   
We term the estimator that solves (\ref{s-eq}) as the SMILE estimator.  A key feature of the SMILE estimator is that we construct $S$ and $\Psi_n(\cdot)$ in a way such that the estimator of $\widehat \theta^2(S)$ and $\widehat \sigma_{\eps}^2(S)$ is robust to the dependence of the genetic effect $\beta$ on $\Sigma$.
In Section \ref{sec:gram}, we specify $\Psi_n$  by connecting it to the   Gaussian variance component model $y \sim N(0, \theta^2 S + \sigma^2_{\eps} I )$.

SMILE does not require exact recovery of the true similarity.  A sufficient condition for SMILE estimators to be unbiased is that the working similarity matrix $S$ satisfies the moment property
\beq
\label{mcond}
\E\Big[ 
     \Trace\!\left\{
       A(S)  ( X \beta \beta\tr X\tr - \theta^{2} S ) 
        \right\}
     \Big]  = 0,
     \eeq 
      for a suitable class of functions $A(\cdot).$ 
A more general  consistency result for SMILE is established through the primal similarity model, a re-parameterized regression induced by the eigen-factors $\PhiS(X)$ of $S=XW(X)X^T = \PhiS \PhiS\tr$. In the primal model, the genetic effects are projected onto the eigenspace of $S$, and represented through a new set of coefficients, 
\beq
\label{eq:primal}
X \beta = \Phi_S \xi, 
\eeq
where $\xi=\xi(X)$ is random, since it is a function of the data $X$,  regardless of whether $\beta$ is random or fixed.  The information of $\theta^2$ is contained in $\xi$. Our framework shows that by choosing an appropriate data dependent $W$,  the SMILE estimator is consistent when the distribution of $\xi$ is asymptotically decoupled from the eigenvalues of the working similarity matrix $S$, i.e., distribution of $\xi$  does not depend the eigenvalues of $S$ asymptotically. 


 We will show in Section \ref{sec:REML} that the special case $S = n XX\tr/\Trace(XX\tr)$ (i.e., $W=\I$) corresponds to the traditional genetic random-effects model (\ref{random0}).  In this case, we study the structure of $(\beta, \Sigma)$ under which SMILE gives a consistent estimation of $(\theta,\sigma_{\eps}^2)$, and show that the distribution of genetic effects $\beta$ needs to be independent of $\Sigma$. 
 In the presence of heterogeneous effects $\beta$  that depend on $\Sigma$ in \eqref{eq:effect},  as discussed earlier, we construct $S=XWX^T$ using a data-dependent weight matrix $W=W(X)$ such that the coefficient $\xiS$ in the primal model \eqref{eq:primal} does not depend on the eigenvalue distribution $S$.


We develop an efficient algorithmic implementation of the proposed  SMILE method for estimating heritability in large-scale biobank GWAS datasets.
We evaluate the performance of SMILE compared with the existing methods in extensive simulation studies.    We apply the method to the UK Biobank (UKBB) data for estimating the heritability of a range of phenotypes.  These empirical studies demonstrate the robustness of the proposed method over the existing methods under various genetic architectures.

\subsection{Organization and notation}

The rest of the paper is organized as follows. Section \ref{sec:method} presents the general SMILE framework and establishes the statistical properties.   Section \ref{sec:REML}  revisits the random-effects model through the SMILE framework, clarifying the applicability and limitations of the classical approach in the heterogeneous genetic architecture.
 Section \ref{sec:method-3} discusses the choice of $W$ for accurate heritability estimation in the setting of heterogeneous genetic effects. 
Section~\ref{sec:implement} develops an efficient implementation that makes SMILE
scalable to large biobank-scale GWAS data.
 Section \ref{sec:simulation} evaluates the finite sample performance of the proposed methods and compares it with several existing methods under different genetic architectures. Section~\ref{sec:real-data} applies SMILE to UK Biobank data for heritability estimation of several traits. We conclude with a discussion in Section \ref{sec:discuss}. The derivation details of our procedure and the proofs of the main theorems are given in the Supplementary Materials. 

We introduce the notation here. The identity matrix is denoted as $\I$  or  $\I_n$ for the $n \times n$ matrix.  The rank of a matrix $W$ is denoted as $\rk(W).$ 
For a matrix $X  \in \R^{n\times p} $, $X_{i\cdot}$,  $X_{\cdot j}$, and $X_{i,j}$ denote respectively the $i$-th row,  $j$-th column, and  $(i,j)$ entry of the matrix $X$. 
For a symmetric matrix $A$, $\lambda_{\min}\left(A\right)$ and $\lambda_{\max}\left(A\right)$  denote respectively the smallest and largest eigenvalue of $A$. 
For a set $\cA$, $\left|\cA\right|$ denotes the cardinality,  and $X_{\cA\cdot}$ denotes the submatrix of $X$ consisting of rows  $X_{ i \cdot}$ with $i\in \cA$ and for a vector $z \in \R^{p}$, $z_{\cA}$ is the subvector  with indices in $\cA$. 
We use $c$ and $C$ to denote generic positive constants that may vary from place to place. For two positive sequences $a_n$ and $b_n$,  $a_n \lesssim b_n$ means $a_n \leq C b_n$ for all $n$ and $a_n \gtrsim b_n $ if $b_n\lesssim  a_n$ and $a_n \asymp b_n $ if $a_n \lesssim b_n$ and $b_n \lesssim a_n$.  A random vector $z$ is  isotropic (up to a constant) if $\E[zz\tr] = c \I.$ A random matrix  $X$  is called isotropic if $ \mathrm{vec}(X)$ is isotropic. 

\section{The SMILE framework}
\label{sec:method}
\label{sec:gram}


\subsection{SMILE estimating equations for $(\theta^2,\sigma_{\epsilon}^2)$}
 \label{sec:method2}

 Given a data-dependent weight matrix $W$ (to be  discussed in Section \ref{sec:method-3}), a general  similarity matrix $S$ after scaling is given as  
\beq
\label{eq:w-gram}
S(X) = \frac{ n   X W X\tr }{  \Trace(  X W X\tr ) }. 
\eeq  
 One can  see   $\Trace(S) = n.$ We use the notation $S(X)$ to emphasize its dependence on the observed data, and thereafter, let $S =  X W X\tr$ with  trace scaled for simplicity. 

Let $V(\theta^2, \sigma^2_{\eps}) = \theta^2 XWX^T + \sigma^2_{\eps} \I = \theta^2 S + \sigma^2_{\eps} \I $. For heritability estimation, the working similarity matrix \(S\) need not recover
the true signal similarity \(X\beta \beta\tr X\tr\) exactly. Instead, we call \(S\) a valid representation if it yields unbiased estimating equations at true parameters. Specifically, for some chosen $n\times n$ positive semi-definite matrix $A=A(S;\theta^2,\sigma^2_{\eps})$,  it satisfies the moment condition
\begin{align}
 \label{sw-sim}
\E\left[ y\tr Ay  - \Trace\left( A V \right) \right] = \E\big[\Trace\big(  A \underbrace{ ( X \beta  \beta\tr X\tr - \theta^2 X W X\tr  ) }_{\text{similarity model}} \big) \big] = 0.
 \end{align}
Based on the observed data, we define the similarity modeling residuals as $$ R(\theta^2, \sigma^2) = yy\tr - V( \theta^2, \sigma^2_{\eps}) = yy\tr - { \theta}^2 S - \sigma^2_{\eps} \I.$$ 
 Setting $A=V^{-1}$ and $A=V^{-2}$,
 the SMILE estimator is defined as the solution to the estimating equations
	\begin{align}
		\label{eq:esteq}
  0=\Psi_n(\theta^2 , \sigma^2_{\eps}; S,y) = \begin{pmatrix}
      \Trace( V^{-1} R ) \\
      \Trace( V^{-2} R )
  \end{pmatrix}  = \begin{cases} 
			\Trace( V^{-1}   y y\tr) - n 
			\\
			\Trace(  V^{-2} y y\tr)  - \Trace(V^{-1})
		\end{cases}. 
	\end{align}
 The estimators  ${\widehat \theta}^2$ and $\widehat \sigma^2_{\eps} $ are obtained by a numerical iterative procedure, with details provided in Section \ref{sec:implement}. In summary, SMILE chooses
\((\theta^2,\sigma^2_{\eps})\) so that the fitted similarity-model residual $ R(\widehat \theta^2,\widehat \sigma^2_{\eps})$ is orthogonal
to the weight matrices \(V^{-1}\) and \(V^{-2}\) under the trace inner product. 

 The similarity-based  equations \eqref{eq:esteq} coincide with the score
equations for the variance components \((\theta^2,\sigma_{\eps}^2)\) under
the conditional Gaussian model
\begin{align}
\label{eq:working-gaus}
y\mid S \sim N\left(0, V \right). 
\end{align}
The log-likelihood, ignoring the constant, and corresponding score equations are 
\begin{align*}
		\label{eq:esteq}
        & \ell( \sigma_{\eps}^2, \theta^2; y, S )   =  - \frac{1}{2} \log| V |    - \frac{1}{2} y \tr V^{-1} y ~ \text{ and } \\
 & \Psi^{\rm G}_n  = \begin{pmatrix}
      \partial \ell / \partial \theta^2 \\
      \partial \ell / \partial \sigma_{\eps}^2
  \end{pmatrix} =  \begin{cases} 
			\Trace( V^{-1}  S V^{-1}  y y\tr) - \Trace( V^{-1} S ) 
			\\
			\Trace(  V^{-2} y y\tr)  - \Trace(V^{-1})
		\end{cases}. 
	\end{align*}
Note $\I = \theta^2 V^{-1} S + \sigma_{\eps}^2 V^{-1}$ and one can easily see that  SMILE estimating equations can be expressed as a linear combination of $\Psi^{\rm G}_n $, i.e.,
\beq
\label{eq:eqv}
\Psi_n = \begin{pmatrix}
    \theta^2 & \sigma^2_{\eps} \\
    0 & 1
\end{pmatrix} \Psi^{\rm G}_n . 
\eeq
Therefore, whenever $\theta^2 > 0$, $\Psi_n = 0$ is equivalent to $\Psi^{\rm G}_n $, meaning that we could apply the same numerical estimation procedure.  This connects the proposed similarity-based framework
to the random-effects framework. For example, when \(S=XX^\top/p\),  \eqref{eq:working-gaus} is  equivalent to the random-effects
model \eqref{random0} with \(\theta^2=\sigma_\beta^2\), and the SMILE
estimator reduce to the corresponding REML estimator for $(\theta^2, \sigma^2_{\eps})$. 

Although the Gaussian model \eqref{eq:working-gaus} provides a convenient interpretation, our  similarity-based consistency theory in Section \ref{sec:method3}   shows that distribution assumptions, specifically  the random-effects assumption, are not required.  For  $S = XX\tr$,  Section \ref{sec:REML} studies the conditions on $X$ and $\beta$ 
that ensure the consistent REML estimator of $(\theta^2,\sigma_{\epsilon}^2)$. The theoretical investigation clarifies the applicability and limitations of the classical random-effects approach. 

We will show in Section \ref{sec:method-3}  that when a proper data-dependent weight matrix $W$ is used in constructing $S=XWX^T$, under some regularity conditions, the SMILE estimator of $(\theta^2,\sigma_{\epsilon}^2)$   that solves (\ref{eq:esteq}) is robust when $\beta$ depends on $\Sigma$, regardless of whether $\beta$ is random or fixed. The consistency of the SMILE estimator 
does not require specifying a model for the dependence of $\beta$ on $\Sigma$, nor does it require an accurate estimation of $\Sigma$.

\subsection{The statistical property of the SMILE estimator}
\label{sec:method3}

This section establishes the theoretical properties of the SMILE estimator solving the estimating equation \eqref{eq:esteq}.  
Our analysis considers a general working similarity matrix \(S=XWX\tr\), allowing \(S\) to be full rank  or rank-deficient.

We introduce  regular assumptions for the signal and the noise, which control the behavior of the quadratic terms in the estimating equations \eqref{eq:esteq}.  For convenience, in the following, we consider the scaled outcome with $\Var(y_i) = 1$ and thus, $h^2 = \theta^2.$ 
\begin{Assumption}
\label{asmp2}
Suppose that the signal $X_{i.}^T\beta$ and the noise $\epsilon_i$ have positive variances $(\theta^2, \sigma^2_{\epsilon})$, and finite fourth moments. The design matrix is generated as $X = Z\Sigma^{1/2}$, where  the entries of $Z$ are i.i.d. random variables with mean $\E[Z_{ij}] = 0$, variance $\Var(Z_{ij}) = 1$, and finite eighth moment $\E[Z_{ij}^8] < \infty$. The covariance matrix has $\Trace(\Sigma) = p.$ 
\end{Assumption}

Let $S = \PhiS \PhiS\tr$ be its rank-$r$ factorization based on the eigen-decomposition, 
where the eigen-factors \(\Phi_S\in\mathbb{R}^{n\times r}\) are eigenvectors of \(S\)
scaled by the square roots of the corresponding positive eigenvalues, and $\Lambda = \mathrm{Diag}(\lam_1, \cdots, \lam_r) = \Phi_S\tr \Phi_S$ is the non-zero eigenvalue matrix of $S$.  

Based on the eigen-factorization of $S$, our theoretical analysis is built on  the primal representation $X \beta  = \Phi_S(X)  \xiS(X)$ in \eqref{eq:primal}, 
where the distribution of $\xi(X)$ is induced by $P(X | S)$ and $\beta$. Then with $A=V^{-1}$ and $A=V^{-2}$,  
the moment equation in  \eqref{sw-sim} can be rewritten as 
\begin{align}
\label{eq-1}
    & \Trace\left\{  V^{-1} ( X \beta  \beta\tr X\tr - \theta^2 S ) \right\}   = \sum^r_{i = 1} \frac{\lam_i}{ (\theta^2 \lambda_i + \sigma_{\eps}^2)  } (\xi^2_i - \theta^2), \\
    & \Trace\left\{  V^{-2} ( X \beta  \beta\tr X\tr - \theta^2 S ) \right\}  = \sum^r_{i = 1} \frac{\lam_i}{ (\theta^2 \lambda_i + \sigma_{\eps}^2)^2  } (\xi^2_i - \theta^2). \notag
\end{align}
Thus, consistency of the estimating equations \eqref{eq:esteq} is driven by the concentration of spectral-weighted averages of \((\xi_i^2-\theta^2)\) around $0$. In other words, the eigenvalues are decoupled from the associated coefficients $\xi$, as shown in the following conditions and  Lemma~\ref{lem0}.


\begin{enumerate}[ 
    label=(C\arabic*),      
    ref=C\arabic*,          
    labelwidth=2.2em,       
    labelsep=0.6em,         
    leftmargin=!,           
    itemindent=0pt,         
    align=left,
    topsep=2pt            ]
  \item\label{cond:C1}
        The non-zero eigenvalues of the similarity matrix $S$ satisfy that  \( C^{-1} \leq r \lambda_i/n \leq C \) for $1 \leq i \leq r$ and $C \geq 1$. 
  \item\label{cond:C2} 
The eigenvalue sequence $(\lambda_1, \ldots, \lambda_r)$ and the coefficients $\xi^2 = ( \xi^2_i, \ldots, \xi^2_r  )\tr$ are asymptotically decoupled, i.e.,  for any bounded function $f(\cdot)$,
\[
\lim_{n, p \to \infty}  \sum^{r}_{i = 1} \left( \frac{\xi^2_i}{\| \xi \| ^2}  -  \frac{1}{r} \right) f(\lambda_i) = 0. 
\]
 \item\label{cond:C3}   $\E[\xi_i^2  \, | \,  \Lambda] = \tau^2$ for $i = 1, \dots, r$, and $\Cov(\xi^2 \, | \, \Lambda) = \Omega$ with \(  \lambda_{\min} ( \Omega ) \geq c' \lambda_{\max} ( \Omega )  \) for some constant $c' > 0$.     
\end{enumerate}

Condition  \eqref{cond:C1} is a regular condition that the non-zero eigenvalues are bounded away from zero and infinity after the trace standardization  $\Trace(S) = n$. Related to the isotropic random-effects, \eqref{cond:C2} suggests that $\xi^2_i/ \| \xi \|^2   \approx 1/r$, and the uncorrelatedness between the eigenvalues of the similarity matrix and coefficients.  Both conditions are satisfied when $r = 1$ or non-zero eigenvalues $(\lambda_1, \ldots, \lambda_r)$ are equal. Section \ref{sec:REML} further discusses these conditions when $S$ is a Gram matrix. 

Condition \eqref{cond:C3}  is introduced for the convergence rate in Theorem~\ref{thm:converg}. It strengthens condition~\eqref{cond:C2}  by assuming a conditional distribution of $\xi^2$ in which the components have equal means and moderate interdependence.  

\begin{Lemma}
\label{lem0}
Suppose that $\lambda_{\max}(S) \leq nC/r$ and $n, p \to \infty$. \\
    (i) If condition~\eqref{cond:C2} holds in probability (or almost surely) and  $f(\cdot)$ is a bounded function, then in probability (or almost surely), 
    \[
        \frac{1}{r} \sum_{i=1}^{r} f(\lambda_i) \left(\xi_i^2 - \theta^2\right) \to 0.
    \]
\noindent
    (ii) If condition~\eqref{cond:C3} holds, then
    \[
        \frac{1}{r} \sum_{i=1}^{r} f(\lambda_i) \left(\xi_i^2 - \theta^2\right) = O_p\big(  \sqrt{ \frac{  \frac{1}{r}\sum^r_{i = 1} f^2(\lam_i) }{ n }} \big).
    \]
\end{Lemma}

Besides, the spectral-weighted average in \eqref{eq-1} suggests that the estimation efficiency of \((\widehat\theta^2,\widehat\sigma_{\eps}^2)\) depends on the eigenvalue distribution of $S$. 
Therefore we  define the quantity 
\beq
\label{eq:cond1}
\Delta_n = n^{-1} \sum^r_{i = 1} (r\lambda_{i}/ n - 1)^2 + (1 - \frac{r}{n}).
\eeq 
When $r = n$,  $\Delta_n = \Var(\lambda_i)$ measures the eigenvalue variation under $\Trace(S) = n$. When $r < n$ and eigenvalues are equal, $\Delta_n = 1- r/n$  measures the number of non-zero eigenvalues. As we will show in Theorem \ref{thm:converg}, a larger $\Delta_n$ means a more informative working similarity  $S$ and thus, a more efficient SMILE estimator.




The following theorem shows that the SMILE estimators that solve \eqref{eq:esteq} are consistent under given conditions. 
 \begin{Theorem}
\label{thm:consis}
Suppose that Assumption~\ref{asmp2} and \eqref{cond:C1}--\eqref{cond:C2} hold for a given similarity matrix  $S$. 
 If $\lim_{n \to \infty} \Delta_n = \Delta > 0$ in probability,  the SMILE estimators that solve \eqref{eq:esteq} are consistent. 
\end{Theorem}
 We will discuss in Sections 3 and 4 the choices of $W$ in $S$ that satisfy the conditions specified in Theorem \ref{thm:consis} to ensure the consistency.
The non-zero value of $\Delta_n$ ensures that the matrix $S$ and $(\gamma S +  I )^{-1}$, where $\gamma=\theta^2/ \sigma^2_{\eps}$ is the signal-noise ratio, are far from the identity matrix, making $\theta^2$ and $\sigma^2_{\eps}$ identifiable. In genetic applications, it means the constructed genetic similarity matrix $S$ is informative in distinguishing the genetic signal and environmental  noise.  

Theorem \ref{thm:converg} provides the convergence rate of the SMILE estimator when the consistency holds.
The results suggest the SMILE estimator is more efficient with a larger $\Delta_n$, which provides practical guidance on the choice of $W$.

\begin{Theorem}
\label{thm:converg}
Suppose that Assumption~\ref{asmp2}, \eqref{cond:C1} and \eqref{cond:C3} hold.  
Then for  sufficiently large $n$, the convergence rates of the  SMILE estimators that solve \eqref{eq:esteq} are, 
\beq
|\widehat \theta^2 - \theta^2 | = O_p\left(\frac{ 1}{ \sqrt{n}} + \frac{1}{ \sqrt{n \Delta_n } } \right)  ~\text{ and }  ~ |\widehat \sigma_{\eps}^2 - \sigma_{\eps}^2 | = O_p\left( \frac{ 1}{ \sqrt{n}} + \frac{1}{ \sqrt{n \Delta_n} } \right). \label{rate} 
\eeq
\end{Theorem}

Theorem \ref{thm:converg} suggests that, when the consistency holds, the SMILE estimator is more efficient as $\Delta_n$ increases. Intuitively, a larger \(\Delta_n\) reflects a more informative working similarity matrix \(S\) in  separating signal and noise.   
When $r < n$, larger \(\Delta_n\) means the low-rank structure in $S$. When \(r=n\),   a large \(\Delta_n\)  corresponds to large eigenvalue dispersion.  In contrast, $S = \I$ with $\Delta_n = 0 $ provides no information.

Lemma \ref{prop:delta_n} shows how the size of $\Delta_n$ depends on the choice of $W$ and the covariance structure of $X_i$. Since a general analysis of data-adaptive \(W\) is infeasible without specifying its specific construction, we focus on the case where \(W\) is fixed. For $S = X W X\tr$ specified in \eqref{eq:w-gram}, denote the corresponding normalized covariance matrix  
\[
\Sigma_W =
\frac{\Sigma^{1/2}W\Sigma^{1/2}}
{\operatorname{tr}(\Sigma^{1/2}W\Sigma^{1/2})}.
\]

\begin{Lemma}
\label{prop:delta_n}
Suppose Assumption \ref{asmp2} holds and $n/p \to c \in (0, \infty)$. Given a fixed symmetric  semi-positive  matrix $W$ and $S = X W X\tr$ specified in \eqref{eq:w-gram} with $\Sigma_W$, it holds almost surely that 
    \beq
    \label{eq:delta_n}
    \lim_{n,p \to \infty} \Delta_n / \left\{ 
    \left( \frac{ \rk(\Sigma_W) \wedge n }{n } \right)^2 { n \Trace(\Sigma_W^2) } + \left( 1 -  \frac{ \rk(\Sigma_W) \wedge n }{n } \right)^2  
    \right\} = 1.  
    \eeq
\end{Lemma}  

\ignore{
\begin{Lemma}
\label{prop:delta_n}
Suppose Assumption \ref{asmp2} holds and $n/p \to c \in (0, \infty)$. Given a pre-specified semi-positive symmetric matrix $W$, for $S = X W X\tr$ specified in \eqref{eq:w-gram} and $\Sigma_W = \Sigma^{1/2} W \Sigma^{1/2}$, it holds almost surely that 
    \[
    \lim_{n,p \to \infty} \Delta_n / \left\{ 
    \left( \frac{ \rk(\Sigma_W) \wedge n }{n } \right)^2 { n \Trace(\Sigma_W^2) } + \left( 1 -  \frac{ \rk(\Sigma_W) \wedge n }{n } \right)^2  
    \right\} = 1.  
    \]
\end{Lemma}
}
 
{  
According to Lemma \ref{prop:delta_n}, for fixed $W$ and $\Sigma_W$ with $r(\Sigma_W) \geq n$, the convergence rate of $\widehat \theta^2$  in \eqref{rate} can be simplified as  
\beq
\label{rate2}
|\widehat \theta^2 - \theta^2 | = O_p\left(\frac{ 1}{ \sqrt{n}} + \frac{1}{ \sqrt{ n^2 \Trace(\Sigma_W^2) } } \right) \lesssim \frac{ 1}{ \sqrt{n}} + \sqrt{\frac{ \rk(\Sigma_W) }{n^2}}, 
\eeq
 where  the Cauchy-Schwarz inequality shows $\Trace(\Sigma^2_W) \geq  1/\rk(\Sigma_W)$.
In summary, Lemma~\ref{prop:delta_n} provides practical guidance for efficient estimation. If the consistency holds, one should select  $W$ that induces a transformed
covariance \(\Sigma_W\) with large spectral dispersion or low-rank structure. 

}

In summary, this section gives a general SMILE estimation framework and discusses the statistical properties, consistency and efficiency, of the SMILE estimator, under the similarity model \eqref{sw-sim}.  However, the remaining questions are when model misspecification happens and how to specify a similarity matrix $S$  to ensure the consistency of the SMILE estimator $(\hat{\theta},\hat{\sigma}_{\epsilon}^2)$. In practice, the choice usually becomes a bias-variance trade-off, as we will discuss in  next two sections.

\section{ SMILE estimator with the Gram similarity matrix  $S=XX^T$}
\label{sec:REML}

 This section examines the assumptions on $X$ and $\beta$ that are
required for unbiased SMILE estimation of $(\theta^2,\sigma_{\epsilon}^2)$ when specifying $S =XX\tr$. 
Specifically, suppose  $X$ is standardized, we have $\Trace(XX\tr) = np$ and 
 \beq
 \label{gram}
S = \frac{ n XX\tr }{ \Trace(XX\tr)} = \frac{XX\tr}{p}.
 \eeq
According to \eqref{eq:eqv}, the SMILE estimating equations (\ref{eq:esteq}) in this case,  $\Psi_n\left(\theta^2 , \sigma^2_{\eps}; S=p^{-1}XX\tr \right)=0$
correspond to an invertible linear transformation of the Gaussian variance-component score equations  under the traditional genetic random-effects model (\ref{random0}). The positive SMILE estimator  reduces to the classical REML estimator assuming homogeneous random genetic effects $\beta \sim N(0, p^{-1}\sigma^2_\beta \I)$. Therefore,  analyzing the SMILE consistency condition under \eqref{gram} 
also clarifies the applicability and limitation of the classical random-effects model based approach in the presence of the heterogeneous genetic architecture and $\beta$ being  either fixed or random effects.  To simplify the notation, we drop the scaling factor and denote $S = XX\tr$ and the estimator as $\widehat\theta^2 (S= XX\tr )$ thereafter.

 We first perform the eigen-decomposition $S=XX\tr = U D^2 U\tr$, which 
corresponds to a compact SVD of $X= U D \Gamma\tr$. The primal form \eqref{eq:primal} for $S = XX\tr$  coincides with the representation of $X\beta$ in terms of principal component regression, 
$$ X \beta =  UD (\Gamma\tr \beta) = \Phi_S \xi,$$
where the eigen-factors $\PhiS  = U D$ and $\xi(X) = \Gamma(X)\tr \beta$.
 Given random $X$,  it follows that
 both $\Gamma(X)$  and 
 $\xi(X)$ in $\R^{r}$ are random  induced by the distribution of $X$, even if $\beta$ is a fixed-effect parameter vector.

 Theorem \ref{thm:consis} shows that consistency of the SMILE estimator using $S=XX^T$  requires the  decoupling condition \eqref{cond:C2} for $\xi = \Gamma\tr \beta$.   A sufficient condition is that $\Gamma\tr \beta =  (\Gamma_{.1}\tr \beta,\cdots,\Gamma_{.r}\tr \beta )\tr$ is an isotropic random vector like standard Gaussian vectors. Since $\Gamma$ is the sample eigenvector matrix of $\widehat \Sigma$, this requires  $\beta$  to project uniformly onto  the eigenvectors of  $\Sigma$. The idea is extended in the following theorem, drawing on the results from random matrix theory \citep{bai2010spectral}.

Let the eigendecomposition of $\Sigma = \sum^p_{i = 1} \nu_i \ell_i \ell\tr_i$ with eigenvalues $ \nu_1 \geq \nu_2 \geq \ldots \geq \nu_p \geq 0$ and  write the coefficients of $\beta$ in the basis of eigenvectors as $( \ell\tr_1 \beta, \ldots, \ell\tr_p \beta ).$ Let $F(t)$ be the spectral distribution of $\Sigma$,  and $G(t)$ be the corresponding weighted spectral distribution  that puts mass $ (\ell\tr_i \beta)^2/ \| \beta \|^2 $ at the place of $\nu_i$, which are 
\[
F( t) =  p^{-1} \sum^{p}_{i = 1} \1 \{ \nu_i \leq t \} \, \text{ and } \quad G( t) =  \frac{1}{\| \beta \|^{2}} \sum^{p}_{i = 1} (\ell\tr_i \beta)^2 \1 \{ \nu_i \leq t \}.
\]

\begin{Theorem}
\label{thm2}
	Suppose that Assumption \ref{asmp2} holds, $\lambda^{+}_{\min}(\Sigma) \geq c' \lambda_{\max} (\Sigma)$ and $n/p \to c \in (0, 1) \cup (1, \infty)$. If the distribution functions $F(t)$ and $G(t)$ are asymptotically equivalent, i.e.,   for any ${z} \in \mathbb{C}^{+}$,
    \beq
    \label{eigen_cond3}
    \lim_{p \to \infty} \frac{1 }{ \| \beta\|^2} \beta\tr (\Sigma - z \I)^{-1} \beta  -  \frac{1}{p} \Trace\left\{ (\Sigma - z \I)^{-1} \right\} = 0,
    \eeq
      then $S=XX\tr/p$ satisfies conditions \eqref{cond:C1}--\eqref{cond:C2} and the SMILE estimators $(\widehat \theta^2, \widehat \sigma_{\eps}^2)$ that solve (\ref{eq:esteq}) are consistent. 
\end{Theorem}


The condition \eqref{eigen_cond3}, corresponding to the Stieltjes transform of $F$ and $G$,  specifies the structure of $(\beta, \Sigma)$  to ensure the decoupling condition \eqref{cond:C2}.  If $\Sigma = I$ or more generally, $\lambda_{\max} (\Sigma) - \lambda^{+}_{\min} (\Sigma) \to 0$, the condition holds uniformly for all $\beta \in \mathbb{R}^p$ since the eigenvalues and the corresponding eigenvectors of $\Sigma$ are asymptotically exchangeable.  
 
 When $\Sigma \neq I$, the condition  \eqref{eigen_cond3} requires that $\beta$ projects uniformly onto the eigenvectors of $\Sigma$, which means
\[
 \frac{ (\ell\tr_i \beta)^2 }{ \| \beta \|^2 } \approx \frac{1}{p} \text{ uniformly for $i$ from $1$ to $p$.}
\]
The condition is violated if \(\beta\) depends on $\Sigma$ in \eqref{eq:effect} and becomes aligned with a subset of eigenvectors. 
This explains the bias of the REML estimator in
GWAS, where genetic effects are often enriched in low-LD or high-LD regions.



 For some outcomes in GWAS, such as lipids, only a limited number of coordinates of $\beta$ have non-zero effects, i.e., $\beta$ is sparse.
 The condition \eqref{eigen_cond3} can also hold for sparse signals, if the active coordinates spread across weakly correlated genomic regions.
 Let \(\mathcal A=\{j:\beta_j\neq 0\}\) be the support of
\(\beta\). A heuristic interpretation is that when the eigenvectors of $\Sigma$ are  isotropic on the active set $\cA$, i.e.,  for $i$ from $1$ to $p$,
\beq
\label{iso-eigen}
 \E[\ell^2_{ij}] = p^{-1} \text{ and } \E[\ell_{ij} \ell_{i k}] = 0  ~ \text{ for $j, k \in \cA$ and $j \neq k$},
\eeq
condition~\eqref{eigen_cond3} holds with 
\begin{gather*}
(\ell\tr_i \beta)^2 = \sum_{j \in \cA} \ell^2_{ij} \beta^2_j + \sum_{\substack{j, k \in \mathcal{A} \\ j \neq k}} \ell_{ij} \ell_{ik} \beta_{j} \beta_{k} \approx \frac{1}{p} \| \beta\|^2_2.
\end{gather*}
For a dense $\beta$ with a large support $\cA$, \eqref{iso-eigen} is restrictive across many coordinates of the eigenvectors. In contrast, a sparse $\beta$ with a small $\cA$ makes the condition more likely to be satisfied, suggesting the robust performance under the sparse genetic  structure in practice, as evaluated in the empirical experiments in Section \ref{sec:simulation}.

\begin{figure}[!ht]
	\centering
	\includegraphics[width=\linewidth]{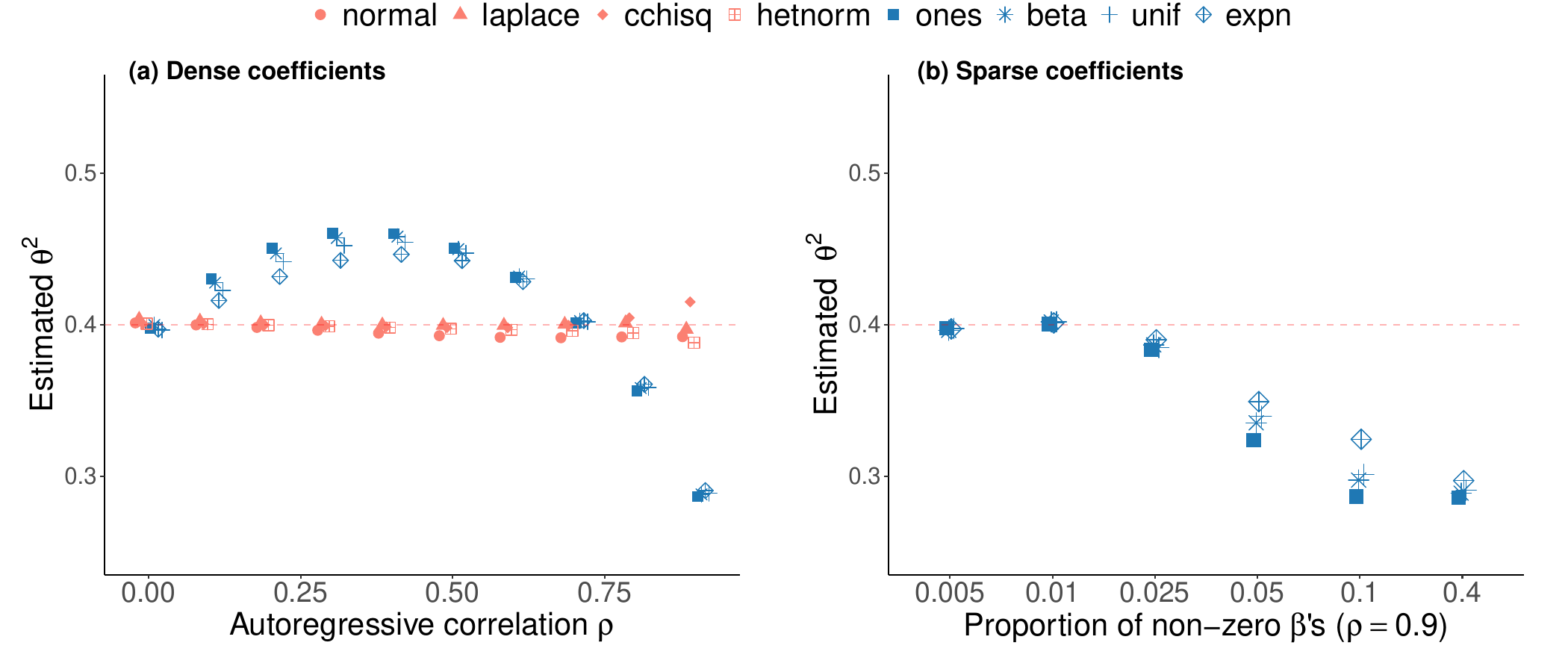}
	\caption{ \it{ \small  The average of the REML estimates $\widehat{\theta}^2$ computed under the classical random-effects model~\eqref{random0},  which is equivalent to the SMILE estimator  $\widehat\theta(S = XX^{\mathsf T})$. The results are obtained from 500 simulated datasets with $(n, p, \theta^2) = (800, 1000, 0.4)$, where $X$ follows a multivariate normal distribution with  autoregressive covariance $\Sigma_{ij} = \rho^{|i - j|}$, and $\rho$ varies from $0$ to $0.9$. The coefficients $\beta$ are generated once and then fixed across simulation replicates. We consider the following approaches to generate $\beta$. In panel (a),  $\beta$ is dense $(|\cA|=p)$ and is generated from several distributions, including mean-zero normal, Laplace, centered $\chi^2(1)$ (cchisq), heterogeneous normal with  increasing variance (hetnorm), and nonzero-mean uniform, beta, exponential (expn), or all-ones (ones) distributions. 
  In panel (b), $x$'s are highly correlated ($\rho$ = 0.9) and  $\beta$  is sparse with evenly spaced support along columns of $X$. The number of non-zero $\beta$'s varies from $5$ to $400$.  } }
 
\label{fig1}
\end{figure} 

 Supporting our discussion, 
Figure \ref{fig1} evaluates the performance of $\widehat{\theta}^{2}\!\bigl(S = XX^{\mathsf T}\bigr)$  using simulation studies, where $\Sigma$ is an autoregressive covariance (AR(1)) with varying correlation strengths and $\beta$ is generated from a range of the regression coefficient distributions. The results show that the estimator is accurate when $x$'s  are independent $(\rho = 0)$ and $\Sigma=I$, or when the entries of $\beta$ are generated as i.i.d. random variables with mean zero. Moreover, even when the mean of $\beta$ is non-zero,  the estimator remains accurate  when $\beta$ is sufficiently sparse with support spread approximately uniformly across coordinates such that its projection onto the oscillatory eigenvectors of the autoregressive $\Sigma$ is close to uniform \citep{sherman2023eigenstructure}.
  
  By contrast, when $\Sigma$ is structured and dense $\beta$ contains a non-zero mean component, the condition in Theorem 3 fails and bias appears. As illustrated in Figure \ref{fig1} (b),  this estimation bias increases with the size of the support.   Thus the key requirement of the SMILE model is not i.i.d. random-effects $\beta$, but the coefficients are evenly distributed across the eigenvectors of $\Sigma$. 
We note such conditions  also appear in high-dimensional ridge-regression prediction analysis, for example, the equidistributed coefficients condition in \cite{hastie2022surprises}. This connection reflects the close relationship between ridge
regression and random-effects modeling.

\ignore{
While $\widehat{\theta}^{2}\!\bigl(S = XX^{\mathsf T}\bigr)$ is equivalent to the REML estimator calculated under the classical random-effects model (\ref{random0}), {\green the results show that the estimator remain unbiased in the settings when $\beta$ does not follow the traditional mean-zero, constant-variance distribution as assumed in the classical random effect model.} {\red This sentence is confusing and contradicts what you want to argue that the REML estimator is generally biased and what is shown in Figure 1. It reads like $\beta$ can have any distribution, the estimate is still unbiased. Also, Figure 1 shows that when $\beta$'s have non-zero means, the estimates are biased as shown in the blue points.} 

 {\green A geometric perspective of the robustness comes from the invariance of the Gram matrix $S=XX\tr$  under the orthogonal transformations or the dimension reduction of the high-dimensional design. Specifically, we have $XX\tr = X' {X'}\tr$ for any 
\[
X' = \PhiS R_{r \times m}~ \text{ with } ~ R_{r \times m} R_{r \times m}\tr = \I_{r}  ~ \text{and} ~ m \geq r.
\] 
Thus, the similarity model  $\E[yy\tr | S=XX\tr]$ stays the same under the generative model of $(y, X')$, making the resulting estimator less dependent on the specific $\beta$. Theorem \ref{thm2} formalizes this property and  generalizes the previous robustness discussion on random-effects  models \citep{dicker2016maximum,jiang2016high,steinsaltz2018statistical}. }
{\red This paragraph is confusing. It contradicts to the discussions of the limitations of the REML estimator, i.e., the REML estimator is not robust and is sensitive the assumption that the distribution of $\beta$ is independent of $X$. Is this paragraph necessary for the key  points you want to make in this section?}
}


As a byproduct, the SMILE framework  specifies the convergence rate of the REML estimator under the high-dimensional genetic random effects model. Specifically, we consider $\rk(\Sigma) > n$ and  $\Delta_n = n \Trace(\Sigma^2)/p^2$.   
\begin{Corollary}
\label{cor:reml}
   Suppose that Assumption \ref{asmp2} holds, $\lambda^{+}_{\min}(\Sigma) \geq c' \lambda_{\max} (\Sigma)$ and $n/p \to c \in (0, 1) \cup (1, \infty)$. In the random-effects model \eqref{random0}, if $\rk(\Sigma) > n$, then we have   
    \beq
|\widehat \theta^2 - \theta^2 | = O_p\left(\frac{ 1}{ \sqrt{n}} + \sqrt{\frac{p^2}{ n^2 \Trace(\Sigma^2) } } \right)  ~\text{ and }  ~ |\widehat \sigma_{\eps}^2 - \sigma_{\eps}^2 | = O_p\left( \frac{ 1}{ \sqrt{n}} + \sqrt{\frac{p^2}{ n^2 \Trace(\Sigma^2) } } \right). 
\eeq
\end{Corollary}

Corollary \ref{cor:reml} specifies  the estimation error of the REML estimator regarding the structure of $\Sigma$ and the dimension $p$. When $\Sigma = \I$, the estimation rate of REML estimator is $\sqrt{p}/n$. For general $\Sigma$, the estimation rate depends on the eigenvalue dispersion $ \Trace(\Sigma^2)$.   When   $\Sigma$ follows a specific autoregressive covariance structure with $\Sigma_{ij} = \rho^{|i - j|}$,
Corollary \ref{cor:reml}  suggests that a stronger autocorrelation (larger $\rho$) yields a smaller estimation variance, as evaluated in  Supplemental Figure \ref{fig:eigenvar}. 


In summary, the consistency of $\widehat \theta^2(S=XX\tr)$ requires that the components of $\beta$ are roughly equally distributed along the eigenvectors of $\Sigma$, which holds when the distribution of the causal genetic variant effects $\beta$ is independent of the LD structure $\Sigma$,  or the SNPs in $X$ are nearly independent $\Sigma \approx I$. %
However, 
natural selection simultaneously influences both the genetic effect sizes $\beta$ and the distribution of genotypes $X$, and therefore $\beta$ may depend on $\Sigma$. Section \ref{sec:method-3}  will propose a genotype similarity representation framework that provides a robust estimation procedure of  ($\theta^2,\sigma_{\epsilon}^2)$ when $\beta$ depends on $\Sigma$ and the genotypes $X$ are correlated.

	 \vspace{-0.1in}
\section{ SMILE estimation  for the heterogeneous effects}
\label{sec:method-3}

 When genetic effects are enriched in low-LD or high-LD regions,  the effect vector $\beta$ depends on  $\Sigma$ and is aligned with the eigenvectors. This dependence  violates the consistency condition when using the standard Gram similarity matrix $S=XX^T$, as in the REML estimator.  SMILE addresses this issue by specifying  $S = X WX\tr$, where $W$ is chosen to 
decouple the eigenvalue distribution of $S$ from the associated regression coefficient $\xi$.

When $W$ is a non-singular fixed weight matrix, define the weighted design, coefficient and re-parameterized model
\[
X_W = X W^{1/2}, \quad \beta_W=W^{-1/2} \beta ~ \text{ and }  X \beta = X_W \beta_W.
\]
Then $S = XWX\tr =  X_W X_W\tr$ is the Gram matrix  in the re-parameterized model. Applying Theorem \ref{thm2} to the re-parameterized model $X_W \beta_W$ suggests that one may choose $W$ such that $\beta_W$ is equally distributed along the eigenvectors of population covariance matrix $\Sigma_{W} = W^{1/2} \Sigma W^{1/2}$ i.e., for any $z \in \mathbb{C}^{+}$,
\begin{align}
    \label{eigen_cond_W}
    \lim_{p \to \infty} \frac{1 }{ \| \beta_W\|^2} \beta_W\tr (\Sigma_W - z \I)^{-1} \beta_W  -  \frac{1}{p} \Trace\left( (\Sigma_W - z \I)^{-1} \right) = 0.
\end{align}
 A natural choice of $W$ that satisfies the weighted spectral condition (\ref{eigen_cond_W})   is 
the true precision matrix  $W=\Sigma^{-1}$ and with $ \Trace( X \Sigma^{-1} X\tr) \approx np$, 
\begin{align}
\label{true}
 S =   X \Sigma^{-1} X\tr/p. 
\end{align}
With this specification, we remove the dependence structure in $X_W$,  i.e., $\Sigma_{W}=\I$.

Nevertheless, since accurately estimating $\Sigma^{-1}$ is challenging in practice, we provide several examples of data-adaptive $W$ that do not require an accurate estimation of the precision matrix $\Sigma^{-1}$.  

\vspace{-0.05in}
  
\subsection{ Decorrelated similarity matrix}
The first example specifies $W$ as the generalized inverse of the sample covariance matrix $\widehat{\Sigma}^{-1}$ and the decorrelated similarity matrix is
\begin{align} 
\label{ginv}
\Smat = \frac{ X \widehat \Sigma ^{-1} X\tr }{ \rk(\widehat \Sigma )}. 
\end{align}
One can take \eqref{ginv}  as the empirical version of \eqref{true}.  Our empirical results show that accurate SMILE estimation  can still be achieved in practice, even though the empirical inverse is different from the true precision matrix.
Corollary \ref{thm:pusedoinv} considers the sample covariance matrix $\widehat \Sigma = n^{-1}X\tr X$ and $\rk(\widehat \Sigma ) = \rk(X)$. 
 
\begin{Corollary}
\label{thm:pusedoinv}
  Under Assumption \ref{asmp2} and $ \rX/n \to c \in (0, 1)$, the similarity matrix $S = nX (X\tr X)^{-1} X\tr /\rX$  satisfies conditions \eqref{cond:C1} -- \eqref{cond:C2} and the SMILE estimators that solve (\ref{eq:esteq}) are consistent with $\Delta_n = 1 - \rX/n$. 
\end{Corollary}

In Corollary \ref{thm:pusedoinv}, $S$  is idempotent (up to a constant) with all non-zero eigenvalues equal to $n/\rX$. Hence $\xi$ is uncorrelated with the eigenvalues. Since $X (X\tr X)^{-1} X\tr   = U U \tr$, the SMILE estimators $(\widehat \theta^2, \widehat \sigma^2_{\eps})$  have simple closed forms
\begin{align}
	\label{sp1}
	\widehat \theta^2 =  \frac{  y\tr U U \tr y - \frac{ \rX }{n} y\tr y }{n- \rX} 
    ~ \text{ and } ~
	&
    \widehat \sigma^2_{\eps}  = \frac{y\tr y - y\tr U U \tr y}{n-\rX}. 
\end{align}
The  estimators above are equivalent to  the
 variance component estimators $\widehat\sigma_{\epsilon}^2 = \hat \epsilon\tr \hat \epsilon/ (n - \rX)$ based on the estimated residuals $\hat \epsilon  = (\I - U U\tr )y$  and $\widehat \theta^2 = y\tr y/n -  \widehat \sigma^2_{\eps}$. 

The size of $\Delta_n$ suggests that estimators are efficient when $p$ is much smaller than $n$ or $X$ is a low-rank design, such as in the local heritability analysis within a chromosome or a genetic region. 
When \(p > n\) and $X$ has full rank, $\rX = n$ and \(\Delta_n=0\). In this case, the decorrelated similarity   $S$ is spectrally uninformative. 

To increase $\Delta_n$ of the decorrelated similarity matrix in the large $p$ setting, one extension is to incorporate the supplementary genotype data.   For example, one may construct a more informative $S$ by pooling the in-sample and external genotypes through $\widehat \Sigma = (X\tr X + X\tr_{E} X_{E})/(n+n_{E})$ where $X_E$ is supplementary genotype data  from the same distribution with sample size $n_E$.  
We note that in this case, there is no analytical expression due to the complexity of $ (\theta^2 S + \sigma^2_{\eps} \I)^{-1}$, and the SMILE estimators are obtained numerically.  

Another way to handle the high-dimensional GWAS data is to use the prior knowledge of the approximate LD block structure, as discussed in the next section.

\subsection{Block-weighted similarity matrix} 
\label{sec:herit}
 To improve computational scalability for analyzing large biobank data  and accounting for the fact that the number of SNPs is often larger than the sample size in GWAS studies, i.e., $p>n$, one can also specify a locally re-weighted similarity matrix by leveraging the structure information of both $\Sigma$ and $\beta(\Sigma)$. Specifically, SNPs across the genome show an approximate LD block structure \citep{pickrell2016detection}.   This suggests that the genetic effects only depend on the local covariance  of nearby variants.   
Therefore  we  divide  $X$ into $M$   blocks $X = ( X_{(1)},\ldots,X_{(M)}  )$, e.g., according to the approximate LD blocks,  and specify $W$ as a  block diagonal matrix of the generalized inverse of each block covariance matrix, i.e., 
\begin{align}
\label{block}
    W = \mathrm{diag} \left (
 	 \widehat \Sigma_{(1)} ^{\dagger},\cdots,
  \widehat \Sigma_{(M)} ^{\dagger} \right ).
\end{align}
Denote the rank of $\Sigma_{(m)}^{-1}$ 
by $r_m$ and $ \rk(W) =  \sum^{M}_{m=1} r_m$. 
The genotype similarity matrix can then be specified as 
\[
S = \frac{ X W X\tr }{  \rk(W) }  = \frac{ \sum^M_{m = 1}  X_{(m)}  \widehat \Sigma^{\dagger}_{(m)}  {X\tr_{(m)}} }{ \sum^{M}_{m=1} r_m }. 
\]
 As suggested in Section \ref{sec:REML}, the weighted spectral condition \eqref{eigen_cond_W} is more likely to hold when 
 the transformation induced by $W$ preserves the local structure of causal effects rather than spreading them across the genome. In particular, if \(\beta\) is sparse, \(\beta_W\) should
remain sparse after transformation.   Thus, a practical principle is to construct $W$ in \eqref{block} using LD blocks.

  Our numerical results in Section \ref{sec:simulation} show that the  SMILE estimator constructed using \eqref{block} allows weak between-block correlations and is fairly robust to the estimated approximate LD block structures \citep{pickrell2016detection} in practice.
A numerical example in the Supplementary Figure \ref{fig:W} shows that even a simple equal block division could greatly reduce the bias.

Regarding the practical construction of the LD blocks, we recommend initiating LD block estimation using the  existing methods, e.g., \cite{pickrell2016detection},  and refining it through empirical evaluation to improve robustness. For traits with enriched effects in long-range LD regions, one may merge nearby blocks within those regions and simulate the likely genetic architecture to evaluate whether estimation performance is robust. 

 We also consider a simpler block-sum estimator $\hs$ that is   derived under the assumption of independent LD blocks by summing the heritability estimators from individual LD blocks by applying \eqref{sp1} to each block,  as given below,  
 \beq
 \label{eq:block-sum}
 \hs = \sum^{M}_{m = 1} \widehat \theta^2( S=X_{(m)}  {\widehat \Sigma_{(m)} }^{\dagger} {X\tr_{(m)}} ).
\eeq
Our numerical results show that this block sum estimator is sensitive to the LD block division and yields a large upward estimation bias due to block dependency, as shown in Supplementary Figure \ref{fig:W}. Therefore, $\hs$ needs to specify large blocks by incorporating all correlated variants in a block, e.g., a whole chromosome. Nevertheless, larger blocks increase $r_m$ and inflate the variance of each blockwise estimate $\widehat \theta^2_m$ in \eqref{sp1}, and therefore require a larger sample size for stable estimation.

\section{Practical implementation to the large-scale data}
\label{sec:implement}
\subsection{ Covariance approximation of the SMILE estimator}
For the SMILE estimator obtained using the estimating equations \eqref{eq:esteq}, 
the first-order Taylor approximation shows that
the asymptotic covariance is
\beq
\label{eq:sandwitch}
\cov\begin{pmatrix}
	\hat \theta^2 \\
	\hat \sigma^2_{\eps}
\end{pmatrix} = \E[\dot{\Psi} | S ]^{-1} \E[\Psi \Psi\tr | S] \E[ \dot{\Psi} | S ]^{-1} ,
	\eeq
	with the non-singular derivative matrix $\dot{\Psi} = \partial \Psi/ \partial (\theta^2, \sigma^2_{\eps})\tr $.  
	 Nevertheless, calculation of $\E[\dot \Psi | S]$ and $\E[\Psi \Psi\tr | S]$, as shown in the supplemental material  \ref{append:likelihood}, involve estimation of the second- and fourth-order moments of $y|\Smat$,  which is unstable in practice. 
     
     A practically effective covariance approximation is obtained by assuming the working Gaussian model \eqref{eq:working-gaus}  i.e., $y|\Smat \sim N(0, \theta^2 S + \sigma^2_{\eps} I ).$  The covariance \eqref{eq:sandwitch} is then simplified  as, when $\theta^2 > 0$,
	\beq
	\label{eq:mle}
    \small
	 \Cov\begin{pmatrix}
		\hat \theta^2 \\
		\hat \sigma^2_{\eps}
	\end{pmatrix} = 
 \frac{2 \theta^4 }{ n \Trace( V^{-2})  - \Trace^2(V^{-1}) } 
 \begin{pmatrix}
	\Trace(V^{-2}) 
		& - \Trace( V^{-1} S V^{-1})   \\
		- \Trace( V^{-1} S V^{-1})   & \Trace( V^{-1} S V^{-1} S )
	\end{pmatrix}.
	\eeq
For large biobank GWAS applications, we use \eqref{eq:mle} for estimating the covariance of the SMILE estimator and constructing its confidence intervals. We provide empirical validation of the covariance approximation in Section \ref{sec:simulation}, and further discussion in Section \ref{sec:discuss}.

\subsection{Numerical implementation}

Given the equivalence between the SMILE estimating equations and the Gaussian score equations under the working Gaussian model
\eqref{eq:working-gaus}, we compute the positive estimates 
$(\widehat \theta^2,\widehat \sigma^2_{\eps})$ of
$\Psi_n(\theta^2,\sigma^2_{\eps})=0$ in \eqref{eq:esteq} by applying
Fisher scoring to the equivalent Gaussian score equations $\Psi^{\rm G}_n $, as given in Algorithm \ref{alg1}.  The estimation procedure only requires iterative updates of  
\beq
\label{eq:simp}
b_y = (\I +  \gamma S)^{-1} y, \quad \tf = \Trace \left\{(\I +  \gamma S)^{-1} \right\} \text{ and } \ts = \Trace \left\{ (\I +  \gamma S)^{-2} \right\}.
\eeq 


Large-scale datasets, such as the UK Biobank data, present new computational challenges. Genotypes, coded as $\{0, 1, 2 \}$, are easy to store, while the elements of $S$  are floating-point numbers and need more storage space. Calculating the $n \times n$ inverse  $(I + \gamma S)^{-1}$  also becomes difficult. 
To avoid computational difficulties, $b_y$ is computed using the conjugate gradient method \citep{nocedal2006numerical}, which only needs the matrix-vector multiplication for \( X
\) and $W$.  This numerical calculation is efficient when $W$ is sparse and $ S=XWX\tr$ has a small condition number.

{\small  
\begin{algorithm}[H]
	\DontPrintSemicolon
	\SetKwInOut{Input}{Input}
	\SetKwInOut{Output}{Output}
	\SetAlgoNoLine
	\KwIn{\( (y, X, W) \) and initial values $ (\theta^2(0), \sigma^2_{\eps}(0) ).$ }
	\KwOut{Estimated $(\hat \theta^2, \hat \sigma^2_{\eps})$ and their estimated covariance.}
	\caption{Estimation and inference of $\theta^2$ with large-scale data }
	
	
	\For{\(k = 0, 1, \ldots\)}{
		Calculate $b_y(k)$ using the conjugate gradient method.
		
		Calculate $\tf(k)$ and $\ts(k)$ with randomized trace estimators.
		

	Obtain $\Psi^{\rm G}(k)$ and the corresponding fisher information matrix $\cI(k)$ under the working Gaussian model as   
    \begin{align*}
    \small
		\Psi^{\rm G} (k)	& = \begin{pmatrix}
                 -\frac{1}{ \gamma \sigma^2_{\eps}} (n - t^{(1)} ) +  \frac{1}{ \sigma^4_{\eps} \gamma } \left( y\tr b_y - b_y \tr b_y  \right),
				\\
				-\frac{1}{\sigma^2_{\eps} } t^{(1)} + \frac{1}{ \sigma^4_{\eps} } b_y \tr b_y, \\
                \end{pmatrix} \\
				\cI(k) & = \begin{pmatrix}
					(\ts - 2\tf + n)/\theta^4   &  ( t^{(1)} - t^{(2)} ) / (\theta^2 \sigma^2_\eps)  \\
					{ ( t^{(1)} - t^{(2)} ) }/ (\theta^2 \sigma^2_\eps)  & t^{(2)}/\sigma^4_{\eps}  
				\end{pmatrix}.
			\end{align*}

		Update the estimators until convergence:
		\[\small
		\begin{pmatrix}
			\theta^2{(k+1)} \\
			\sigma^2_{\eps}(k+1)  
		\end{pmatrix} = \begin{pmatrix}
			\theta^2{(k)} \\
			\sigma^2_{\eps}(k)  
		\end{pmatrix} + \cI^{-1}(k)  \Psi^{\rm G} (k).   
		\]
	}

    \label{alg1}
\end{algorithm}
}

\ignore{
\small  
	\begin{algorithm}[H]
		\DontPrintSemicolon
		\SetKwInOut{Input}{Input}
		\SetKwInOut{Output}{Output}
		\SetAlgoNoLine
		\KwIn{\( (y, X, W) \) and initial values $ (\theta^2(0), \sigma^2_{\eps}(0) ).$ }
		\KwOut{Estimated $(\hat \theta^2, \hat \sigma^2_{\eps})$ and their estimated covariance.}
		\caption{Estimation and inference of $\theta^2$ with large-scale data }
		
		
		\For{\(k = 0, 1, \ldots\)}{
			Calculate $b_y(k)$ using the conjugate gradient method.
			
			Calculate $\tf(k)$ and $\ts(k)$ with randomized trace estimators.
			
			
			Update the estimators until the convergence:
			\[\small
			\begin{pmatrix}
				\theta^2{(k+1)} \\
				\sigma^2_{\eps}(k+1)  
			\end{pmatrix} = \begin{pmatrix}
				\theta^2{(k)} \\
				\sigma^2_{\eps}(k)  
			\end{pmatrix} + \begin{pmatrix}
					\frac{y\tr b_y}{ \theta^2(k) } - n & \frac{ t^{(1)} }{\sigma^2_{\eps}(k) }  \\
					\frac{ \tf - \ts }{\theta^2(k)   \sigma^2_{\eps} (k) } & \frac{ \ts }{ \sigma^4_{\eps}(k) }  
\end{pmatrix}^{-1}   \begin{pmatrix}
					\frac{y\tr b_y}{\sigma^2_{\eps}(k)} - n \\
					\frac{b_y\tr b_y}{\sigma^4_{\eps}(k) } - \frac{t^{(1)}}{\sigma^2_{\eps}(k)} 
				\end{pmatrix} .   
			\]
		}
		
		\label{alg1}
	\end{algorithm}
}

The trace terms $(\tf, \ts)$ involve the eigenvalue of $S$. When $p < n$, they can be calculated from the $p \times p$ matrix $WX\tr X$ instead. When both $p$ and $n$ are large, the trace terms are approximated through a  Monte Carlo trace estimator by $\E[z\tr V^{-1} z] = \Trace(V^{-1}) $  and $\E[ z\tr V^{-1} V^{-1} z  ] = \Trace(V^{-2})$ for a Gaussian random vector $z\sim N(0, \I)$ or a Rademacher random vector $z \sim Unif\{\pm 1\}^n$ \citep{hutchinson1989stochastic},  and $ V^{-1} z$ can be efficiently calculated using the conjugate gradient method.

\section{Simulation study using the UKBB genome-wide genotype data}
\label{sec:simulation}

This section performs simulation studies to evaluate the finite sample performance of the proposed SMILE method and compares it with existing methods. 
To mimic real  GWAS data, we simulate phenotypic/outcome data $Y$ using the UK Biobank genotype array data under various genetic architectures. We sample a set of $N = 30{,}000$ unrelated British individuals of European ancestry with $p = 593{,}300$ SNPs that have minor allele frequency (MAF) greater than $0.01$. In each simulation setting, we repeat the experiments $300$ times, and in each repetition, we randomly select 12,000 individuals to generate the outcome.

The continuous outcome is generated from the linear model $ y_i =  X_i^T \beta + \varepsilon_i$ with $\Var(y_i) = 1$ and   heritability  $h^2 = \theta^2 = 0.2.$ The error term \(\varepsilon_i\) is assumed to follow a normal distribution with $\E(\varepsilon_i)  = 0$ and $\sigma^2_{\eps} = 1 - \Qbeta$. To simulate sparse and non-sparse genetic architectures, we select a causal variant set $\cA$ with varying non-zero elements  $|\cA| \in \{ 10, 10^2, 10^3, 10^4, 10^5 \}$.

Considering possible dependence between genetic effects \(\beta\) and the genotype
distribution, we simulated four genetic architectures:  (a) randomly located
causal variants (randomly selected $\cA$) with normal effects $\beta_j \sim N(0, 1)$; (b) randomly located causal variants with
larger effects assigned to low-LD and low-MAF variants based on their LD score \citep{bulik2015ld}; (c) randomly located causal
variants with larger effects assigned to high-LD and high-MAF variants; and (d)
region-dependent effects concentrated in the MHC region, with effect sizes aligned
with the local LD eigenstructure. The effect sizes are rescaled so that \(h^2=0.2\). Full details of the data-generating mechanisms are given in the Supplementary Material \ref{supp_sec:setup}.

\ignore{
Considering the potential dependency between the genetic effects $\beta$ and the  distribution of genotypes $X$, four different effect distributions are simulated as follows: 

\begin{itemize}	
	\item[(a)] {\bf Random location and random normal effects:} $\cA$ is selected randomly and  non-zero coefficients $\beta_j \sim N(0, 1)$ for $j \in \cA.$ The coefficients are then re-scaled so that $\Qbeta = 0.2.$ 
	
	\item[(b)] {\bf Random location and low-LD dependent effects:} Randomly selected $\cA$ and variants with low LD and low MAF tend to have large effects.    Specifically, for the $j$th variant,  we define the weight $\omega_j = \{f_j(1- f_j)\}^{-0.75}/\ell_j,$ where $f_j$ is the effect  allele frequency and $\ell_j$ is the corresponding LD score \citep{bulik2015ld}. Then $ \beta_j$ is a realization from the normal distribution $ N(0, \omega^2_j ) $  for $j \in \cA.$ 

	\item[(c)] {\bf Random location and high-LD dependent effects:}  Variants with high LD and high MAF tend to have large genetic effects, with the weight $\omega_j = \{f_j(1 - f_j)\}^{0.75} \times \ell_j.$ 
	
	\item[(d)] {\bf Region-dependent location and effects:}  $\cA$ is sampled from variants in MHC region (Chr6:$2.55$M - $3.35$M) with the sampling probability $\P(\beta_j \neq 0) = 1/[1 + \exp\{2 (\ell_j - \bar \ell ) \}]$ where $\bar \ell$ is the median of LD score. Once $\cA$ is determined, the elements of $\beta_\cA$ are assigned according to the entries of the $1000$th eigenvector of the LD matrix in the MHC region.

\end{itemize}
}

\ignore{
 To calculate the SMILE estimator,  we first standardize the genotype data and specify $W$ as a block diagonal matrix that is the inverse of the LD matrix, as discussed in Section \ref{sec:herit}.  Specifically, we modify LD blocks from  \cite{pickrell2016detection} by merging the LD blocks within the long-range LD regions specified in \cite{price2008long} to capture the main LD structure. Some LD blocks are weakly correlated with each other, and are allowed by our method. Finally, we have 1683 blocks in total, and vary the block sizes, where the block of the MHC region on the chromosome 6   
 contains the largest number of  $7443$ SNPs.   
   In each block, we specify the generalized inverse of the LD matrix by inverting the eigenvalues that collectively explain more than $99.5\%$ of the total variance. 
   
   The calculation of  $W$ in \eqref{block} could be conducted using the genotypes of all $30K$ subjects, or the $12K$ in-sample individuals that match to the outcomes.  Their performances are similar, as we reported in the Supplementary Table \ref{tab:supp-cover-1} and \ref{tab:supp-cover-2}. In what follows, we report the results based on the $W$ using the genotypes of 30K subjects.}

To compute the SMILE estimator, we construct \(W\) as a block-diagonal matrix of
generalized inverses of within-block LD matrices, as described in
Section~\ref{sec:herit}. The block partition modifies the LD blocks of
\cite{pickrell2016detection} and yields 1683 blocks in total. Details are given in
Supplementary Section~\ref{supp_sec:setup}. 
 We compare the performance of SMILE with three existing methods: GCTA \citep{yang2010common} based on the classical random-effects model, LDAK-thin \citep{speed2012improved} based on the weighted random-effects model and LDMS \citep{yang2015genetic} based on the stratified random-effects model with SNPs stratified into 4 LD groups. 

\begin{figure}
\centering
 \includegraphics[width=0.8\linewidth]{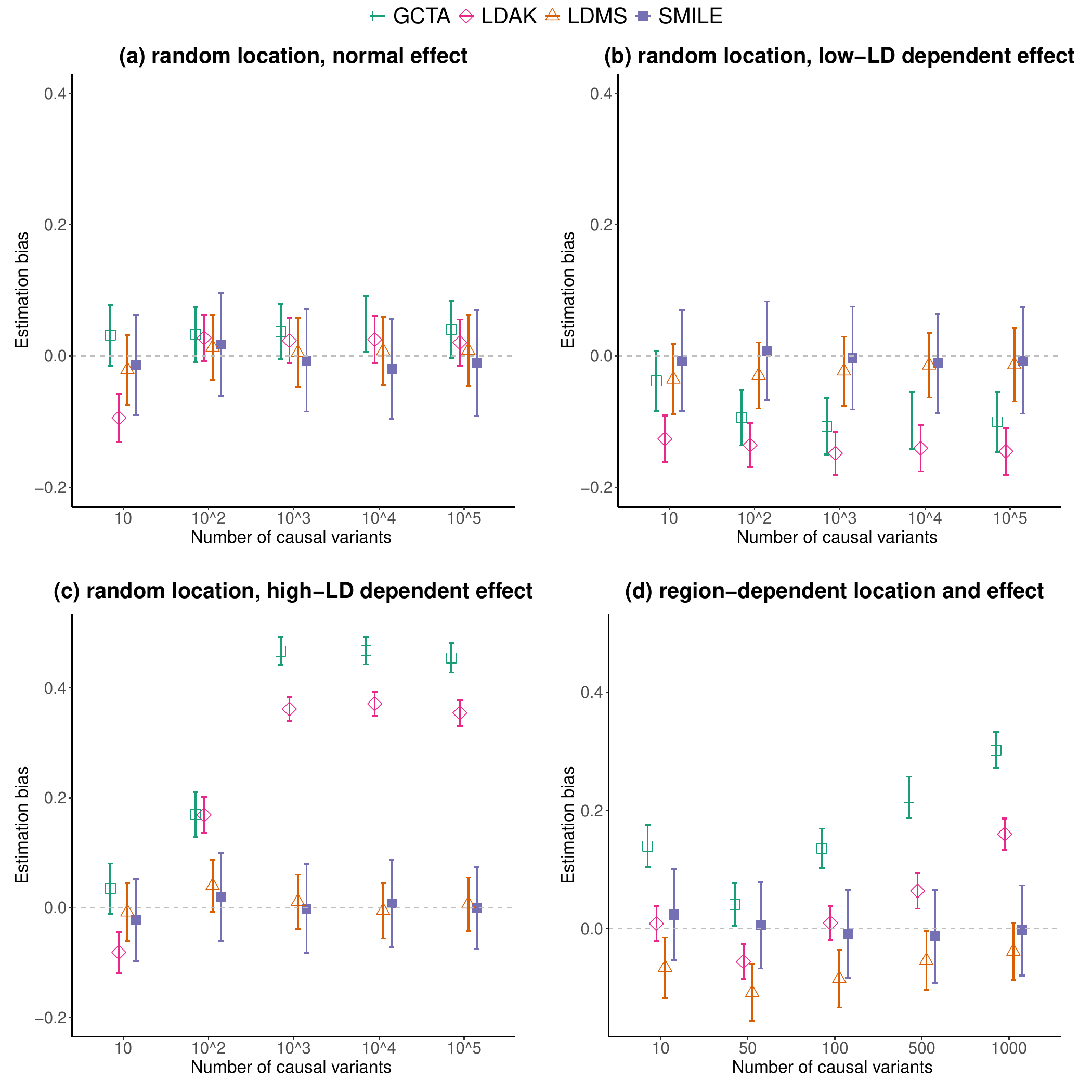} 
	\caption{	\small \it{ Estimation bias under four genetic architectures. Points are average
biases over 300 replicates, and error bars denote empirical standard errors.} }
\label{fig:est} 
\end{figure}

 \ignore{
\begin{table}
\centering
\caption{ 	
	 Comparison of the coverage probabilities and the confidence interval (CI) lengths in   different simulation settings.  Each row reports the empirical coverage probabilities, with the nominal level $0.95.$  "Normal" denotes the setting where $\beta_j$'s are normally distributed. "L-LD" denotes the setting where the large effects are enriched in low-LD regions, and "H-LD" denotes the setting where the large effects are enriched in high-LD regions. "Region" denotes the setting where the effect ($\beta$) locations and sizes are region-dependent. The setting column represents the number of non-zero $\beta$'s. 
}
\label{tab:cover}
	\resizebox{\textwidth}{!}{
		\small
\begin{tabular}{l l r r r r r r r r}
	\toprule
	& & \multicolumn{4}{c}{Normal} & \multicolumn{4}{c}{L-LD} \\
	\midrule
	 & $|\cA|$ & Bias & Emp. SE & Est. SE & Cover & Bias & Emp. SE & Est. SE & Cover \\
	\midrule
	\Proposed & 10 & -0.014 & 0.078 & 0.076 & 0.950 & -0.007 & 0.078 & 0.077 & 0.943\\
 & $10^2$ & 0.017 & 0.078 & 0.079 & 0.953 & 0.008 & 0.077 & 0.075 & 0.967\\
 & $10^3$ & -0.007 & 0.078 & 0.078 & 0.960 & -0.003 & 0.078 & 0.078 & 0.950\\
	\midrule
	GCTA & 10 & 0.032 & 0.046 & 0.043 & 0.867 & -0.038 & 0.046 & 0.043 & 0.847 \\
	& $10^2$  & 0.033 & 0.042 & 0.043 & 0.903 & -0.094 & 0.042 & 0.042 & 0.387 \\
  & $10^3$ & 0.038 & 0.042 & 0.042 & 0.837 & -0.107 & 0.041 & 0.043 & 0.283\\
	\midrule
	LDAK & 10 & -0.094 & 0.037 & 0.035 & 0.250 & -0.126 & 0.036 & 0.034 & 0.047 \\
	& $10^2$ & 0.028 & 0.035 & 0.036 & 0.893 & -0.136 & 0.033 & 0.034 & 0.020 \\
  & $10^3$ & 0.023 & 0.035 & 0.034 & 0.883 & -0.148 & 0.034 & 0.033 & 0.007\\
	\midrule
	LDMS & 10 & -0.021 & 0.052 & 0.053 & 0.927 & -0.035 & 0.052 & 0.054 & 0.873\\
 & $10^2$ & 0.013 & 0.052 & 0.049 & 0.940 & -0.030 & 0.052 & 0.050 & 0.895\\
 & $10^3$ & 0.005 & 0.052 & 0.053 & 0.933 & -0.023 & 0.052 & 0.053 & 0.910\\
	\toprule
	& & \multicolumn{4}{c}{H-LD} & \multicolumn{4}{c}{Region} \\
	\midrule
	 & $|\cA|$ & Bias & Emp. SE & Est. SE & Cover & Bias & Emp. SE & Est. SE & Cover \\
	\midrule
SMILE & 10 & -0.022 & 0.078 & 0.075 & 0.950 & 0.024 & 0.077 & 0.077 & 0.957\\
 & $10^2$ & 0.020 & 0.078 & 0.080 & 0.953 & -0.009 & 0.078 & 0.075 & 0.953\\
 & $10^3$ & -0.001 & 0.077 & 0.081 & 0.940 & -0.002 & 0.078 & 0.076 & 0.950\\
 \midrule
GCTA & 10 & 0.035 & 0.043 & 0.046 & 0.867 & 0.140 & 0.041 & 0.036 & 0.043\\
 & $10^2$ & 0.170 & 0.043 & 0.041 & 0.023 & 0.136 & 0.040 & 0.034 & 0.043\\
 & $10^3$ & 0.467 & 0.036 & 0.026 & 0.000 & 0.302 & 0.039 & 0.031 & 0.000\\
 \midrule
LDAK & 10 & -0.081 & 0.035 & 0.037 & 0.393 & 0.009 & 0.033 & 0.029 & 0.973\\
 & $10^2$ & 0.169 & 0.036 & 0.033 & 0.003 & 0.010 & 0.033 & 0.028 & 0.983\\
 & $10^3$ & 0.362 & 0.031 & 0.022 & 0.000 & 0.160 & 0.033 & 0.027 & 0.000\\
\midrule
LDMS & 10 & -0.008 & 0.052 & 0.053 & 0.940 & -0.065 & 0.051 & 0.051 & 0.777\\
 & $10^2$ & 0.040 & 0.051 & 0.047 & 0.893 & -0.084 & 0.051 & 0.048 & 0.617\\
 & $10^3$ & 0.011 & 0.050 & 0.049 & 0.950 & -0.038 & 0.050 & 0.048 & 0.880\\
\bottomrule
\end{tabular}
}
\end{table}
}

\begin{table}
\centering
\caption{ 	
Simulation performance under four genetic architectures. Bias and SE are
reported in units of \(10^{-2}\).  Coverage is reported in percent ($\%$) with the nominal level $0.95$.  "Normal" denotes the setting where $\beta_j$'s are normally distributed. "L-LD" denotes the setting where the large effects are enriched in low-LD regions, and "H-LD" denotes the setting where the large effects are enriched in high-LD regions. "Region" denotes the setting where the effect ($\beta$) locations and sizes are region-dependent. The setting column represents the number of non-zero $\beta$'s. 
}
\label{tab:cover}

	\resizebox{0.88\textwidth}{!}{%
		\tiny
\begin{tabular}{@{}l l cccccccc@{}}
\toprule
	& & \multicolumn{4}{c}{Normal} & \multicolumn{4}{c}{L-LD} \\
	\midrule
	 & Setting & Bias & Emp. SE & Est. SE & Cover & Bias & Emp. SE & Est. SE & Cover \\
	\midrule 
    SMILE & 10 & -1.4 & 7.8 & 7.6 & 95.0 & -0.7 & 7.8 & 7.7 & 94.3 \\ 
    & \(10^2\) & 1.7 & 7.8 & 7.9 & 95.3 & 0.8 & 7.7 & 7.5 & 96.7 \\ 
    & \(10^3\) & -0.7 & 7.8 & 7.8 & 96.0 & -0.3 & 7.8 & 7.8 & 95.0 \\ 
    \midrule 
    GCTA & 10 & 3.2 & 4.6 & 4.3 & 86.7 & -3.8 & 4.6 & 4.3 & 84.7 \\ 
    & \(10^2\) & 3.3 & 4.2 & 4.3 & 90.3 & -9.4 & 4.2 & 4.2 & 38.7 \\ 
    & \(10^3\) & 3.8 & 4.2 & 4.2 & 83.7 & -10.7 & 4.1 & 4.3 & 28.3 \\ 
    \midrule 
    LDAK & 10 & -9.4 & 3.7 & 3.5 & 25.0 & -12.6 & 3.6 & 3.4 & 4.7 \\ 
    & \(10^2\) & 2.8 & 3.5 & 3.6 & 89.3 & -13.6 & 3.3 & 3.4 & 2.0 \\ 
    & \(10^3\) & 2.3 & 3.5 & 3.4 & 88.3 & -14.8 & 3.4 & 3.3 & 0.7 \\ 
    \midrule 
    LDMS & 10 & -2.1 & 5.2 & 5.3 & 92.7 & -3.5 & 5.2 & 5.4 & 87.3 \\ 
    & \(10^2\) & 1.3 & 5.2 & 4.9 & 94.0 & -3.0 & 5.2 & 5.0 & 89.5 \\ 
    & \(10^3\) & 0.5 & 5.2 & 5.3 & 93.3 & -2.3 & 5.2 & 5.3 & 91.0 \\
	\toprule
	& & \multicolumn{4}{c}{H-LD} & \multicolumn{4}{c}{Region} \\
	\midrule
	 & Setting & Bias & Emp. SE & Est. SE & Cover & Bias & Emp. SE & Est. SE & Cover \\
	\midrule 
SMILE & 10 & -2.2 & 7.8 & 7.5 & 95.0 & 2.4 & 7.7 & 7.7 & 95.7 \\ 
& \(10^2\) & 2.0 & 7.8 & 8.0 & 95.3 & -0.9 & 7.8 & 7.5 & 95.3 \\ 
& \(10^3\) & -0.1 & 7.7 & 8.1 & 94.0 & -0.2 & 7.8 & 7.6 & 95.0 \\ 
    \midrule 
GCTA & 10 & 3.5 & 4.3 & 4.6 & 86.7 & 14.0 & 4.1 & 3.6 & 4.3 \\ 
& \(10^2\) & 17.0 & 4.3 & 4.1 & 2.3 & 13.6 & 4.0 & 3.4 & 4.3 \\ 
& \(10^3\) & 46.7 & 3.6 & 2.6 & 0.0 & 30.2 & 3.9 & 3.1 & 0.0 \\ 
\midrule 
LDAK & 10 & -8.1 & 3.5 & 3.7 & 39.3 & 0.9 & 3.3 & 2.9 & 97.3 \\ 
& \(10^2\) & 16.9 & 3.6 & 3.3 & 0.3 & 1.0 & 3.3 & 2.8 & 98.3 \\ 
& \(10^3\) & 36.2 & 3.1 & 2.2 & 0.0 & 16.0 & 3.3 & 2.7 & 0.0 \\ 
\midrule LDMS & 10 & -0.8 & 5.2 & 5.3 & 94.0 & -6.5 & 5.1 & 5.1 & 77.7 \\ 
& \(10^2\) & 4.0 & 5.1 & 4.7 & 89.3 & -8.4 & 5.1 & 4.8 & 61.7 \\ 
& \(10^3\) & 1.1 & 5.0 & 4.9 & 95.0 & -3.8 & 5.0 & 4.8 & 88.0 \\
\bottomrule
\end{tabular}
}
\end{table}

\textbf{{ Estimation bias}}:  Figure \ref{fig:est} compares the estimation biases of the  four methods. SMILE  provides approximately unbiased estimates of heritability in each setting,  insensitive to the underlying genetic architectures or the sparsity of effects.   As expected, GCTA, LDAK and LDMS perform well when the effect distribution is close to the homogeneous random-effects assumption. 
 However, when the distribution of $\beta$ depends on the LD, Figures \ref{fig:est} (b) - (d) show that the random-effects methods, GCTA and LDAK, give biased estimates due to the assumption violation.  The partition-based method LDMS has a smaller estimation bias than other methods when $\cA$ is randomly selected but gives a larger bias for region-dependent $\cA$.   
 
 In Table \ref{tab:cover}, another interesting observation is that in (b) - (c), the estimation bias of GCTA and LDAK increases as the size of $\cA$ grows, irrespective of the underlying effect distribution (L-LD or H-LD).   Our framework explains the phenomenon. When $|\cA|$ is small,  randomly selected variants in $\cA$ are approximately uniformly distributed across the genome, and the weighted spectral condition is more likely to hold, regardless of the effect size distribution. In contrast, in MHC regions where the variants are highly correlated, the weighted spectral condition is violated. Therefore, when $\cA$ is region-dependent in (d), the bias gets larger. 

The smaller bias of SMILE comes at the cost of a larger standard error in some settings.  This reflects the bias-variance trade-off induced by the choice of $W$ where decorrelating or locally reweighting the genotype can compress the eigenvalue spectrum of $S$, thereby reducing  $\Delta_n$ in Theorem~\ref{thm:converg}.

\textbf{{Variance estimation and confidence interval}}:  Table \ref{tab:cover} compares the  variances,  coverage probabilities, and the confidence interval lengths of different estimators. SMILE consistently shows better coverage probability performance, close to the nominal level. The empirical and estimated standard errors are also consistent for different effect distribution settings.
The performances of GCTA, LDAK, and LDMS vary widely depending on the settings of how the $\beta_j$'s are generated. When the $\beta_j$'s are not normally distributed, their coverage probabilities are considerably lower than those of  SMILE due to the bias in these estimators, especially in the region-dependency setting.

\section{Heritability Analysis of the UK Biobank GWAS Data}
\label{sec:real-data}
We compute the heritability for a range of phenotypes using the UK Biobank data.   We focus on common variants by excluding the SNPs with the MAFs less than $0.01$. We follow the standard GWAS QC protocol by removing the SNPs with the genotype missingness proportion larger than $0.05$ and the SNPs that fail the Hardy–Weinberg test at a significance threshold $10^{-6}$, and excluding the individuals with the missing genotype proportion exceeding $0.05$. We only consider the SNPs on the autosomes. This gives the total number of SNPs $p = 560{,}965.$ Our analysis focuses on unrelated individuals of the  European ancestry with $n \approx 350{,}000$.

We consider a range of complex traits, biomarkers and blood cell counts. The complex traits include height, body mass index (BMI), systolic blood pressure (SBP), diastolic blood pressure (DBP), high-density lipoprotein cholesterol (HDL), low-density lipoprotein cholesterol (LDL) and HbA1C. 
The biomarkers include apolipoprotein A (APOA), apolipoprotein B (APOB) and cystatin C.  Blood cell count variables include white blood cells (WBC), red blood cells (RBC), platelets (PLT), 
and neutrophils (NEUT). The outcomes are pre-processed by  
adjusting for age, sex, and the top 20 ancestry PCs as covariates in linear regression and taking the residuals for the subsequent heritability estimation.
For the biomarker outcomes, we also adjust BMI as an additional covariate. For the cell counts, we apply a log transformation to the outcome before covariate adjustment. 

We notice that  heritability estimation is sensitive to outcome outliers, e.g., extreme outcome values cause larger $\Var(y_i)$ and lead to a smaller heritability estimation. To reduce the outlier effects, for biomarker analysis, we only keep the residual observations within 5 standard deviations. For cell count analysis, we conduct the inverse normal transformation on the residuals. A complete evaluation of different outcome processing or transformations for heritability estimation is given in the Supplementary Material Section  \ref{sec:stability}.

 {SMILE is implemented by specifying $W$ using the LD block approach described in Section \ref{sec:simulation}. To demonstrate that the proposed method is robust to the block division, for complex traits with extensive genetic architectures, we compare SMILE with the block summation estimator   $\hs$ in \eqref{eq:block-sum}, which estimates the heritability for each block and calculates the total heritability by summing the $1683$ block specific heritability estimates, with the variance calculated as \cite{hou2019accurate}.  
Due to the computational limitations of GCTA and LDMS, the comparison among GCTA, LDMS, and SMILE is performed on the same random subset of $n=160{,}000$ individuals. We additionally report SMILE estimates and the block summation estimator using the full unrelated European-ancestry sample of approximately $n=350{,}000$ individuals. } 
 
\begin{figure}
\centering
\includegraphics[width=0.8\linewidth, height = 0.75 \textheight]{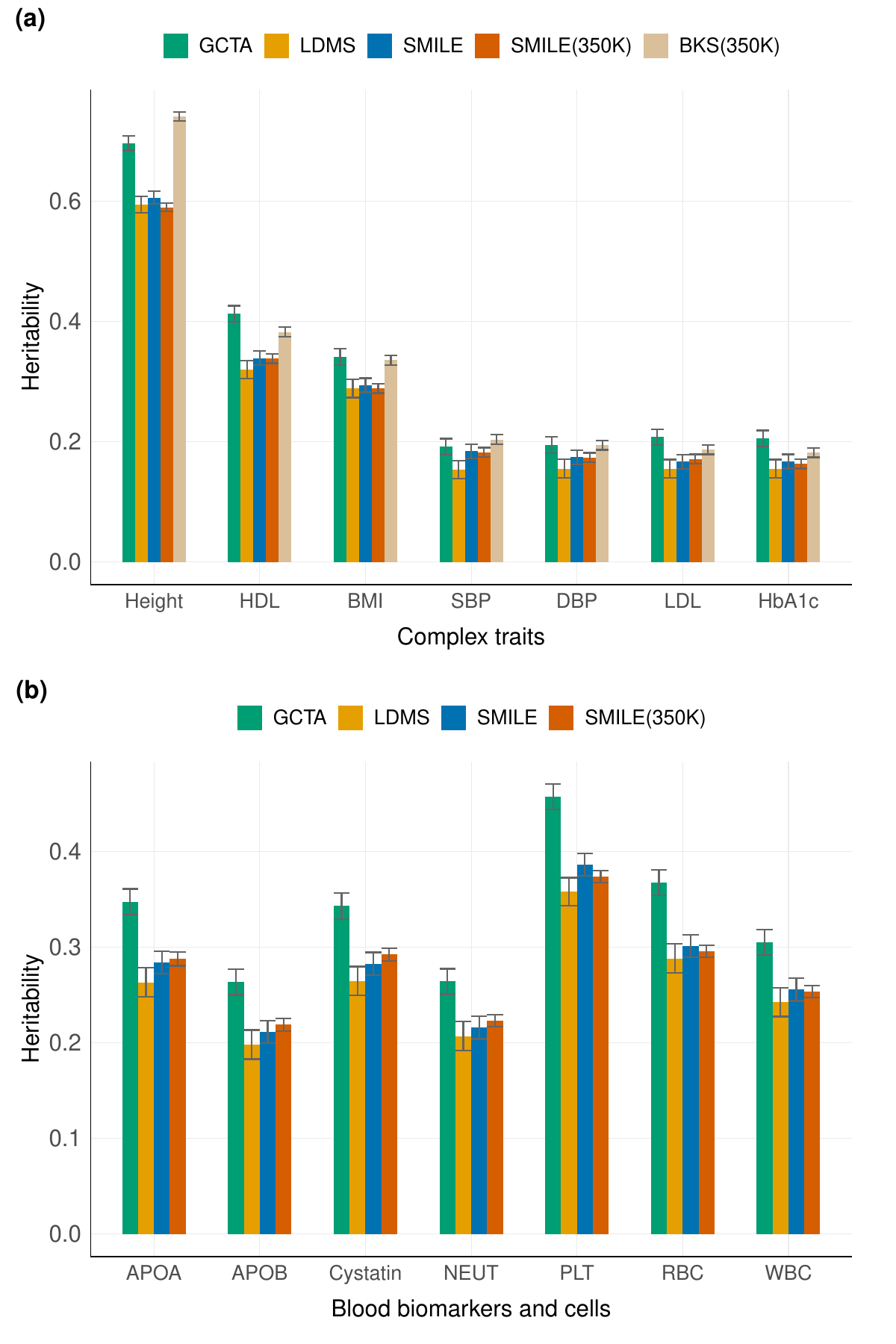}
\caption{ \small \it{Heritability estimation of (a) Complex traits and  (b) Biomarkers and blood cell counts based on the UKBB dataset, where error bars represent the $95\%$ confidence intervals.  GCTA, LDMS and SMILE are evaluated with $n = 160K$ and $p = 560K.$ SMILE and the  block summation estimator (BKS), are also evaluated with the larger dataset $n = 350K.$ } }
\label{fig:heri}
\end{figure}

Figure \ref{fig:heri}(a) shows that height  has the highest heritability ($h^2=0.59$), and HbA1c
has the lowest heritability ($h^2=0.16$). Different methods give a similar ranking for the heritability of 7 outcomes. Nevertheless,  GCTA gives larger heritability estimates than LDMS and SMILE, especially for height and HDL. The larger GCTA estimates are also observed in Figure \ref{fig:heri}(b) for biomarkers and blood cell counts.  
It is not surprising to observe the simple block-sum estimate  $\hs$ overestimates heritability and gives higher estimates because of the block dependency, especially for height, which has 
a highly polygenic genetic architecture.

With respect to the comparison between LDMS and SMILE, the two methods achieve almost the same estimates for height and BMI which are highly polygenic with evenly distributed genetic effects \citep{li2023accurate}. However, the estimates of other outcomes differ, as observed in LDL or SBP.   Meanwhile, the significant difference between LDMS and SMILE is also observed for biomarkers such as APOA, APOB, and cystatin C, due to the unevenly distributed genetic architecture of these outcomes \citep{li2023accurate}. 


\section{Discussion}
\label{sec:discuss}
SMILE is motivated by the practical challenge that causal variant effects are difficult to select or model because many effects are weak, correlated through LD, and heterogeneous under natural selection. To address these limitations, SMILE provides a similarity representation framework for heritability estimation, which directly  models the 
genetic similarity directly through a weighted genotype Gram matrix. 
The framework includes classical
random-effects and fixed-effects methods as special cases, while
avoiding accurate estimation of \(\beta\) or the full precision matrix.  We establish
consistency and convergence-rate results, develop a scalable implementation for
biobank-scale data. 
Simulations and UK Biobank analyses show that SMILE is robust across diverse
genetic architectures.



This work leaves several problems for future research. Inference for SMILE is an important direction. The key is to  quantify the distribution of $y|S.$  
In Section \ref{sec:implement}, we use the working Gaussian model $y|S \sim N(0, \theta^2 S + \sigma^2_{\eps} I)$ for covariance approximation (\ref{eq:mle}). Supplementary
Figure~\ref{fig:eigenvar} evaluates this approximation  for \(S=XX\tr\), which is almost exact when $\Sigma = I$ since $\Gamma$ is Haar-distributed and  $ \xi = \Gamma\tr \beta $ is approximately normal \citep{bai2010spectral}. When $\Sigma$=AR(1),  the approximated covariance gets closer to the empirical variance as the dimension $p$ increases.  Recent advances on
eigenvector distributions in random matrix theory may provide tools for developing
a rigorous inference theory
\citep{johnstone2018pca,ding2024eigenvector}.

Although  SMILE does not require a full characterization of $\beta$, incorporating estimated coefficient information could
improve efficiency and guide better choices of \(W\).  Computationally, the accuracy
of randomized trace estimation depends on the spectrum of \(S\).   Reducing the
condition number through \(W\) improves computation but may compress the spectrum
and reduce statistical efficiency. Thus, choosing \(W\) involves a practical trade-off
among bias, variance, and computation. 

SMILE currently requires individuals from the same population without population stratification so that the inverse of the sample covariance matrix can effectively eliminate the in-sample dependence structure due to population structure. A future work is to extend SMILE to heritability estimation for admixture populations, and for binary case-control phenotypes.

\begin{center}
	{\large \textbf{Supplementary Materials}}
\end{center}

In the Supplementary Material, we provide proof of theorems, propositions and lemmas, derivation details of the proposed statistical method SMILE and additional numerical results.
Comprehensive stability analysis of SMILE under outcome transformation, outlier removal and computation precision is also included.  The R package SMILE is available at https://github.com/JianqiaoWang/SMILE.


\bibliographystyle{abbrvnat}
\bibliography{Heri}

@article{li2023accurate,
  title={Accurate and efficient estimation of local heritability using summary statistics and the linkage disequilibrium matrix},
  author={Li, Hui and Mazumder, Rahul and Lin, Xihong},
  journal={Nature Communications},
  volume={14},
  number={1},
  pages={7954},
  year={2023},
  publisher={Nature Publishing Group UK London}
}

@article{yang2017concepts,
  title={Concepts, estimation and interpretation of SNP-based heritability},
  author={Yang, Jian and Zeng, Jian and Goddard, Michael E and Wray, Naomi R and Visscher, Peter M},
  journal={Nature Genetics},
  volume={49},
  number={9},
  pages={1304--1310},
  year={2017},
  publisher={Nature Publishing Group}
}

@article{janson2017eigenprism,
  Author = {Janson, Lucas and Barber, Rina Foygel and Candes, Emmanuel},
  Journal = {Journal of the Royal Statistical Society: Series B (Statistical Methodology)},
  Number = {4},
  Pages = {1037--1065},
  Publisher = {Wiley Online Library},
  Title = {EigenPrism: inference for high dimensional signal-to-noise ratios},
  Volume = {79},
  Year = {2017}}

@article{dicker2014variance,
  Author = {Dicker, Lee H},
  Journal = {Biometrika},
  Number = {2},
  Pages = {269--284},
  Publisher = {Oxford University Press},
  Title = {Variance estimation in high-dimensional linear models},
  Volume = {101},
  Year = {2014}}

@article{verzelen2018adaptive,
  Author = {Verzelen, Nicolas and Gassiat, Elisabeth and others},
  Journal = {Bernoulli},
  Number = {4B},
  Pages = {3683--3710},
  Publisher = {Bernoulli Society for Mathematical Statistics and Probability},
  Title = {Adaptive estimation of high-dimensional signal-to-noise ratios},
  Volume = {24},
  Year = {2018}}

@article{pickrell2016detection,
  title={Detection and interpretation of shared genetic influences on 42 human traits},
  author={Pickrell, Joseph K and Berisa, Tomaz and Liu, Jimmy Z and S{\'e}gurel, Laure and Tung, Joyce Y and Hinds, David A},
  journal={Nature Genetics},
  volume={48},
  number={7},
  pages={709--717},
  year={2016},
  publisher={Nature Publishing Group}
}

@article{bulik2015ld,
  title={{LD} Score regression distinguishes confounding from polygenicity in genome-wide association studies},
  author={Bulik-Sullivan, Brendan K and Loh, Po-Ru and Finucane, Hilary K and Ripke, Stephan and Yang, Jian and Patterson, Nick and Daly, Mark J and Price, Alkes L and Neale, Benjamin M and Schizophrenia Working Group of the Psychiatric Genomics Consortium and others},
  journal={Nature Genetics},
  volume={47},
  number={3},
  pages={291--295},
  year={2015},
  publisher={Nature Publishing Group}
}

@article{yang2010common,
  title={Common SNPs explain a large proportion of the heritability for human height},
  author={Yang, Jian and Benyamin, Beben and McEvoy, Brian P and Gordon, Scott and Henders, Anjali K and Nyholt, Dale R and others},
  journal={Nature Genetics},
  volume={42},
  number={7},
  pages={565--569},
  year={2010},
  publisher={Nature Publishing Group}
}

@article{hou2019accurate,
  title={Accurate estimation of {SNP}-heritability from biobank-scale data irrespective of genetic architecture},
  author={Hou, Kangcheng and Burch, Kathryn S and Majumdar, Arunabha and Shi, Huwenbo and Mancuso, Nicholas and Wu, Yue and Sankararaman, Sriram and Pasaniuc, Bogdan},
  journal={Nature Genetics},
  volume={51},
  number={8},
  pages={1244--1251},
  year={2019},
  publisher={Nature Publishing Group}
}

@article{jiang2016high,
  title={On high-dimensional misspecified mixed model analysis in genome-wide association study},
  author={Jiang, Jiming and Li, Cong and Paul, Debashis and Yang, Can and Zhao, Hongyu and others},
  journal={Annals of Statistics},
  volume={44},
  number={5},
  pages={2127--2160},
  year={2016},
  publisher={Institute of Mathematical Statistics}
}

@article{cai2018semi,
  title={Semisupervised inference for explained variance in high dimensional linear regression and its applications},
  author={Cai, T Tony and Guo, Zijian},
  journal={Journal of the Royal Statistical Society Series B},
  volume={82},
  number={2},
  pages={391--419},
  year={2020},
  publisher={Royal Statistical Society}
}

@book{bai2010spectral,
	title={Spectral analysis of large dimensional random matrices},
	author={Bai, Zhidong and Silverstein, Jack W},
	volume={20},
	year={2010},
	publisher={Springer}
}

@article{bryson2021marchenko,
	title={Marchenko--Pastur law with relaxed independence conditions},
	author={Bryson, Jennifer and Vershynin, Roman and Zhao, Hongkai},
	journal={Random Matrices: Theory and Applications},
	volume={10},
	number={04},
	pages={2150040},
	year={2021},
	publisher={World Scientific}
}

@article{loh2015efficient,
	title={Efficient Bayesian mixed-model analysis increases association power in large cohorts},
	author={Loh, Po-Ru and Tucker, George and Bulik-Sullivan, Brendan K and Vilhjalmsson, Bjarni J and Finucane, Hilary K and Salem, Rany M and Chasman, Daniel I and Ridker, Paul M and Neale, Benjamin M and Berger, Bonnie and others},
	journal={Nature Genetics},
	volume={47},
	number={3},
	pages={284--290},
	year={2015},
	publisher={Nature Publishing Group US New York}
}

@article{yang2015genetic,
	title={Genetic variance estimation with imputed variants finds negligible missing heritability for human height and body mass index},
	author={Yang, Jian and Bakshi, Andrew and Zhu, Zhihong and Hemani, Gibran and Vinkhuyzen, Anna AE and Lee, Sang Hong and others},
	journal={Nature Genetics},
	volume={47},
	number={10},
	pages={1114--1120},
	year={2015},
	publisher={Nature Publishing Group US New York}
}

@article{finucane2015partitioning,
	title={Partitioning heritability by functional annotation using genome-wide association summary statistics},
    author={Finucane, Hilary K and Bulik-Sullivan, Brendan and Gusev, Alexander and Trynka, Gosia and Reshef, Yakir and Loh, Po-Ru and others},
	journal={Nature Genetics},
	volume={47},
	number={11},
	pages={1228--1235},
	year={2015},
	publisher={Nature Publishing Group US New York}
}

@article{reid2016study,
	title={A study of error variance estimation in lasso regression},
	author={Reid, Stephen and Tibshirani, Robert and Friedman, Jerome},
	journal={Statistica Sinica},
	pages={35--67},
	year={2016},
	publisher={JSTOR}
}

@article{gazal2017linkage,
	title={Linkage disequilibrium--dependent architecture of human complex traits shows action of negative selection},
	author={Gazal, Steven and Finucane, Hilary K and Furlotte, Nicholas A and Loh, Po-Ru and Palamara, Pier Francesco and others},
	journal={Nature Genetics},
	volume={49},
	number={10},
	pages={1421--1427},
	year={2017},
	publisher={Nature Publishing Group US New York}
}

@article{johnstone2018pca,
	title={PCA in high dimensions: An orientation},
	author={Johnstone, Iain M and Paul, Debashis},
	journal={Proceedings of the IEEE},
	volume={106},
	number={8},
	pages={1277--1292},
	year={2018},
	publisher={IEEE}
}

@inproceedings{dicker2016maximum,
  title={Maximum likelihood for variance estimation in high-dimensional linear models},
  author={Dicker, Lee H and Erdogdu, Murat A},
  booktitle={Artificial Intelligence and Statistics},
  pages={159--167},
  year={2016},
  organization={PMLR}
}

@article{pritchard2001rare,
	title={Are rare variants responsible for susceptibility to complex diseases?},
	author={Pritchard, Jonathan K},
	journal={The American Journal of Human Genetics},
	volume={69},
	number={1},
	pages={124--137},
	year={2001},
	publisher={Elsevier}
}

@article{zeng2018signatures,
	title={Signatures of negative selection in the genetic architecture of human complex traits},
	author={Zeng, Jian and De Vlaming, Ronald and Wu, Yang and Robinson, Matthew R and Lloyd-Jones, Luke R and others},
	journal={Nature Genetics},
	volume={50},
	number={5},
	pages={746--753},
	year={2018},
	publisher={Nature Publishing Group US New York}
}

@article{haseman1972investigation,
	title={The investigation of linkage between a quantitative trait and a marker locus},
	author={Haseman, JK and Elston, RC},
	journal={Behavior Genetics},
	volume={2},
	number={1},
	pages={3--19},
	year={1972},
	publisher={Springer}
}

@article{price2008long,
	title={Long-range LD can confound genome scans in admixed populations},
	author={Price, Alkes L and Weale, Michael E and Patterson, Nick and Myers, Simon R and Need, Anna C and Shianna, Kevin V and Ge, Dongliang and Rotter, Jerome I and Torres, Esther and Taylor, Kent D and others},
	journal={The American Journal of Human Genetics},
	volume={83},
	number={1},
	pages={132--135},
	year={2008},
	publisher={Elsevier}
}

@article{speed2012improved,
	title={Improved heritability estimation from genome-wide SNPs},
	author={Speed, Doug and Hemani, Gibran and Johnson, Michael R and Balding, David J},
	journal={The American Journal of Human Genetics},
	volume={91},
	number={6},
	pages={1011--1021},
	year={2012},
	publisher={Elsevier}
}

@article{steinsaltz2018statistical,
	title={Statistical properties of simple random-effects models for genetic heritability},
	author={Steinsaltz, David and Dahl, Andrew and Wachter, Kenneth W},
	journal={Electronic journal of statistics},
	volume={12},
	number={1},
	pages={321},
	year={2018},
	publisher={NIH Public Access}
}

@article{wang2024regression,
  title={A Regression-Based Approach to Robust Estimation and Inference for Genetic Covariance},
  author={Wang, Jianqiao and Li, Sai and Li, Hongzhe},
  journal={Journal of the American Statistical Association},
  volume={119},
  number={548},
  pages={2585--2597},
  year={2024},
  publisher={Taylor \& Francis}
}

@article{hutchinson1989stochastic,
  title={A stochastic estimator of the trace of the influence matrix for Laplacian smoothing splines},
  author={Hutchinson, Michael F},
  journal={Communications in Statistics-Simulation and Computation},
  volume={18},
  number={3},
  pages={1059--1076},
  year={1989},
  publisher={Taylor \& Francis}
}

@article{ding2024eigenvector,
  title={Eigenvector distributions and optimal shrinkage estimators for large covariance and precision matrices},
  author={Ding, Xiucai and Li, Yun and Yang, Fan},
  journal={arXiv preprint arXiv:2404.14751},
  year={2024}
}

@article{gilmour1995average,
	ISSN = {0006341X, 15410420},
	URL = {http://www.jstor.org/stable/2533274},
	author = {Arthur R. Gilmour and Robin Thompson and Brian R. Cullis},
	journal = {Biometrics},
	number = {4},
	pages = {1440--1450},
	publisher = {International Biometric Society},
	title = {Average Information {REML}: An Efficient Algorithm for Variance Parameter Estimation in Linear Mixed Models},
	urldate = {2025-09-14},
	volume = {51},
	year = {1995}
}

@article{wang2022estimation,
  title={Estimation of genetic correlation with summary association statistics},
  author={Wang, Jianqiao and Li, Hongzhe},
  journal={Biometrika},
  volume={109},
  number={2},
  pages={421--438},
  year={2022},
  publisher={Oxford University Press}
}

@article{hartley1967maximum,
	title={Maximum-likelihood estimation for the mixed analysis of variance model},
	 author = {H. O. Hartley and J. N. K. Rao},
	journal = {Biometrika},
	number = {1/2},
	pages = {93--108},
	publisher = {[Oxford University Press, Biometrika Trust]},
	title = {Maximum-Likelihood Estimation for the Mixed Analysis of Variance Model},
	urldate = {2025-09-14},
	volume = {54},
	year = {1967}
}

@book{nocedal2006numerical,
  title     = {Numerical Optimization},
  author    = {Nocedal, Jorge and Wright, Stephen J.},
  edition   = {2nd},
  year      = {2006},
  publisher = {Springer},
  series    = {Springer Series in Operations Research and Financial Engineering},
  address   = {New York},
  isbn      = {9780387303031}
}

@article{hastie2022surprises,
  title={Surprises in high-dimensional ridgeless least squares interpolation},
  author={Hastie, Trevor and Montanari, Andrea and Rosset, Saharon and Tibshirani, Ryan J},
  journal={Annals of Statistics},
  volume={50},
  number={2},
  pages={949},
  year={2022}
}

@article{bai2007asymptotics,
	author = {Z. D. Bai and B. Q. Miao and G. M. Pan},
	title = {{On asymptotics of eigenvectors of large sample covariance matrix}},
	volume = {35},
	journal = {The Annals of Probability},
	number = {4},
	publisher = {Institute of Mathematical Statistics},
	pages = {1532 -- 1572},
	year = {2007},
	doi = {10.1214/009117906000001079},
	URL = {https://doi.org/10.1214/009117906000001079}
}

@article{yengo2022saturated,
  title   = {A saturated map of common genetic variants associated with human height},
  journal = {Nature},
  year    = {2022},
  volume  = {610},
  number  = {7933},
  pages   = {704--712},
  author  = {Yengo, Lo{\"i}c and Vedantam, Sailaja and Marouli, Eirini and Sidorenko, Julia and Bartell, Eric and others}
}

@article{smit2025polyg, 
title={Polygenic prediction of body mass index and obesity through the life course and across ancestries}, volume={31}, 
number={9}, 
journal={Nature Medicine}, 
author={Smit, Roelof A. J. and Wade, Kaitlin H. and Hui, Qin and Arias, Joshua D. and Yin, Xianyong and Christiansen, Malene R. and others}, year={2025}, month=sept, pages={3151–3168} }

@inproceedings{sherman2023eigenstructure,
  title={On the Eigenstructure of the {AR} (1) Covariance},
  author={Sherman, Peter J},
  booktitle={2023 IEEE Statistical Signal Processing Workshop (SSP)},
  pages={6--10},
  year={2023},
  organization={IEEE}
}

\end{document}


\title{ Supplementary Material for ``Heritability estimation using genetic similarity representation''}
\author{Jianqiao Wang and Xihong Lin}
\date{}
\maketitle	

\tableofcontents

\appendix

\section{Proofs of Main Theorems and Lemmas}

Throughout the supplement, $\Psi_n$ denotes the SMILE moment-basis
estimating function defined in the main text using
$A=V^{-1}$ and $A=V^{-2}$. We use $\mathcal{U}_n$ to denote the
equivalent Gaussian score-basis estimating function. 
As in standard variance components analysis \citep{hartley1967maximum}, we reformulate the estimating equations in terms of the signal-to-noise ratio $\gamma = \theta^2/\sigma^2_{\eps}$ and $\sigma^2_{\eps}$ as, 
\begin{align*}
& \psi_{1} = -\frac{1}{ \sigma^2_{\eps}}  y\tr (I + \gamma S)^{-1} y  -n = 0,
	\\
	 &  
\psi_{2 } =	-\frac{1}{\sigma^2_{\eps} } \Trace\{ (I + \gamma S)^{-1}  \} + \frac{1}{ \sigma^4_{\eps} } y\tr (I + \gamma S)^{-2}  y = 0.
\end{align*}
and separate $\gamma$ and $\sigma^2_{\eps}$. Eliminating $\sigma^2_{\eps}$ yields the following estimating equation for $\gamma$:
\beq
\label{est-gamma}
\Psi_n(\gamma)   = \frac{y\tr (I + \gamma S)^{-2} y }{ \Trace\{ (I + \gamma S)^{-1}  \} } - \frac{ y\tr (I + \gamma S)^{-1} y - y\tr (I + \gamma S)^{-2} y }{ n - \Trace\{ (I + \gamma S)^{-1}  \} } = 0
\eeq
Introducing  $\kappa_i = (1 + \gamma \lambda_i)^{-1}$ and $ \bar{\kappa} =  n^{-1}\Trace\{ (\I + \gamma S)^{-1} \}$, we can rewrite the estimating equation for $\hat \gamma$ as 
\begin{align}
\text{\eqref{est-gamma}} & = \frac{ n^{-1} y\tr H(\gamma) y }{ \bar \kap (1 - \bar \kap)  } ,   \quad \text{ where } \\
 H(\gamma) =  & { (I + \gamma S)^{-2} - n^{-1}\Trace\{ (I + \gamma S)^{-1} \} (I + \gamma S)^{-1} } \label{eq:H} 
\end{align}
The eigenvalues of $H(\gamma)$ are given by $h_i = \kappa_i^2 - \kappa_i \bar{\kappa}$ for $i = 1, \dots, n$, and
\begin{gather*}
	\Trace(H) =  \sum^n_{i = 1} \kap_i ( 	\kappa_i - \bar{\kappa})=  \sum^n_{i = 1} (\kappa_i - \bar \kappa)^2. 
\end{gather*} 

Given that $\theta^2$ and $\sigma^2_{\eps}$ are positive constants in Assumption 1, $\gamma$ lies in a compact space $\Theta \subset (0, \infty)$.  
We first establish  the consistency of $\hat \gamma$ as the solution to \eqref{est-gamma}. Consistency of $\widehat \sigma^2_{\eps} $ then follows from the linear combination of estimating equations $ \Psi_{\theta^2} + \Psi_{\sigma^2_{\eps}}/ \gamma = 0 $, which yields that  
\begin{align}
\label{est}
\widehat \sigma^2_{\eps} (\hat \gamma ) = \frac{1}{n} { y\tr (I + \widehat\gamma S)^{-1}  y } \text{ and }  \widehat \theta^2 (\hat \gamma ) =  \widehat \sigma^2_{\eps} (\widehat \gamma )  \widehat \gamma.
\end{align}
For clarification, we denote $(\theta^2_{0}, \sigma^2_{0}, \gamma_{0} )$ as the true parameters, and we denote $\Lambda$ the diagonal eigenvalue matrix of $S$.

\subsection{Proof of Theorem 1}
\paragraph{Step 1:} 
Note that 
\beq
\label{yHy}
y\tr H y = (\PhiS \xi + \eps)\tr H (\PhiS \xi + \eps) .
\eeq
  By replacing $\xi\tr \PhiS\tr H \PhiS \xi$ with $ \theta_0^2 \Trace( H S)  $, we define the approximation of $\Psi_n$ as 
  $$
\widetilde \Psi_n(\gamma) =  \frac{ (\theta_0^2 \Trace( H S) + 2 \eps\tr H \PhiS \xi + \eps\tr H \eps) }{ \bar \kappa ( 1 - \bar \kappa) } 
$$  
and decompose $\Psi_n$ into three parts,
\begin{align*}
\Psi_n(\gamma)  & = \underbrace{  \E[ \widetilde \Psi_n(\gamma) | \Lambda ] }_{T_1}  + \underbrace{ \Psi_n(\gamma) - \widetilde \Psi_n(\gamma)}_{T_2}  +   \underbrace{ \widetilde \Psi_n(\gamma) - \E[ \widetilde \Psi_n(\gamma) | \Lambda ] }_{T_3} \\
\text{ with } \quad T_1 & = \frac{ \sigma^2_{0}  \Trace\big( H ( \I + \gam_0 S)  \big)}{ \bar \kappa (1 - \bar \kappa)} 
\\
T_2  & =   \frac{ \xi\tr \PhiS\tr H \PhiS \xi - \theta_0^2 \Trace( H S) }{ \bar \kappa ( 1 - \bar \kappa) } \\
T_3  & = \frac{ 2 \eps\tr H \PhiS \xi + \eps\tr H \eps - \sigma^2_{0}  \Trace(H) }{ \bar \kappa ( 1 - \bar \kappa) } 
\end{align*} 
 Here $T_1 = \E[ \widetilde \Psi_n(\gamma) | D ]$ is the deterministic component, $T_2 = \Psi_n(\gamma) - \widetilde \Psi_n(\gamma)$ represents the approximation error, and $T_3 = \widetilde \Psi_n(\gamma) - \E[ \widetilde \Psi_n(\gamma) | D ]$ represents the random part. 

We first show   $ T_1 = 0$ if and only if $\gamma = \gamma_0$ as ${n,p \to \infty}$.  Since $h_i = \kap_i ( \kap_i - \bar \kap )$,  
\begin{gather*}
	\kappa_i (1 + \gamma_0 \lam_i) = 1 + (\gamma_0 - \gamma) \lam_i \kappa_i \\
	 h_i (1 + \gamma_0 \lam_i) = (\kappa_i - \bar{\kap} ) + (\gamma_0 - \gamma) h_i \lam_i.
\end{gather*}
It follows that  
\begin{gather*}
\Trace\big( H ( \I + \gam_0 S)  \big) = (\gamma_0 - \gamma) \Trace\big( H S  \big) \\
T_1 = \frac{ \sigma^2_{0} ( \gam_0 - \gamma) }{  \bar \kap ( 1 - \bar \kap) } \Trace\big( H S  \big).
\end{gather*}
Applying  \eqref{l-r2} in Lemma \ref{lem1}, we obtain
 \begin{align} 
 \Trace\big( H S  \big) \asymp -r \Delta_n \quad \text{and} \quad |T_1| \gtrsim   \sigma^2_{0} |\gam_0 - \gamma| n \Delta_n. 
 \label{eq:yHy}  
 \end{align} 
Therefore, uniformly over the compact space 
$
\cA_{\varepsilon} = \left\{ \gamma \in \Theta: |\gamma - \gamma_0| > \varepsilon  \right\}, 
$
we have 
\(
|T_1(\gamma)| \geq c_{\eps} n \Delta_n.
\)
If $\lim_{n \to \infty} \Delta_n = \Delta > 0$, the deterministic part satisfies $T_1 = 0$ if and only if $\gamma = \gamma_0$.

Next, we show that as $n \to \infty$, the remainder terms are asymptotically negligible, i.e., 
\begin{gather}
\label{sup_bound}
\sup_{ \left\{ \gamma \in \Theta: |\gamma - \gamma_0| > \varepsilon  \right\} } ( T_2 + T_3)/ T_1  = o_P(1) \\ 
\text{ where } ~~	\frac{T_2}{T_1} = \frac{ \xi^{\mathsf T} \Phi_S^{\mathsf T} H \Phi_S \xi - \theta_0^2 \Trace(HS) }{ \sigma_0^2 (\gamma_0 - \gamma)\Trace(HS)}, \quad
	 \frac{T_3}{T_1} = \frac{ 2 \epsilon^{\mathsf T} H \Phi_S \xi + \epsilon^{\mathsf T} H \epsilon - \sigma_0^2 \Trace(H) }{ \sigma_0^2 (\gamma_0 - \gamma)\Trace(HS)}.
\end{gather}
Therefore  the goal is to show that 
\[
\sup_{\gamma \in \Theta} \frac{ \xi^{\mathsf T} \Phi_S^{\mathsf T} H \Phi_S \xi - \theta_0^2 \Trace(HS) }{ \Trace(HS)} = o_p(1) \text{ and } \quad  \sup_{\gamma \in \Theta} \frac{ 2 \eps\tr H \PhiS \xi + \eps\tr H \eps - \sigma^2_{0}  \Trace(H)}{ \Trace(HS)} = o_p(1).
\]
The key step is to separate $\gamma$ from the distribution of $\xi$ and $\eps$. According to the push-through identity, 
$$ 
\PhiS\tr H \PhiS = (I + \gamma  \Lambda )^{-2} \Lambda - n^{-1} \Trace( (I + \gamma  \Lambda)^{-1} )  (I + \gamma  \Lambda )^{-1} \Lambda,
$$
we let
 $$
 A_n(\gamma) = \frac{1}{r} \{  \xi^{\mathsf T} \Phi_S^{\mathsf T} H \Phi_S \xi - \theta_0^2 \Trace(HS) \} = \frac{1}{r} \sum^r_{i = 1} w_i(\gamma) (\xi^2 - \theta_0^2),
 $$
 where $w_i(\gamma) = \lambda_i h_i(\gamma)$ and  one can show that $ |w_i(\gamma) | \leq 1$.
 We show that the size of $A_n(\gamma)$ is controlled uniformly. 
Taking the derivative regarding to $\gamma$, one has 
\begin{gather*}
\dot{\kappa}_i(\gamma) = -\lam_i \kap^2_i, ~\, |\dot{ \bar\kappa} | = | \frac{1}{n} \sum^r_{j = 1} \lambda_j \kappa^2_j |, ~ \, 
\dot{w}_i(\gamma) = \lambda_i (2 \kappa_i \dot{\kappa}_i - \dot \kappa_i  \bar \kappa - \dot{\bar \kappa} \kappa_i  ) 
\end{gather*}
Under Condition \ref{cond:C1} and $\gamma \in \Theta$, one can show that 
\begin{gather*} 
|\kappa_i| \leq 1,~ |\lam_i h_i| \leq C, ~  |\lam_i \kap_i| \leq C, \text{ and } \lam_i^2 \kappa^2_i \leq C.
\end{gather*}
Therefore
\[
\sup_{i, \gamma} | w_i(\gamma) | \leq C' ~ \text{ and } ~ \sup_{i,\gamma} | \dot{w}_i(\gamma) | \leq C. 
\]
According to Lemma \ref{lem1}, for any fixed $\gamma$, we have $A_n(\gamma) = o_p(1)$. To make this uniform, take a finite grid $\gamma_1, \cdots, \gamma_m$ of $\Theta$ with size at most $\eta$. Then 
\begin{align*}
\sup_{\gamma \in \Theta} |A_n(\gamma)| & \leq \max_{ 1 \leq j \leq m } | A_n(\gamma_j)| + C \eta  \frac{1}{r} \sum^r_{i = 1} |\xi^2_i - \theta^2_0| 
\end{align*}
where we use the fact that 
\begin{gather*}
| A_n(\gamma) - A_n(\gamma') | \leq \frac{1}{r} \sum^r_{i = 1} | w_i(\gamma) - w_i(\gamma') | |\xi^2_i - \theta^2_0| \leq C |\gamma - \gamma'| \frac{1}{r} \sum^r_{i = 1} |\xi^2_i - \theta^2_0|. 
\end{gather*}
It means for any fixed $\delta > 0$, we have 
\[
\Pr( \sup_{\gamma \in \Theta} |A_n(\gamma)| > \delta ) \leq  \sum^m_{i = 1} \Pr( | A_n(\gamma_j)| > \frac{\delta}{2} )   + \Pr( C \eta  \frac{1}{r} \sum^r_{i = 1} |\xi^2_i - \theta^2_0| > \frac{\delta}{2} ).
\]
The first term goes to zero for fixed $m$ and $n \to \infty$ since $A_n(\gamma) = o_p(1)$. For the second term, since 
\(
r^{-1} \sum^r_{ i = 1} |\xi^2 - \theta_0^2| \leq { \| \xi \|^2 }/{r}+ \theta^2_0 = O_p(1)
\), for any $\rho > 0$, one can choose $\eta$ small enough such that 
\[
\Pr( C \eta  \frac{1}{r} \sum^r_{i = 1} |\xi^2_i - \theta^2_0| > \frac{\delta}{2} ) \leq \rho.
\]
In this way, we show that $\sup_{\gamma \in \Theta} |A_n(\gamma)| = o_p(1)$. Correspondingly, 
\[
\sup_{\gamma \in \Theta} \frac{ \xi^{\mathsf T} \Phi_S^{\mathsf T} H \Phi_S \xi - \theta_0^2 \Trace(HS) }{ \Trace(HS)} = o_p(1).
\]

For $T_3/ T_1$, we define 
\[
R_n(\gamma) =  2 \eps\tr H(\gamma) \PhiS \xi + \eps\tr H(\gamma) \eps - \sigma^2_{0}  \Trace(H(\gamma)).
\]
Due to 
  \begin{gather*}
\Var\left( \eps\tr H \eps | S \right)  \lesssim \sigma^4_0 \Trace( H^2 ) ~ ~\text{and} ~~
\Var(\epsilon\tr  H  X \beta | S, \, \xi ) =  \sigma^2_0 \xi\tr \PhiS\tr H^2 \PhiS \xi.  
 \end{gather*}
By  Lemma \ref{lem-s2},
$$
\xi\tr \PhiS\tr H^2 \PhiS \xi = \sum^r_{i = 1} \xi^2_i \lam_i h_i^2  \approx \theta^2_0 \Trace( H^2 S) 
$$ and uniformly in $\gamma \in \Theta$ 
\[
  \frac{ \Trace(H^2) + \Trace(H^2 S) }{ \Trace^2(HS) } = O(\frac{1}{n \Delta_n}).
\]
Hence, for each fixed $\gamma$, 
\beq
\label{t3}
\frac{ R_n(\gamma) }{\Trace(HS) } = O_p( \frac{ \Trace^{1/2}(H^2) + \Trace^{1/2}(H^2 S) }{ \Trace(HS) } ) = O_p( \frac{1}{ \sqrt{ n \Delta_n } } ).
\eeq
Next, we show the uniform convergence of $R_n(\gamma)$ through the equicontinuity. For $\gamma, \gamma' \in \Theta$,
\[
|R_n(\gamma) - R_n(\gamma')| \leq 2 | \eps\tr \{ H(\gamma) - H(\gamma') \} X \beta | + | \eps\tr \{ H(\gamma) - H(\gamma') \} \eps  -\sigma^2_0 \Trace\{ H(\gamma) - H(\gamma') \} |
\]
According to 
\[
\| H(\gamma) - H(\gamma') \|_2 \leq \max_{i = 1,..,r} |h_i(\gamma) - h_i(\gamma')| \leq C\frac{r}{n} |\gamma - \gamma'|
\]
one has 
\begin{gather*}
 \frac{1}{r}| \eps\tr \{ H(\gamma) - H(\gamma') \} X \beta | \leq \frac{1}{r} \| \eps \| \| X \beta \|  \| H(\gamma) - H(\gamma') \|_{2} \leq  C \frac{\| \eps \| \| X \beta \|}{n} | \gamma - \gamma'| = O_p( | \gamma - \gamma'| ) \\
 \frac{1}{r} \eps\tr \{ H(\gamma) - H(\gamma') \} \eps \leq C \frac{\| \eps \|^2}{n} | \gamma - \gamma'| \\
 \frac{1}{r} \sigma^2_0 \Trace \{ H(\gamma) - H(\gamma') \}  \leq  C | \gamma - \gamma' |.
\end{gather*}
Therefore
\begin{align*}
\sup_{ |\gamma - \gamma' | \leq \eta }\frac{1}{r} |R_n(\gamma) - R_n(\gamma')| = O_p(\eta) 
\end{align*}
Similarly, we obtain,
 \beq
 \label{t4}
\sup_{\gamma \in \Theta} \frac{ R_n(\gamma)}{ \Trace(HS)} = o_p(1).
 \eeq
Then  $T_3/ T_1 \cip 0$ for any $|\gamma -  \gamma_0| > \epsilon > 0$ and $\lim_{n \to \infty} \Delta_n > 0$.

\ignore{
To show the pointwise convergence in \eqref{sup_bound-2} for each $\gamma$, we apply the Chebyshev inequality   
\[
\P\left( \frac{ | \Psi_n(\gamma) - \E[\Psi_n (\gamma) | S] | }{  \E[\Psi_n (\gamma) | S]  } > \delta | S \right) \leq \frac{\delta^{-2}}{ | \gamma - \tgam |^2 }  \Var( Q_{\gamma} | S) 
\]
Next we  show $\Var( Q_{\gamma} | S) = \sigma^{-4}_0 \Var( y\tr H y |S)/ \Trace^2(H S ) =  o_p(1).$ Specifically, we  control the deviations of 
\beq
y\tr H y - \E[y\tr H y | S] = \underbrace{ \beta\tr X\tr H  X \beta - \E[\beta\tr X\tr H  X \beta| S]}_{T_1} + 2 \underbrace{\epsilon\tr  H  X \beta}_{T_2} + \underbrace{ \epsilon\tr  H \epsilon - \E[ n^{-1}\epsilon\tr  H \epsilon | S] }_{T_3}.
\eeq 
Here $T_1$ and $T_3$ are  main variance terms  and  $ T_1 = \sum^r_{i=1} \lambda_i h_i \left(\xi_i^2 - \E[\xi^2_i|S] \right)$  based on the representation  $X \beta = \PhiS \xi =  U_W \tr D_W \xiS$. The variance of $T_1$  is bounded by  Assumption 2 and   $\Var(\xi^2_i | S) = O(\theta_0^4 r/n)$ shown in Lemma \ref{prob-order}. The variance of the quadratic term $T_3$ is quantified by  Lemma \ref{lem:sub-gaussian}. Then
\begin{align} 
& \Var( y\tr H y |S)  \lesssim \Var(T_1|S) + \Var(T_3|S) \lesssim  
\Var(\xi_i^2|S) \sum^r_{i = 1} h^2_i \lam^2_i  + \sigma_0^{4} \sum^n_{i = 1} h^2_i \label{eq2-1} \\
 &    \Var(Q_{\gamma} | S)  \lesssim \frac{ \Var( y\tr H y |S)  }{ \sigma_0^{4} \big( \sum^n_{i = 1} \lam_i h_i \ \big)^2 } \lesssim \frac{ \sum^n_{i = 1} h_i^2  }{ ( \sum^n_{i = 1} \lam_i h_i)^2 } +  \Var(\xi_i^2|S) \frac{  \sum^r_{i=1} \lam^2_i h^2_i }{ ( \sum^n_{i = 1} \lam_i h_i)^2 } \label{eq2}. 
\end{align}
Therefore for \eqref{eq:yHy}, we apply  \eqref{l-r1} in Lemma \ref{lem1} and for bounded $\gamma\in (0, C]$,  there is  as $n \to \infty$ and $\lim_{n \to \infty} \Delta > 0$, we have
$$
\Var(Q_{\gamma} |S) = O( \frac{\gamma^2_0 (\gamma + 1)^2}{ n \Delta} + \frac{ r (\gamma + 1)^2 }{ n^2 \Delta} ) = o_p(1).
$$ 
Note $\Var(Q_{\gamma}) = \E[ \Var(Q_{\gamma} |S) ] +  \Var( \E[ Q_{\gamma} |S ] ) = \E[ \Var(Q_{\gamma} |S) ] $ with $\E[ Q_{\gamma} |S ] = 0.$ Then the dominated convergence theorem yields that  
$$
\lim_{n \to \infty} \Var(Q_{\gamma}) = \lim_{n \to \infty} \E[ \Var(Q_{\gamma} |S) ]   = \E[ \lim_{n \to \infty} \Var(Q_{\gamma} |S) ] = o_p(1).
$$

The uniform convergence of \eqref{sup_bound}  is seen from the fact that $\gamma$ only changes the deterministic  $h_i = \kappa^2_i - \kappa_1 \bkap$ and is irrelevant to the distribution of random $\eps$ or $\xi$. With $\check \kappa = r^{-1} \sum^r_{i = 1} (1 + \gamma \lam_i)^{-1}$ and $\cq = r^{-1} \sum^r_{i = 1} \lam_i / (1 + \gamma \lam_i)$, there is
\[
h_i =  \lam_i \kappa^2_i \frac{1}{n}\sum^n_{j = 1}\frac{\gamma (\lam_j - \lam_i)}{1 + \gamma \lam_j } = \gamma \left(   \kappa_i^2  \frac{r}{n}  \cq  -  ( \frac{r \check{\kap}}{n} + \frac{n - r}{n} )  \kappa^2_i \lambda_i   \right) 
\]  and 
\begin{align*}
T_1 & = - \left( \frac{r \gamma \check \kappa}{n} +  \frac{ (n-r) \gamma }{n} \right) \underbrace{ \sum^r_{i=1} \lambda^2_i \kappa_i^2 \left(\xi_i^2 - \E[\xi^2_i|S] \right) }_{A_1} 
+ \frac{r \check{q} \gamma}{n} \underbrace{ \sum^r_{i=1} \lambda_i \kappa^2_i \left(\xi_i^2 - \E[\xi^2_i|S] \right) }_{B_1}   
 \end{align*}
Random functions $A_1$ and $B_1$  are decreasing and continuous functions of $\gamma > 0$ on the compact space $\Theta$, which gives a uniform control of 
\begin{align*}
\sup_{\gamma \in \Theta}  A_1 = O( r/\sqrt{n}) ~ \text{ and} ~ \sup_{\gamma \in \Theta}  B_1 = O( r/\sqrt{n}).
\end{align*}
Based on  $\sum^r_{i = 1} \lam_i h_i  = O(r \gamma \cq^2 \Delta) $, we have 
\[
\frac{\left( \frac{r \gamma \check \kappa}{n} +  \frac{ (n-r) \gamma }{n} \right)}{ \sum^r_{i = 1} \lam_i h_i } = O( \frac{1}{n} + \frac{n-r}{nr}   ), 
\text{ and } 
\frac{r \cq \gamma }{ \sum^r_{i = 1} \lam_i h_i } = O( \frac{1}{n }).   
\]
Then $\sup_{\gamma \in \Theta} |T_1|/ \sum^r_{i = 1} \lam_i h_i = o_p(1)$ follows.  The other terms can be decomposed into monotone functions and analyzed similarly, which yields the uniform control.  
 }
\ignore{
i.e.,   $\sup_{ |\gamma - \gamma_0| > \epsilon }  $

, \eqref{sup_bound} is shown by checking that the variance  $\Var( y\tr H y |S)$ is uniformly bounded, i.e., for every $\eps > 0$, 
\[
\sup_{ |\gamma - \gamma_0| > \epsilon } \frac{ \Var( y\tr H y |S) }{ (\E[y\tr H y | S])^2 } 
\leq \frac{C}{n \epsilon}.
\]}
\ignore{
Therefore for $\vec{a} = (a_1, a_2, \ldots, a_r)$ and $\vec{b} = (b_1,b_2, \ldots, b_n)$ with $\|\vec a\| = 1$ and $\|\vec b\| = 1$, we have
\begin{align*}
\sup_{\gamma} | y\tr H y - \E[y\tr H y | S] | \lesssim_P  & \sup_{\gamma }  (\sum^r_{i = 1} \lambda_i^2 h_i^2)  \sup_{ \vec{a} }   \sum^r_{i = 1} a_i  \left(\xi_i^2 - \E[\xi^2_i|S] \right) + \\ & \sup_{\gamma }  (\sum^n_{i = 1} h_i^2)   \sup_{ b } \sum^r_{i = 1} b_i  \left(e_i^2 - \E[e^2_i] \right).
\end{align*}
}

\ignore{and
 \[
\sup_{ |\gamma - \tgam| > \epsilon } \frac{ \left\{ \sum^n_{i = 1} \sigma^2_0 h_i^2   \right\}  }{  \big( \sum^n_{i = 1} \lam_i h_i \ \big)^2 }  
\leq \frac{C}{n}.
\]
}
In summary, based on the event with probability tending to one,
\[
\sup_{ \left\{ \gamma \in \Theta: |\gamma - \gamma_0| > \varepsilon  \right\} } | T_2 + T_3 |/ |T_1|  < \frac{1}{2},
\]
we have for every $|\gamma - \gamma_0| > \epsilon$, $|\Psi_n(\gamma)| \geq \frac{1}{2} |T_1(\gamma)| > 0$.  Therefore for any solution of the estimation equation $\widehat \gamma \in \Theta$, there is 
\[
\P( |\widehat \gamma - \gamma_0 | \geq \epsilon ) \to 0.
\]
This proves the consistency of $\widehat \gamma$. 

\subsection{Proof of Theorem \ref{thm:converg}}
Lemma \ref{lem0}  characterizes the convergence rate of quadratic forms under a moment condition of $\xi^2 = ( \xi^2_i, \ldots, \xi^2_r  )\tr$. Therefore the consistency of $(\widehat \theta^2, \widehat \sigma^2_{\eps})$ is guaranteed. 

\paragraph{Convergence rate of $\hat \gamma$.}  
By \eqref{eq:yHy} and the estimating equation $\Psi(\widehat \gamma) = T_1 (\widehat \gamma ) + T_2(\widehat \gamma ) + T_3 (\widehat \gamma ) = 0$, there is 
\begin{align*}
-T_1 = T_2  + T_3 \Rightarrow   \widehat{\gamma} - \gam_0 = Q({\widehat \gamma}) 
\end{align*}
where
\begin{align*}
Q({ \gamma})  &=  \frac{ y\tr H(\gamma) y - \theta_0^2 \Trace( H({ \gamma}) S) - \sigma^2_0 \Trace(H({ \gamma})) }{ \sigma^2_0  \Trace(H({ \gamma}) S ) }  \\
& = \underbrace{ \frac{ A_n(\gamma) }{\sigma^2_0  \Trace(H({ \gamma}) S )} }_{Q_1} +  \underbrace{ \frac{ R_n(\gamma)  }{ \sigma^2_0  \Trace(H({ \gamma}) S )} }_{Q_2}.
\end{align*}
Note that 
\[
Q(\widehat \gamma) = Q( \gamma_0 ) + Q(\widehat \gamma) - Q(\gamma_0) = Q( \gamma_0 ) \left(1 +  \frac{Q(\widehat \gamma) - Q(\gamma_0)}{ Q(\gamma_0) } \right).
\]
Since $\widehat{\gamma} \cip \gamma_0$, the continuous mapping theorem implies $Q(\widehat{\gamma}) / Q(\gamma_0) \cip 1$. Thus, it suffices to control the order of $Q(\gamma_0)$. The analysis in \eqref{t3} shows,
\[
Q_2(\gamma_0) = O_p\left( \frac{1}{\sqrt{n} } \frac{ \Trace^{1/2}(H^2) + \Trace^{1/2}(H^2 S)}{ \Trace(HS) } \right)  =  O_p(  \frac{1}{ \sqrt{ n \Delta_n }} ).
\]
Furthermore, Lemma~\ref{lem0} gives
\[
Q_1 (\gamma_0)  = O_p\left( \sqrt{ \frac{r}{n} \frac{ \Trace(H^2 S^2) }{ \Trace^2(H S)  }   } \right) = O_p\left( \frac{1}{ \sqrt{ n \Delta_n} }  \right).
\]
Hence,
$$
 |\widehat \gamma - \gamma_0|= O_p\left( \frac{1}{ \sqrt{ n \Delta_n} }  \right).
$$

\paragraph{Convergence rates of $\widehat \theta^2$ and $\widehat \sigma_{\eps}^2$.}
Note that 
\beq
\label{eq:sigma0}
\widehat \sigma_{\eps}^2 ( \widehat \gamma ) - \sigma_0^2 =  n^{-1}\left[ y\tr (I + \widehat \gamma S)^{-1} y -  y\tr (I + \gam_0 S)^{-1} y \right]  + n^{-1} y\tr (I + \gam_0 S)^{-1} y  - \sigma^2_0 . 
\eeq
According to $y\tr (I + \gam_0 S)^{-1} y  = (\PhiS \xi + \eps)\tr (I + \gam_0 S)^{-1} (\PhiS \xi + \eps) $, we can show
\[
| n^{-1} y\tr (I + \gam_0 S)^{-1} y - \sigma_0^2 | \lesssim_P  \frac{1}{n^{1/2}},
\]
where we use the fact that  $S (I + \gam_0 S)^{-1} $ and  $ (I + \gam_0 S)^{-1} $ both have bounded eigenvalues. 
Moreover, since
\begin{align*}
(I + \gam_0 S)^{-1} - (I + \widehat \gamma S)^{-1} & = (I + \gam_0 S)^{-1} ( \widehat{\gamma} - \gam_0 ) S  (I + \widehat \gamma S)^{-1}, 
\end{align*}
we have 
\[
n^{-1}|y\tr \left[ (I + \gam_0 S)^{-1}  -  (I + \widehat \gamma S)^{-1} \right] y | 
\leq {|\gamma_0 - \widehat \gamma|} \frac{1}{n} y\tr y. 
\]
Combining two terms in \eqref{eq:sigma0} and note $\| y \|^2/n = O_p(1)$, there is 
\[
|\widehat \sigma^2_{\eps} - \sigma^2_{0} | \lesssim_P   
 \frac{ 1 }{\sqrt{   n } } +   \frac{ 1 }{ \sqrt{n \Delta_n } }.
\]
The  estimation rate for $\widehat \theta^2 = \widehat \sigma^2_{\eps} \widehat \gamma $ then follows from
\[
| \widehat \sigma^2_{\eps} \widehat \gamma  - \sigma^2_0 \gamma_0 | \lesssim_P \sigma^2_{0} (\hgam - \gam_0 ) + \gam_0 ( \widehat \sigma^2_{\eps} - \sigma^2_{0} ) = O_p( \frac{1}{\sqrt{n}} +   \frac{1}{\sqrt{n \Delta}} ). 
\]

\subsection{Proof of Theorem \ref{thm2}}

Theorem \ref{thm2} is established using the existing random matrix results, which are reviewed in Section \ref{sec:rmt}. First we show Condition \eqref{cond:C1} holds almost surely. For $S = XX\tr/p =  p^{-1} Z \Sigma Z\tr$, according to Lemma \ref{thm-s-extrem-eigen} in Section \ref{sec:rmt}, it holds almost surely that 
\begin{align*}
	\lambda^{+}_{\min}(S) & \geq   \lambda_{\min}^{+} (p^{-1} Z Z\tr) \lambda_{\min}^{+} (\Sigma) = (1 - \sqrt{n/p})^2  \lambda_{\min}^{+} (\Sigma) \\ 
	\lambda_{\max}(S) & \leq   \lambda_{\max} (p^{-1} Z Z\tr) \lambda_{\max} (\Sigma) = (1 + \sqrt{n/p})^2  \lambda_{\max}^{+} (\Sigma) .
\end{align*} 
Therefore, with $ C^{-1} \leq \lambda^{+}_{\min} (\Sigma) / \lambda_{\max} (\Sigma)  \leq C$ and $n/p \to c \in (0,1)\cup (1, \infty)$,  Condition \eqref{cond:C1} holds almost surely.

Condition \eqref{cond:C2} follows directly from Theorem 	\ref{thm-s-eigen} in Section \ref{sec:rmt}, which shows that if 
 for any ${z} \in \mathbb{C}^{+}$,
\beq
\lim_{p \to \infty} \frac{1 }{ \| \beta\|^2} \beta\tr (\Sigma - z \I)^{-1} \beta  -  \frac{1}{p} \Trace\left( (\Sigma - z \I)^{-1} \right) = 0,
\eeq
then for bounded function $f(t)$ and $v = (v_1, v_2, \ldots, v_n) \tr = \Gamma\tr \beta/ \| \beta \| $, one has 
\[
\sum^n_{j = 1} |v^2_j| f(\lambda_j) - \frac{1}{n} \sum^n_{j = 1} f(\lam_j) \cia 0.
\] 
Finally according to the proof in Theorem \ref{prop:delta_n},  
\begin{gather*}
	\Delta_n / \left\{ \frac{ \rk^2(\Sigma) \wedge n^2 }{n p^2} \Trace(\Sigma^2) + \left( 1 -  \frac{ \rk(\Sigma) \wedge n }{n } \right)^2  \right\} \cia 1.
\end{gather*}
One can check that $ \Trace^2(\Sigma) \leq p \Trace(\Sigma^2) $ and therefore  $n \Trace(\Sigma^2)/ p^2  \geq n/p$. The consistency then follows from $\Delta_n > 0$.

\subsection{Proof of Lemma 1}


For (i), it is easy to check that $n^{-1}\beta\tr X\tr X \beta \cia \theta_0^2$ according to the strong law of large number.  
Given the representation $X \beta = \PhiS  \xiS$ and $\PhiS\tr \PhiS =  \Lambda$, 
\[
n^{-1}\beta\tr X\tr X \beta = n^{-1} \xiS \tr \Lambda  \xiS =  n^{-1} \sum^r_{i=1} \lambda_i \xi_i^2 = \frac{ \| \xi \|^2 }{r} . \underbrace{ \frac{r}{n} \sum^r_{i = 1} \lambda_i \xi^2_i / \|\xi \|^2  }_{ T' }.
\]
Note $\sum^r_{i = 1} \lambda_i = n$ and $ \lam_i \leq n C/ r$, then applying (C2) yields that  $ T' \to 1.$  Then one can check that $$r^{-1} \| \xi \|^2 = 	n^{-1}\beta\tr X\tr X \beta/ T'  \to \theta_0^2$$ with $n^{-1}\beta\tr X\tr X \beta$ converging to $\theta^2_0.$ Correspondingly, the convergence of $\xi\tr A \xi =  r^{-1}\sum^r_{i = 1} \xi^2_i f(\lam_i)$ can be checked by
\[
r^{-1} \{ \xi\tr A \xi - \theta^2_0 \Trace(A) \}  =  \frac{ \| \xi \|^2 }{r} \sum^r_{i = 1} ( \frac{ \xi^2_i }{\| \xi \|^2} - \frac{1}{r} ) f(\lam_i) +  (\frac{ \| \xi \|^2 }{r} - \theta^2_0) \frac{\Trace(A) }{r}. 
\]
Combining the condition (C2), bounded $\Trace(A)/r$ and $\| \xi \|^2/r  \to \theta_0^2$, we get the desired results.

For (ii), note that $\tau^2 = \E[ \xi_i^2\, | \, \Lambda] =  \E[ n^{-1}\beta\tr X\tr X \beta \, | \, \Lambda] $ and 
\begin{align}
	& n^{-1}\beta\tr X\tr X \beta - \theta^2_0  
	=   n^{-1} \sum^r_{i=1} \lambda_i (\xi_i^2 - \tau^2 ) +   \tau^2 - \theta^2_0,   
	\\
	&  \sum^r_{i=1} f(\lambda_i) (\xi_i^2 - \theta^2_0 ) = \sum^r_{i=1} f(\lambda_i) (\xi_i^2 - \tau^2 ) +   ( \tau^2 - \theta^2_0) {\Trace(A)}. \label{lm-rate-1} 
\end{align}		
Let 
\[
M_{\Lambda} = \E[ (n^{-1}\beta\tr X\tr X \beta - \theta^2_0 )^2 \, | \,  \Lambda].
\]
Since
\[
\E[ M_{\Lambda} ] =  \Var(n^{-1}\beta\tr X\tr X \beta) \leq C\frac{\theta_0^4}{n},
\]
Markov inequality gives that $\E[ M_{\Lambda} ] = O_p(n^{-1/2}) $ and in particular, one has 
that
\begin{gather*}
    |\tau^2 - \theta^2_0| = | E[ n^{-1}\beta\tr X\tr X \beta - \theta^2_0  ] | \leq M^{1/2}_{\Lambda} = O_p( n^{-1/2} ) \\  ( \tau^2 - \theta^2_0) {\Trace(A)} = O_p( n^{-1/2} \Trace(A) ).
\end{gather*}
Define $ a = (a_1, \ldots, a_r)\tr$  and  $ b = (b_1, \ldots, b_r)\tr$ with
$$a_i = \frac{f(\lambda_i)}{ \left( \sum^r_{i = 1} f^2(\lambda_i) \right)^{1/2}}, \, \quad  b_i = \frac{\lambda_i}{  (\sum^r_{i = 1} \lambda^2_i)^{1/2} } $$  
Then
$$
\sum^r_{i=1} f(\lambda_i) \xi_i^2 = \sqrt{ \sum^r_{i = 1} f^2(\lambda_i) } \sum^r_{i=1} a_i \xi_i^2 \quad \text{ and } \quad  \sum^r_{i=1} \lam_i \xi_i^2 =  (\sum^r_{i = 1} \lambda^2_i )^{1/2} \sum^r_{i=1} b_i \xi_i^2.
$$  
The bounded  $ \lam_{\min} (\Omega) \geq c' \lam_{\max}(\Omega) $ in $\Cov(\xi^2 | S) = \Omega$ ensures that  for $\| a \| = \| b \| = 1$, 
\beq
\Var( a\tr \xi^2 | S)  \asymp  \Var( b\tr \xi^2 | S) = \frac{ \Var( \sum^r_{i=1} \lambda_i \xi_i^2   | S )  }{ \sum^r_{i = 1} \lam^2_i } = O_p \left( \frac{r}{n} \theta^4_0  \right).
\label{lm-rate-2}
\eeq
Condition \eqref{cond:C1} ensures that $\sum^r_{i = 1} \lam^2_i \asymp n^2/r $.  Note $\Var( n^{-1} \sum^r_{i=1} \lambda_i \xi_i^2   | S ) = \Var( n^{-1} \beta\tr X\tr X \beta  | S ) $, it follows that
\[
\Var( a\tr \xi^2 | S) = O_p \left( \frac{r}{n} \theta^4_0  \right).
\]
Hence, by Chebyshev's inequality  
\[
\sum^r_{i=1} f(\lambda_i) (\xi_i^2 - \tau^2 )  = \sqrt{ \sum^r_{i = 1} f^2(\lambda_i) } O_p( \Var^{1/2}( a\tr \xi^2 | S)   )  =   O_p( \sqrt{\frac{r \sum^r_{i = 1} f^2(\lambda_i) }{n} } ).
\]
Note $\Trace(A) \leq \sqrt{ r\Trace(A^2) }$, then  
\[
\sum^r_{i=1} f(\lambda_i) (\xi_i^2 - \theta^2_0 ) =  O_p\left( \sqrt{ \frac{ r \sum^r_{i = 1} f^2(\lambda_i) }{n} } \right).
\]

\subsection{Proof of Lemma  \ref{prop:delta_n}}

We first prove the theorem based on the Gram matrix case that $S = XX\tr/p = p^{-1} Z \Sigma Z\tr$ with $\Trace(\Sigma) = p$.  The result is then extended directly to the general weighted Gram matrix $S= X W X\tr = Z \Sigma_{W} Z\tr$ with $\Sigma_{W} = \Sigma^{1/2} W \Sigma^{1/2}$.

\paragraph{Gram matrix $S = XX\tr/p =  p^{-1} Z \Sigma Z\tr $.} We will show  $\rk(S) = \rk(p^{-1} Z \Sigma Z\tr) \cia n \wedge \rk( \Sigma )$ based on
\begin{align}
	\label{ev-prod}
	\lambda^{+}_{\min}(S) & =   \lambda^{+}_{\min}( p^{-1} Z\tr Z \Sigma) \geq \lambda_{\min}^{+} (p^{-1} Z Z\tr) \lambda_{\min}^{+} (\Sigma) \\ 
	\lambda^{+}_{\max}(S) & =   \lambda^{+}_{\max}( p^{-1} Z\tr Z \Sigma) \leq \lambda_{\max}^{+} (p^{-1} Z Z\tr) \lambda_{\max}^{+} (\Sigma).
\end{align}
Lemma \ref{thm-s-extrem-eigen} in \cite{bai2010spectral} shows that  
$\lambda^{+}_{\min} (Z\tr Z/p) = \lambda_p$ if $p \leq  n$ and $\lambda^{+}_{\min} (Z\tr Z/p) = \lambda_{n}$ if $p > n$ where 
\begin{align}
	\label{ev-prod2}
	\lambda_{\max}(ZZ\tr/p) \cia (1 + \sqrt{n/p})^2 ~~ \text{ and } ~~ \lambda^{+}_{\min}(ZZ\tr/p) \cia (1- \sqrt{n/p})^2.
\end{align}
Thus,
\[
\rk(ZZ\tr/p) \cia p \wedge n ~\text{ and } ~\rk(S) = \rk(Z\tr Z \Sigma) \cia  n \wedge \rk(\Sigma). 
\]
Use $XX\tr XX\tr  = Z \Sigma Z\tr Z \Sigma Z\tr$ and  $ Z\tr Z = \sum^n_{i = 1} Z\tr_{i.} Z_{i.}$,
\begin{align}
	\label{prop1-term0}
	\Delta_n =  \frac{r^2}{n^3} \sum^n_{i = 1} \lam^2_i + \frac{n - 2r}{n} =  \frac{r^2}{n^2} (\frac{1}{n} \sum^n_{i = 1} \lam^2_i -1) + \frac{ (n - r)^2 }{n^2}, ~ \text{ and }   \\
n^{-1}\sum_{i} \lambda^2_i =  \frac{  \Trace(XX\tr XX\tr )   }{np^2} =\frac{1}{n p^2} \sum_{i, j \leq n} \Trace( \Sigma  Z_{i.}\tr Z_{i.} \Sigma  Z_{j.}\tr Z_{j.}   ) = A + B
\end{align}
where
\[
A    = \frac{1}{np^2} \sum_{i \neq j} \Trace( \Sigma  Z_{i.}\tr Z_{i.} \Sigma  Z_{j.}\tr Z_{j.}   )  ~ \text{ and } ~  
B  = \frac{1}{np^2}   \sum_{i} \Trace( \Sigma  Z_{i.}\tr Z_{i.} \Sigma  Z_{i.}\tr Z_{i.}   ) =  \frac{1}{np^2}   \sum_{i}  (Z_{i.} \Sigma  Z_{i.}\tr)^2.
\]
Note $\E[ \Trace( \Sigma  Z_{i.}\tr Z_{i.} \Sigma  Z_{j.}\tr Z_{j.} ) ] = \Trace(\Sigma^2)$ and independent components of $Z_{i.} = (z_1, \ldots, z_p)$ suggest that 
\begin{align*}
	\E[(Z_{i.} \Sigma  Z_{i.}\tr)^2] & = \E[ \sum_{a_1, a_2, a_3, a_4}   z_{ a_1}  z_{ a_2} z_{ a_3} z_{ a_4} \Sigma_{a_1, a_2}  \Sigma_{a_3, a_4} ] \\ 
	& = \E[ \sum^n_{j = 1}   z^4_{j} \Sigma^2_{j, j} ] + \sum_{s \neq t}   \E[   z^2_{ s}  z^2_{t} ] (\Sigma_{s,s}  \Sigma_{t, t} +  \Sigma^2_{s,t} + \Sigma^2_{t,s} ), 
\end{align*}
where each $z_{j}$ must appear at least twice to make the term  $z_{ a_1}  z_{ a_2} z_{ a_3} z_{ a_4}$ non-zero. The bounded moments of $\E^{1/k}[Z^k_{ij}] \leq C$ and $\Trace(\Sigma) = p$ then yield 
\begin{align}
	&\E[A] = \frac{n(n-1)}{np^2} \Trace(\Sigma^2) , \quad \E[B] =   \frac{ n}{np^2} \left( 2 \Trace(\Sigma^2) + \Trace^2(\Sigma) + (C-3) \sum^n_{j = 1} \Sigma^2_{j, j}  \right), ~~ \\
	& \E[A + B - 1] = \frac{n+1}{p^2} \Trace(\Sigma^2) + \frac{ \Trace^2(\Sigma) }{p^2} +\frac{ C-3 }{p^2} \sum^n_{j = 1} \Sigma^2_{j, j} -1  = \frac{n}{p^2} \Trace(\Sigma^2) \left( 1 + O\left( \frac{1}{n} \right) \right).  \label{lm-mom-term1}
\end{align}
Next we show the almost sure convergence of $A + B$ by considering
\[
\Var(A+ B) \leq 2 \left( \Var(A) + \Var(B)  \right).
\]
Specifically, we note that
\begin{align*}
	\Var(A) & =  \frac{1}{n^2 p^4}  \E[ \left\{\sum_{i \neq j} \Trace( \Sigma  \left( Z_{i.}\tr Z_{i.} - \I \right) \Sigma   \left( Z_{j.}\tr Z_{j.} - \I \right)  ) \right\}^2  ] \\
	& = \frac{1}{n^2 p^4}  \E[ \sum_{i \neq j} \sum_{s \neq t} \left\{  \Trace( \Sigma  \left( Z_{i.}\tr Z_{i.} - \I \right) \Sigma   \left( Z_{j.}\tr Z_{j.} - \I \right)  ) \right\} \left\{  \Trace( \Sigma  \left( Z_{s.}\tr Z_{s.} - \I \right) \Sigma   \left( Z_{t.}\tr Z_{t.} - \I \right)  ) \right\}  ] \\
	& = \frac{2 }{n^2 p^4} \E[ \sum_{i \neq j} \left\{  \Trace( \Sigma  \left( Z_{i.}\tr Z_{i.} - \I \right) \Sigma   \left( Z_{j.}\tr Z_{j.} - \I \right)  ) \right\} \left\{  \Trace( \Sigma  \left( Z_{i.}\tr Z_{i.} - \I \right) \Sigma   \left( Z_{j.}\tr Z_{j.} - \I \right)  ) \right\}  ] \\
	&= \frac{2}{p^4} \E[ (Z_{i.} \Sigma Z_{j.}\tr )^4 - \Trace^2(\Sigma^2) ]    \\
	\Var(B) & = \frac{ n }{n^2 p^4} \Var\left( (Z_{i.} \Sigma  Z_{i.}\tr)^2  \right).
\end{align*}
To give a bound of $ \Var( A )$ and $\Var( B )$, 
let $M = Z_{i.} \Sigma Z\tr_{i.} - \Trace(\Sigma)$, then we have 
\begin{align*}
	\left( Z_{i.} \Sigma Z\tr_{i.} \right)^4 & = \left( M + \Trace(\Sigma) \right)^4 = M^4  + 4 M^3 \Trace(\Sigma)  + 6 M^2 \Trace^2(\Sigma)^2 
	+ 4 M \Trace^3(\Sigma)  + \Trace^4(\Sigma). \\
	\left( Z_{i.} \Sigma Z\tr_{i.} \right)^2
	&= M^2 + 2 M \Trace(\Sigma)  + \Trace^2(\Sigma).
\end{align*} 
According to Lemma \ref{lem:sub-gaussian} for bounding $\E[M^2]$, \( \Trace(\Sigma^4) \leq \Trace^2(\Sigma^2) \) and \( \Trace(\Sigma^3) \leq p \Trace(\Sigma^2) \), according to some calculations, we arrive that
\begin{align*}
	\Var ( ( Z_{i.} \Sigma Z_{i.}\tr )^2 )  & = \E[ (Z_{i.} \Sigma Z_{i.}\tr )^4 ] -  \E^2[ (Z_{i.} \Sigma Z_{i.}\tr )^2 ]  \\ 
	&  = \E[M^4] + 4 \E[M^3] \Trace(\Sigma)  + 6 \E[M^2] \Trace^2(\Sigma)  + \Trace^4(\Sigma)  - (\E M^2 + \Trace^2(\Sigma) )^2  \\
	& = O( \Trace^2(\Sigma^2) +  \Trace^2(\Sigma) \Trace(\Sigma^2)   ). \\
	\Var\left( (Z_{i.} \Sigma Z_{j.}\tr )^2 \right) & = O( \Trace^2(\Sigma^2) ).
\end{align*}
which means that
\begin{align}
	\label{lm-mom-term2}
	\Var(A+ B) & \asymp \frac{1}{p^4} \Var\left( (Z_{i.} \Sigma Z_{j.}\tr )^2 \right) + \frac{ 1 }{ np^{4}} \Var ( ( Z_{i.} \Sigma Z_{i.}\tr )^2 ) \\ 
	&   \asymp \frac{1}{p^4} { \Trace^2(\Sigma^2)  } + \frac{ 1 }{ np^{2}} \Trace(\Sigma^2)   = \frac{1}{n^2} \left(   \frac{ n^2 }{p^4} { \Trace^2(\Sigma^2)  } + \frac{ n }{ p^{2}} \Trace(\Sigma^2) \right).
\end{align}
Collecting the terms in \ref{lm-mom-term1} and \ref{lm-mom-term2},  for any $\eps > 0$, we have $n^{-1}\sum_{i} \lambda^2_i - 1  = A+ B- 1$ and 
\begin{align*}
	P\left( | \frac{ n^{-1}\sum_{i} \lambda^2_i - 1 }{ n \Trace(\Sigma^2)/p^2 }  -  1  | > \eps \right) & \leq \frac{1}{ \eps^2 }  \frac{ \Var(A+ B) + \E[ ( A+ B -1  -  n \Trace(\Sigma^2)/ p^2 )^2  ] }{ ( n \Trace(\Sigma^2)/ p^2  )^2  }. \\
	& \lesssim \frac{1}{\eps^2} \frac{1}{ n^2} (1 + \frac{ p^2 }{ n \Trace(\Sigma^2) }) \\ 
	& \lesssim \frac{1}{ n^2} (1 + \frac{ p }{ n})
\end{align*}
Apply the Borel-Cantelli Lemma and $p/n \to c \in (0, \infty)$, 
\beq 
\label{delta-limit}
\left\{  n^{-1}\sum_{i} \lambda^2_i - 1 \right\} / \left\{ n\Trace (\Sigma^2)/p^2 \right\} \cia 1 .\eeq 
Finally with $r \cia n \wedge r(\Sigma)$, we plug \eqref{delta-limit} into \eqref{prop1-term0} and show that 
\begin{gather*}
	\Delta_n / \left\{ \frac{ \rk^2(\Sigma) \wedge n^2 }{n p^2} \Trace(\Sigma^2) + \left( 1 -  \frac{ \rk(\Sigma) \wedge n }{n } \right)^2  \right\} \cia 1
\end{gather*}

\paragraph{Weighted Gram matrix $S = X WX\tr$.} Given a pre-specified $W$, we may write the general weighted Gram matrix as 
$$ S= \frac{ n X W X\tr}{ \Trace( X W X\tr) } = \frac{Z \Sigma_{W} Z\tr} { \Trace( n^{-1} Z \Sigma_{W} Z\tr) } $$ where 
$\Sigma_{W} = \Sigma^{1/2} W \Sigma^{1/2}$. Note  $n^{-1} Z \Sigma_{W} Z\tr / \Trace(  \Sigma_{W} ) \cia 1$. In the following, we directly consider $ S = n^{-1} Z \Sigma_{W} Z\tr $ with $\Trace( \Sigma_{W}  ) = 1$.

Consider  \eqref{prop1-term0} where
\begin{align*}
\E[  n^{-1}\sum_{i} \lambda^2_i - 1 ] & = \E[  n^{-1}\Trace( Z \Sigma_{W} Z\tr Z \Sigma_{W} Z\tr  ) - 1 ]   \\
& = (n+1) \Trace(\Sigma_W^2) +  \Trace^2(\Sigma_W) +\frac{ C-3 }{p^2} \sum^n_{j = 1} \Sigma^2_{W,j j} -1 \\
&  = n \Trace(\Sigma_W^2) \left( 1 + O\left( \frac{1}{n} \right) \right) \\
	\Var( n^{-1}\sum_{i} \lambda^2_i - 1 )  & \asymp  { \Trace^2(\Sigma_W^2)  } + \frac{ 1 }{ n } \Trace(\Sigma_W^2)  
\end{align*}
This means that 
\begin{align*}
	P\left( | \frac{ n^{-1}\sum_{i} \lambda^2_i - 1 }{ n \Trace(\Sigma_W^2) }  -  1  | > \eps \right) & \leq \frac{1}{ \eps^2 }  \frac{ \Var(A+ B) + \E[ ( A+ B -1  - n \Trace(\Sigma_W^2) )^2  ] }{ ( n \Trace(\Sigma_W^2)  )^2  }. \\
	& \lesssim \frac{1}{\eps^2} \frac{1}{ n^2} (1 + \frac{ 1 }{ n \Trace(\Sigma_W^2) })
\end{align*}
Note according to Cauchy-Schwarz,  $\Trace^2(\Sigma_W) \leq p \Trace(\Sigma^2_W) $ then if $p/n \to c \in (0, \infty)$
\[
\left\{ n^{-1}\sum_{i} \lambda^2_i - 1 \right\} / \left(n \Trace(\Sigma^2_W) \right)  \cia 1.
\]
and
\begin{gather*}
	\Delta_n / \left\{ \frac{ \rk^2(\Sigma_W) \wedge n^2 }{n^2}  { n\Trace(\Sigma_W^2) } + \left( 1 -  \frac{ \rk(\Sigma_W) \wedge n }{n } \right)^2  \right\} \cia 1
\end{gather*}

\section{Proof of Auxiliary Lemmas}

	\ignore{
	 \subsection{Proof of Proposition 2}
	  Let the eigendecomposition of $\Sigma = \sum^p_{i = 1} \nu_i \ell_i \ell\tr_i$ with eigenvalues $ \nu_1 \geq \nu_2 \geq \cdots \geq \nu_p \geq 0$ and  write the coefficients of $\beta$ in the basis of eigenvectors as $( \ell\tr_1 \beta, \cdots, \ell\tr_p \beta ).$
	 \begin{align*}
	  \frac{1 }{ \| \beta\|^2} \beta\tr (\Sigma - z \I)^{-1} \beta  -  \frac{1}{p} \Trace\left( (\Sigma - z \I)^{-1} \right) 
	 =  \sum^p_{i = 1} (\frac{ (\ell_i\tr \beta)^2 }{ \| \beta \|^2 } - \frac{1}{p} ) \frac{1}{\lambda_i - z}. 
	 \end{align*}
	 where with $z = a + b i  \in \mathbb{C}^{+}$ and Cauchy-Shwartz inequality
	 \[
	 |\sum^p_{i = 1} (\frac{ (\ell_i\tr \beta)^2 }{ \| \beta \|^2 } - \frac{1}{p} ) \frac{1}{\lambda_i - z} | \leq \frac{\sqrt{p}}{b} \left(  \sum^p_{i = 1} \left( \frac{ (\ell_i\tr \beta)^2 }{ \| \beta \|^2 } - \frac{1}{p} \right)^2   \right)^{1/2}.
	 \]
	 Denote the support of $\beta$  by  $\cA$ and $ \ell_i \tr \beta = \ell_{i, \cA}\tr \beta_{A}.$ Then  Proposition 2 holds if for all $1 \leq i \leq p$
	\[
	 \frac{ (\ell_i\tr \beta)^2 }{ \| \beta \|^2 } - \frac{1}{p} =  \| \ell_{i, \cA} \|^2 \frac{ ( \ell_{i,\cA}\tr \beta_{\cA} )^2 }{ \| \ell_{i,\cA} \|^2  \| \beta_{\cA} \|^2 } - \frac{1}{p} = o(p^{-1}). 
	\]
Therefore one can check that with
\[
\| \ell_{i,\cA} \|^2 = (1 + o(1))\frac{s_{\cA}}{ p}, \text{ and } \frac{ ( \ell_{i,\cA}\tr \beta_{\cA} )^2 }{ \| \ell_{i,\cA} \|^2  \| \beta_{\cA} \|^2 } =   \left(1 + o(1)\right) \frac{1}{s_A},
\]
	
	
	
	\ignore{To show \eqref{dev0}, we consider a similar decomposition like  \eqref{eq2-1}. According to Assumption 2, there is
		$$
		\Var(T_1|S)\lesssim \Var(\xi_i^2|S) \sum^r_{i = 1} \kappa^2_i \lam^2_i \leq  \theta_0^2 \sum^r_{i = 1} \kappa_i^2
		$$ 
		For $T_b$, we know that  $\Var( T_b ) = n^{-2} \sigma^2_{\eps} \beta\tr X\tr  H^2(S)  X \beta$, which converges to $n^{-2} \sigma^2_{\eps} \theta^2 \Trace( H^2 S )$ and  Lemma \ref{lem:sub-gaussian} shows that.  
		\begin{align*} 
			\Var( T_c )  \lesssim n^{-2} \sigma^4_0 \Trace( (I + \gamma S)^{-2} )
		\end{align*}
	}
\end{proof}

}

\subsection{Proof of Lemma \ref{lem-s2}}

\ignore{
\begin{Lemma}
	\label{lem1}
	For the similarity matrix $S$ with rank $r$, define
	\beq
	\label{eq:Lambda}
	\Lambda=    \left(  \frac{ r^2/ n^2 }{ (r/n + \gamma)^2 }  \frac{1}{r} \sum^r_{i = 1} ( \tilde \lambda_i - 1 )^2   + \frac{n - r}{n}   \right) 
	\eeq
	Under Assumption \ref{asmp1}, we have 
	\begin{align}
		\sum^n_{i = 1} h_i \lambda_i & \asymp  r \gamma\check{q}^2  \Lambda  ~ \text{ and } ~
		\frac{ n^{-1}\sum^r_{i = 1} h_i \lambda_i }{ (1 - \bkap) \bkap }  \asymp \frac{\Lambda}{ \gamma(1 - r/n) + r/n }  , \label{l-r2} 
		\\
		\frac{\sum^r_{i = 1} h^2_i \lambda^2_i}{ (\sum^n_{i = 1} h_i \lambda_i)^2  } & = O\left( \frac{1}{r \Lambda } \right)   \\
		\frac{\sum^r_{i = 1} h^2_i \lambda_i}{ (\sum^n_{i = 1} h_i \lambda_i)^2  } & = O\left( \frac{1}{n \Lambda } \right)  \\
		\frac{\sum^n_{i = 1} h^2_i}{ (\sum^n_{i = 1} h_i \lambda_i)^2  }  & = O\left( \frac{ r }{ n^2 \Lambda} + \frac{  \gamma^2 }{n \Lambda } \right). \label{l-r1}
	\end{align}
Besides, $\Lambda$ is connected to $\Delta = n^{-1} \sum^n_{i = 1} (r\lambda_{i}/ n - 1)^2$ by
\begin{align}
	\label{l-r3}
	\Lambda  \geq \frac{1}{2} \left[ \frac{r}{n} \frac{\Delta}{ (1 + \gamma)^2 } + (1 - \frac{r}{n}) \frac{\Delta}{C' + 1}  \right] \gtrsim \frac{\Delta}{(\gamma + 1)^2}.
\end{align}
\end{Lemma}
}

\begin{Lemma}
	\label{lem-s2}
	\label{lem1}
	Consider a similarity matrix $S$ of rank $r$ with eigenvalues $\{ \lam_i \}^r_{i = 1}$ ($\lam_i = 0$ for $i > r$) and  $H(\gamma)$ defined in \eqref{eq:H}. Suppose Assumption ~\ref{asmp2} and condition ~\eqref{cond:C1} hold, and $ C^{-1} \leq \gamma \leq C$ for some constant $C > 0$. Then 
	\begin{align}
		|\Trace(HS)| \asymp  r \Delta_n,  \text{ and }  \quad ~ 
		  (1 - \bkap) \bkap  \asymp  \frac{r}{n} . \label{l-r2} 
	\end{align}
Besides,
		\begin{gather*}
		\frac{ \Trace(H^2 S^2) }{ \Trace^2(HS) } = O\left( \frac{1}{r \Delta_n } \right), \quad 
		\frac{ \Trace(H^2 S) }{ \Trace^2(HS) }   = O\left( \frac{1}{n \Delta_n } \right),  \\
		\frac{ \Trace(H^2) }{ \Trace^2(HS) }   = O\left( \frac{ r }{ n^2 \Delta_n} + \frac{ 1 }{n \Delta_n } \right). \label{l-r1}
	\end{gather*}

\end{Lemma}

\begin{proof}
Recall $\kappa_i = (1 + \gamma \lambda_i)^{-1}$ and $h_i = \kap_i^2 - \kap_i \bar \kap.$  Write $\tilde \lambda_i = r/n \lambda_i$. From condition ~\eqref{cond:C1},  $ M^{-1}_1 \leq \tilde \lambda_i \leq M_1 $ for $1 \leq  i \leq r$  and $\lam_i = 0$ for $i > r + 1.$ Using $\Trace(S) = n$ gives $\sum^r_{i = 1} \tilde \lambda_i = r.$ 
 Define
	\begin{gather*}
	q_i = \kappa_i \lam_i  =  \frac{ \tilde \lambda_i} {  r/n + \gamma \tilde \lambda_i }  ~ \text{  where } ~  \frac{1}{ \gamma + r/(n M_1) }  \leq q_i \leq \frac{1}{ \gamma + rM/n } \\
		\cq = r^{-1}\sum^r_{i = 1} q_i \asymp \frac{1}{ \gamma + r/n}. 
	\end{gather*}
	A direct calculation shows that  
	\begin{align*} 
	& 1 - \bar \kappa  = \frac{r \gam}{n} \check q \asymp \frac{r}{n},  ~\text{ and }~ \bar \kappa \asymp (1 - \frac{r}{n}) + \frac{r^2}{n^2} \\
	& \bar \kappa (1 - \bar \kappa  ) \asymp \frac{r}{n} + \frac{r^2}{n^2} ( \frac{r}{n} - 1) \asymp \frac{r}{n}.
	\end{align*}  For $\Trace(HS) =  \sum^r_{i = 1} h_i \lam_i$, one can check that 	$|h_i \lambda_i| \leq 1$ and  
\begin{align}
	h_i \lambda_i & = \frac{ \lambda_i}{ 1 + \gamma \lambda_i } (  \frac{ 1}{ 1 + \gamma \lambda_i } - n^{-1} \sum^n_{i=1} \frac{ 1}{ 1 + \gamma \lambda_i }  )  =  \gamma q_i ( \frac{r}{n} \check{q} - q_i ) 
	 \notag \\
	\Trace(HS) & = -{\gamma}\sum^r_{i = 1} q_i (q_i - r/n \check q )  = -{\gamma} \left(  \sum^r_{i = 1} (q_i - \check q)^2 + \frac{ r( n - r) }{n}  \check q^2 \right)  \label{l-t1}
\end{align} 
The term $\sum^r_{i = 1} ( q_i - \check q )^2$ is quantified based on the sample variance formula 
\begin{align*} 
	r^{-1}\sum^r_{i = 1} ( q_i - \check q )^2 & = \frac{1}{2r^2} \sum_{i} \sum_{j} (\frac{  \tilde \lambda_i }{ r/n + \gamma \tilde \lambda_i } - \frac{  \tilde \lambda_j  }{ r/n + \gamma \tilde \lambda_j})^2 
	\\ 
	& = \frac{1}{2r^2} \frac{ r^2 }{ n^2 } \sum_{i} \sum_{j} \frac{ ( \tilde \lam_i - \tilde \lam_j)^2 }{ (r/n + \gamma \tilde\lam_i )^2 (r/n + \gamma \tilde\lam_j )^2 } 
\end{align*}
which means 
\[
 \frac{ r^2 }{ n^2 } \frac{\delta^2}{ (r/n + \gamma M^{-1}_1 )^4 }  \geq  \frac{1}{r}\sum^r_{i = 1} ( q_i - \check q )^2    \geq \frac{ r^2 }{ n^2 } \frac{\delta^2}{ (r/n + \gamma M_1 )^4 } 
\]
where 
\[
 \delta^2 = \frac{1}{ (2r^2)}  \sum_{i, j} ( \tilde \lam_i - \tilde \lam_j)^2 = r^{-1}\sum^r_{i = 1} (\tilde\lam_i - 1)^2.
\]
With $\cq \asymp 1/(\gamma + r/n)$, it then follows that 
\begin{gather*}
	\label{l-t2}
    \frac{1}{r}\sum^r_{i = 1} ( q_i - \check q )^2  \asymp \check q^2 \frac{r^2}{n^2} \frac{1}{ (r/n + \gamma)^2 }  \delta^2  
    \\
	\eqref{l-t1} \asymp r \gamma \check{q}^2 \left( \frac{ r^2/ n^2 }{ (r/n + \gamma)^2 }  \delta^2 + \frac{n - r}{n} \right) \asymp r \Delta_n.  
\end{gather*}
where we use the fact that with  bounded  $\gamma$ and $\Delta_n = r\delta^2/n  + (n-r)/n \leq C' + 1$,
\begin{gather*}
\frac{ r^2/ n^2 }{ (r/n + \gamma)^2 }  \delta^2 + \frac{n - r}{n}  \geq \frac{1}{2} \left[ \frac{r}{n} \frac{\Delta_n}{ (r/n + \gamma)^2 } + (1 - \frac{r}{n}) \frac{\Delta_n}{C' + 1}  \right] 
\\
C''\leq \frac{1}{2} \left[ \frac{r}{ n(1 + \gamma)^2 } + (1 - \frac{r}{n}) \frac{1}{C' + 1}  \right] \leq \frac{ \frac{ r^2/ n^2 }{ (r/n + \gamma)^2 }  \delta^2 + (n - r)/{n} }{ r\delta^2/n  + (n-r)/n } \leq \max\left\{ \frac{r}{n (r/n + \gamma)^2 }   , 1 \right\} \leq C.
\end{gather*}
For terms in \eqref{l-r1}, their sizes are quantified in the same way as  \eqref{l-t1}. Specifically,
\begin{align} 
\Trace(H^2 S^2)  & =   \gamma^2 \sum^r_{i = 1} q^2_i ( q_i -  r \check q / n)^2 \lesssim  \gamma^2 \check q^2 \sum^r_{i = 1} q_i ( q_i -  r \check q / n) \lesssim \Trace(HS) 
\end{align}
Using that each eigenvalues $\lam_i \asymp n/r$, we have 
\begin{align}
\Trace(H^2 S)  & \asymp \frac{r}{n} \Trace(H^2 S^2) \lesssim \Trace(HS).
\end{align}
Finally, one has
\begin{align}
\Trace(H^2 ) & =  \gam^2 \sum^n_{i = 1} \frac{1}{ (1 + \gamma \lam_i)^2 } \left( q_i -  \frac{r}{n} \check q  \right)^2  \notag 
	\\
	& \lesssim {\gamma^2}  \left( \frac{1}{ (1 + \gam n/ r )^2 } \sum^r_{i = 1}  q_i ({ q_i -  r \check q / n }) + \left( 1 - \frac{1}{ (1 +  \gam n /r )^2 }  \right) \frac{r^2  (n-r) }{n^2} \check q^2  \right), \label{l-t3} \\
    & \lesssim  \frac{1}{ (1 + \gam n/ r )^2 } \Trace(HS) + \frac{r^2 (n-r)}{n^2}
\end{align} 
where $ 1 - (1 + \gamma n/r)^{-2}  \asymp  \frac{\gamma n/r}{ ( 1 + \gamma n/r ) } \asymp \gamma \cq.$ After some simplifications, we have 
\begin{align*}
	\frac{ \Trace(H^2 S^2) }{ \Trace^2(H S)  } = \frac{ {\sum^r_{i = 1} h^2_i \lam^2_i   }  }{ (\sum^r_{i = 1} h_i \lambda_i)^2 }  & \lesssim  \frac{ 1 }{ r \Delta_n }  \\
	\frac{ \Trace(H^2) }{ \Trace^2(H S)  } =  \frac{ \sum^r_{i = 1} h^2_i   }{ (\sum^r_{i = 1} h_i \lambda_i)^2 } 
	& \lesssim \frac{ r }{ n^2 \Delta_n} + \frac{ (n-r)  }{n^2 \Delta_n ^2 }  \lesssim \frac{ r }{ n^2 \Delta_n} + \frac{1 }{n \Delta_n  }. 
\end{align*}
where the last inequality uses the fact that $(n -r)/n \leq \Delta_n$. 

\end{proof} 

\begin{Lemma}
\label{lem:Hderiv}
Under the conditions of Lemma~\ref{lem1}, let
\[
\dot H(\gamma)=\frac{\partial H(\gamma)}{\partial\gamma},
\qquad
D_\gamma=-\Trace\{H(\gamma)S\}.
\]
If \(\Delta_n\to\Delta>0\), then uniformly over \(\gamma\in\Theta\),
\[
\frac{\Trace\{\dot H(\gamma)^2S\}}{D_\gamma^2}
+
\frac{\Trace\{\dot H(\gamma)^2\}}{D_\gamma^2}
=
O\left(\frac1n\right).
\]
\end{Lemma}

\begin{proof}
Differentiating
\[
h_i(\gamma)=\kappa_i^2-\bar\kappa_\gamma\kappa_i,
\qquad
\kappa_i=(1+\gamma\lambda_i)^{-1},
\]
gives
\[
\dot h_i
=
-2\lambda_i\kappa_i^3
-
\dot{\bar\kappa}_\gamma\kappa_i
+
\bar\kappa_\gamma\lambda_i\kappa_i^2,
\]
where
\[
\dot{\bar\kappa}_\gamma
=
-\frac1n\sum_{j=1}^r\lambda_j\kappa_j^2.
\]
Writing \(\alpha_n=r/n\), Condition~\eqref{cond:C1} and
\(C^{-1}\le\gamma\le C\) imply, uniformly for \(i\le r\),
\[
\lambda_i\kappa_i\le C,\qquad
\lambda_i\kappa_i^2\le C\alpha_n,\qquad
\lambda_i^2\kappa_i^3\le C\alpha_n,
\]
and
\[
|\dot{\bar\kappa}_\gamma|\le C\alpha_n^2.
\]
Thus
\[
|\dot h_i|\le C\alpha_n,\qquad i\le r,
\]
and for \(i>r\),
\[
\dot h_i=-\dot{\bar\kappa}_\gamma,
\qquad
|\dot h_i|\le C\alpha_n^2.
\]
Consequently,
\[
\Trace(\dot H^2S)
=
\sum_{i=1}^r\dot h_i^2\lambda_i
\lesssim
r\alpha_n^2\frac1{\alpha_n}
=
\frac{r^2}{n},
\]
and
\[
\Trace(\dot H^2)
\lesssim
r\alpha_n^2+n\alpha_n^4
\lesssim
\frac{r^3}{n^2}+\frac{r^4}{n^3}.
\]
Since Lemma~\ref{lem1} gives \(D_\gamma\asymp r\Delta_n\), and
\(\Delta_n\to\Delta>0\), we obtain
\[
\frac{\Trace(\dot H^2S)}{D_\gamma^2}
=
O\left(\frac1n\right),
\qquad
\frac{\Trace(\dot H^2)}{D_\gamma^2}
=
O\left(\frac1n\right).
\]

Finally,
\[
w_i(\gamma)=\lambda_i h_i(\gamma),
\qquad
\dot w_i(\gamma)=\lambda_i\dot h_i(\gamma).
\]
For \(i\le r\), \(\lambda_i\asymp n/r=\alpha_n^{-1}\), while
\(|\dot h_i|\le C\alpha_n\). Hence \(|\dot w_i|\le C\). The bound
\(|w_i|\le C\) follows from the same calculation in Lemma~\ref{lem1}.
\end{proof}

\ignore{
\begin{Lemma}
	\label{prob-order}
	Under the primal similarity representation \eqref{eq:primal}, Assumption \eqref{asmp2} and Condition \eqref{cond:C2}, for the specified matrix $S$ with rank $r$, there is
	\beq 
	\label{var_bound}
\E[	\Var( \xi^2_i | S) ]\leq {r \theta^4_0 }/{n}. 
	\eeq
\end{Lemma}
}
\ignore{
\begin{proof}
	Recall the total variance formula
	\[
	\Var(n^{-1}\beta\tr X\tr X \beta) =   \Var( n^{-1}\E[\beta\tr X\tr X \beta | S]) + \E( \Var[n^{-1}\beta\tr X\tr X \beta | S]) \leq C\frac{\theta_0^4}{n}. 
	\]
	Given the representation $X \beta = \PhiS_{n \times r} \xiS_{r \times 1}$ with $\PhiS = {U_W}_{n \times r} {D_W}_{ r \times r} $, and $S$ with $r$ non-zero eigenvalues  $\lambda_i$ for $i =1, \ldots, r$, there is $\PhiS\tr \PhiS =  D^2_W$ and 
	\[
	\beta\tr X\tr X \beta - \E[ \beta\tr X\tr X \beta | S] = \sum^r_{i=1} \lambda_i \left(\xi_i^2 - \E[\xi^2_i |S] \right).
	\]
	Applying Assumption \eqref{asmp2} and with $\lam_i \asymp n/r$, 
	\begin{align*}
		\Var\left(n^{-1}\E[\beta\tr X\tr X \beta | S] \right) &  =  \Var\left( n^{-1}\sum^r_{i=1} \lambda_i \xi^2_i |S  \right)  = \Var( \xi^2_i | S)  \frac{\sum^{r}_{i = 1} \lambda^2_i}{n^2}  \asymp  \frac{\Var( \xi^2_i | S)}{r}. 
	\end{align*}
	The results then followed by $	\E[ \Var\left(n^{-1}\E[\beta\tr X\tr X \beta | S] \right)] \leq  \Var(n^{-1}\beta\tr X\tr X \beta)$ and  
	\begin{align*}
	r^{-1} \E[ \Var( \xi^2_i | S) ] \leq C \theta_0^4 /n.
	\end{align*}
	\ignore{To show \eqref{dev0}, we consider a similar decomposition like  \eqref{eq2-1}. According to Assumption 2, there is
		$$
		\Var(T_1|S)\lesssim \Var(\xi_i^2|S) \sum^r_{i = 1} \kappa^2_i \lam^2_i \leq  \theta_0^2 \sum^r_{i = 1} \kappa_i^2
		$$ 
		For $T_b$, we know that  $\Var( T_b ) = n^{-2} \sigma^2_{\eps} \beta\tr X\tr  H^2(S)  X \beta$, which converges to $n^{-2} \sigma^2_{\eps} \theta^2 \Trace( H^2 S )$ and  Lemma \ref{lem:sub-gaussian} shows that.  
		\begin{align*} 
			\Var( T_c )  \lesssim n^{-2} \sigma^4_0 \Trace( (I + \gamma S)^{-2} )
		\end{align*}
	}
\end{proof}

\subsection{Preliminary results in the random matrix theory}
This section lists some well-studied theoretical results on the sample covariance matrix. We first introduce some concepts and definitions.   Suppose $X= Z\Sigma^{1/2}$ where the entries of $Z_{ij}$ of the $Z$ are i.i.d. random variables with
 $\E[Z_{ij}] = 0$, $\Var(Z_{ij}) = 1$ and $\E[Z^4_{ij}] < \infty$. The data dimension to sample size ratio $p/n \to c > 0$ when $n \to \infty.$  Suppose that the empirical spectral distribution of $\Sigma$ converges to a non-random limit as $p \to \infty$. 

Given a symmetric positive semidefinite matrix $A \in \mathbb{R}^{d \times d}$, we define the its spectral distribution as the empirical distribution of its eigenvalues. This is denoted as $F_{A}$, and writing $\lam_i(A), i = 1, \ldots, d$ for the eigenvalues of $A$, we have 
\[
F_{A}(t) = \frac{1}{d} \sum^d_{ i = 1} \1\{ \lam_i(A) \leq t \}. 
\] 
It is resolvent is given by 
\[
G_{A}(z) = ( A - z \I)^{-1}
\]
where $z$ is a complex variable defined away from all the real eigenvalues of $A$.
Another tool  to introduce is Stieltjes transform defined for the distribution $F$ supported on $[0, \infty)$
\[
m_F(z) = \int^{\infty}_{0} \frac{1}{\lambda - z} dF(\lambda), ~~ z \in \mathbb{C} \setminus \mathbb{R}^{+}.
\]
For the matrix $A$, one can check that $m_{F_{A}}(z) =  {d}^{-1} \Trace( G_{A}(z) ).$ 


\ignore{
There is also  a close connection between the Stieltjes transform of a measure F and its moments, which we denote by
\[
\mu_{k}(G) = \int \lam^k d F(\lam), k = 0, 1, 2 \cdots.
\]
By Taylor expansion, 
\[
\frac{1}{ s - z} = -\frac{1}{z} \frac{1}{ 1 - s/z} = - \frac{1}{z} \sum^{\infty}_{k = 0} (s/z)^k
\]
}

\begin{Assumption}
	\begin{enumerate}
		\item  Suppose $X= Z\Sigma^{1/2}$ where $Z_{ij}$ are i.i.d. random variables with mean $\E[Z_{ij}] = 0$, $\Var(Z_{ij}) = 1$ and $\E[Z^4_{ij}] < \infty$. 
		
		\item The sample size $n \to \infty$ while the dimensionality $p \to \infty$ as well, such that the aspect ratio $p/n \to \alpha > 0$.
		
		\item The spectral norm of $\Sigma$ is bounded in $p$ and its spectral distribution $F_{\Sigma}$ converges to a proper probability distribution $H$ supported on $[0, \infty)$, called the population spectral distribution (PSD).
	\end{enumerate}
\end{Assumption}

\ignore{
}

Next we define the sample covariance matrix \(
\widehat \Sigma = n^{-1} X\tr X. 
\) Recall the ESD of $\widehat \Sigma$ is 
\[
F_{\widehat \Sigma}(t) = \frac{1}{d} \sum^d_{ i = 1} \1\{ \lam_i(\widehat \Sigma) \leq t \}. 
\]
The following theorem proves that with probability $1$, $ F_{\widehat \Sigma}$ weakly converges to a nonrandom probability measure $F_{\alpha, H}$ where $p/n \to \alpha$ is the aspect ratio.
\begin{Theorem}[Yao et al, 2018]
	Under Assumption RMT, then almost sure  $ F_{\widehat \Sigma}$ weakly converges to a nonrandom probability measure $F_{\alpha, H}$.  The Stieltjes transform of the spectral measure of $\widehat \Sigma$ satisfies 
	\[
	m_{\widehat \Sigma} (z) = \frac{1}{p} \Trace\left( (\widehat \Sigma - z \I_{p \times p})^{-1} \right) \text{  converges to $m(z)$ }
	\] 
	both almost surely and in expectation, for any $ z \in \mathbb{C}^{+}$. Here $m(z)$ is implicitly defined by the equation 
	\[
	m(z) = \int \frac{1}{ t(1 - \alpha - \alpha z m(z)  ) - z } dH(t), z \in \mathbb{C}^{+}.
	\] 
\end{Theorem}

Denote the spectral decomposition of $\widehat \Sigma = \Gamma D \Gamma\tr$, where $ D$ is the diagonal eigenvalue matrix and $\Gamma = (\Gamma_{ij})$ is a unitary matrix consisting of the orthonormal eigenvectors of $\widehat \Sigma$. Let $u$ be an arbitrary nonrandom unit vector with $\| u \| = 1$ and that $v = (v_1, v_2, \ldots, v_n) \tr = \Gamma\tr u.$ 
Note the weighted empirical distribution is defined by 
\[
F_{u, \widehat \Sigma}(t) =  \sum^d_{ i = 1} v_i^2 \1\{ \lam_i(\widehat \Sigma) \leq t \}. 
\]
and
\beq
\label{m-v}
\frac{p}{n} \left( m_{\widehat \Sigma  }( z) + \frac{1}{z} \right) = m_{ S  }( z) + \frac{1}{z} ~~ z \in \mathbb{C} \setminus \mathbb{R}^{+}. 
\eeq
The next theorem specifies the setting when the weighted empirical distribution $F_{u, \widehat \Sigma}(t)$ is asymptotically equivalent to the $F_{\widehat \Sigma}(t).$

\begin{Theorem}(Bai et al, 2007)
	\label{thm-s-eigen}
	Under Assumption RMT and denote $H(t)$ as a proper limit population spectral distribution function of $\Sigma$. Then for a fixed unit vector $u \in \mathbb{R}^{p}$ with $u\tr (\Sigma - z \I)^{-1} u \to  m_{F_{H}}(z)$, where $m_{F_{H}}(z) $ denotes the Stieltjes transform of $H(t)$, it follows that for 
    \[
    F_{u, \widehat \Sigma}(t)  \to F_{\alpha,H} \quad \text{ a.s.} 
    \]
    Consequently, if $f(t)$ is a bounded function and $v = (v_1, v_2, \ldots, v_n) \tr = \Gamma\tr u$, one has 
    \[
    \sum^n_{j = 1} |v^2_j| f(\lambda_j) - \frac{1}{n} \sum^n_{j = 1} f(\lam_j) \to 0 , a.s.
    \]
\end{Theorem}
}

\ignore{
\paragraph{Moment property}
The next lemma establishes the connection between moments and the limiting distribution $F_{\alpha, H}.$
\begin{Lemma}[Lemma 2.16]
	Let
	$a_{j}= \int x^{j}\,dF_{y,H}(x)$ and 
	$b_{j}= \int t^{j}\,dH(t)$ for $j\ge 1$
	be, respectively, the $j$-th moments of the LSD $F_{\alpha,H}$ and the PSD $H$.
	Then
	\begin{equation}
		a_{j}
		= \;\alpha^{-1}\!
		\sum_{(i_{1},\dots,i_{j})}
		\alpha^{\,i_{1}+i_{2}+\cdots+i_{j}}\,
		(b_{1})^{i_{1}} (b_{2})^{i_{2}}\cdots(b_{j})^{i_{j}}
		\,\varphi^{(j)}_{i_{1},i_{2},\dots,i_{j}},
	\end{equation}
	where the sum is taken over all integer partitions
	\[
	(i_{1},\dots,i_{j}):\qquad
	j = 1\,i_{1}+2\,i_{2}+\cdots+j\,i_{j}, 
	\]
	and
	\begin{equation}
		\varphi^{(j)}_{i_{1},i_{2},\dots,i_{j}}
		= \frac{j!}{i_{1}!\,i_{2}!\cdots i_{j}!\,
			\bigl(j+1-(i_{1}+i_{2}+\cdots+i_{j})\bigr)!}
	\end{equation}
	is the corresponding multinomial coefficient.
\end{Lemma}

With $\Trace(\Sigma) = p$, for the first two moments, we have 
\[
a_1 = b_1 = 1, a_2 = \beta_2 + \alpha b_1^2,  a_2 - a^2_1 = \Trace(\Sigma^2) + (\alpha -1) 
\]
}

\ignore{
Finally, we have
\[
\Var(\left(Z_{i.} \Sigma Z_{i.}\tr )^2 \right)  = \E[ (Z_{i.} \Sigma Z_{i.}\tr )^4 ] -  \E^2[ (Z_{i.} \Sigma Z_{i.}\tr )^2 ] = O(\Trace(\Sigma^4) + \Trace^2(\Sigma^2) + \Trace(\Sigma^3) \Trace(\Sigma)  + \Trace^2(\Sigma) \Trace(\Sigma^2)   ).
\]
where
\[
( \sum^r_{ k = 1 } \widetilde\lambda_k u^2_{k} )^4  =  ( \sum^r_{ k = 1 } \widetilde\lambda_k (u^2_{k} - 1) + \Trace(\Sigma) )^4 = 
\]
\[
\Var(\left(Z_{i.} \Sigma Z_{i.}\tr )^2 \right) = \E[ ( \sum^r_{ k = 1 } \widetilde\lambda_k u^2_{k} )^4  ] - \E^2[ ( \sum^r_{ k = 1 } \widetilde\lambda_k u^2_{k} )^2  ] = 
\]
\[
\Var(Z_{i.} \Sigma Z_{j.}\tr)  = 2\Trace(\Sigma^2) + \Trace^2(\Sigma)
\]
Based on the moment property of Gaussian variables, one can check 
\[
\Var(A+ B)  \asymp p^{-4} \left( { \Trace^2(\Sigma^2) + \Trace(\Sigma^4)  } \right).
\]
by 
\begin{align*}
\Var\left( (Z_{i.} \Sigma Z_{i.}\tr )^2 \right) = C_1 \Trace(\Sigma^4) + C_2 \Trace^2( \Sigma^2), ~ \text{ and } ~
\Var\left( (Z_{i.} \Sigma Z_{j.}\tr )^2 \right) = C_{3}\Trace(\Sigma^4) + C_{4}\Trace^2( \Sigma^2).
\end{align*}
}


\begin{Lemma}
	\label{lem:sub-gaussian}
	Let $z = (z_1, \ldots, z_n )\tr$ be vectors of independent  random variables such that $\E[z_i] = 0$, $\E[z^2_i] = 1$ and finite $8$th moments. Then for a symmetric semi-positive matrix $A$,
	\begin{align*}
	\E[ (z\tr A z - \Trace(A) )^2] & = O\left( \Trace(A^2) \right) \\
		\E[ (z\tr A z - \Trace(A) )^3] & = O\left( \Trace^{3/2}(A^2) \right) \\
\E[ (z\tr A z - \Trace(A) )^4] & = O\left( \Trace^2(A^2) \right).
	\end{align*}
Let  $v =  (v_1, \ldots, v_n )\tr$ be another vector of independent  random variables such that $ \E[v_i] = 0$, $\E[v^2_i] = 1$ and finite $4$th moments.  For the following bi-linear terms, 
\begin{align*}
	\E[ (z\tr A v)^2] & = \Trace(A^2) \text{ and }\\
	\E[ ((z\tr A v)^2 - \Trace(A^2) )^2] & = O\left( \Trace(A^4) + \Trace^2(A^2) \right) 
\end{align*}
\end{Lemma}

The lemma could be easily derived when $z$ is a sub-Gaussian random vector, using the Hanson-Wright inequality for the quadratic forms.  Lemma 1 in \cite{bai2007asymptotics} shows that  for any $p \geq 2$, we have
	\begin{align*}
		\E\left[ \left| z\tr A z - \Trace(A) \right|^p \right] & \le K_p\left( \left( \E[|z_1|^4] \Trace(A^2) \right)^{p/2} + \E[z_1^{2p}] \Trace\left( A^p \right) \right) 
	\end{align*}
where $K_p$ is a constant. The results for the quadratic term then follows with $\Trace(A^p) \leq \Trace^p(A).$ 

The following proof only requires the finite 8th moments assumption instead of the sub-Gaussian assumption.

\begin{proof}
	
	We directly consider $\E[ (z\tr A z - \Trace(A) )^4]$. The treatment for $\E[ (z\tr A z - \Trace(A) )^3]$ and $\E[ (z\tr A z - \Trace(A) )^2]$ are the same. The analysis is based on the diagonal terms and off-diagonal terms separately where 
	\[
	z\tr A z - \Trace(A) = \sum^n_{i = 1} A_{ii} (z^2_i - 1) +  \sum^n_{i \neq j} A_{ij} z_i z_j.
	\]  
	Therefore
	\[
	\E[ (	z\tr A z - \Trace(A) )^4 ] \lesssim  \E[ (\sum^n_{i = 1} A_{ii} (z^2_i - 1))^4 ] + \E[ ( \sum^n_{i \neq j} A_{ij} z_i z_j)^4 ]
	\]
	where
	\begin{align*}
	 \E[ (\sum^n_{i = 1} A_{ii} (z^2_i - 1))^4 ] & = \E[ \sum_{ j,k,l,m}   (z^2_i - 1) (z^2_j - 1) (z^2_h - 1)(z^2_k - 1)  A_{jj}  A_{kk}  A_{ll} A_{mm}   ] \\
	 & = O( \sum^n_{ i = 1} A^4_{ii} +  \sum^n_{ i , j} A^2_{ii} A^2_{jj}  ) 
	\end{align*}
The off-diagonal terms require more attentions 
\[
 \E[ ( \sum^n_{i \neq j} A_{ij} z_i z_j)^4  = 
\E[ \sum_{ \substack{ i_1, j_1\ldots, i_4, j_4 \\
	i_1 \neq j_1, ..., i_4 \neq j_4 } }   z_{ i_1}  z_{j_1} ... z_{i_4} z_{ j_4}  A_{i_1, j_1}  A_{i_2, j_2} A_{i_3, j_3} A_{i_4, j_4} ] 
\]	
Still, the indices among the 8 coordinates must appear at least twice (i.e., paired) to make the  term non-zero, related to form a 4-edge multigraph, without self-loops due to $i_k \neq j_k$.The following scenarios are considered for $(i_1, j_1\ldots, i_4, j_4)$: (a) two vertices  $(j, k)$  and $A^4_{jk}$   (b) three vertices with $j,k,l$ and $A_{jk} A_{kl} A_{ l j} A_{j k} $ or $ A_{kj} A_{jk} A_{jl} A_{lj}$    (d) four different indices $j \neq  k \neq l \neq m$, which  
gives a cycle like  \( A_{jk} A_{kl} A_{lm}  A_{mj} \) (e) two separate cycle like \( A_{jk} A_{kj} A_{lm} A_{ml} \). 

Note that these terms are bounded by 
\[
\Trace(A^4) = \sum_{j,k,l,m} A_{jk} A_{kl} A_{lm}   A_{mj},  \text{ and } \Trace^2(A^2) = \sum_{j,k,l,m} A_{jk} A_{kj} A_{lm}   A_{ml}.
\] 
Therefore collecting the diagonal and off-diagonal terms, we arrive that 
\[
\E[ (z\tr A z - \Trace(A) )^4]  = O\left( \Trace^2(A^2) \right) = O( \Trace^2(A^2) ).
\]

For the bi-linear term, there is 
\begin{gather*}
	\E[  (z\tr A v)^2 - \Trace(A^2) )^2] \lesssim \E T^2_1 + \E T^2_2 ~~ \text{ where} \\
T_1 =  \sum_{i, j} (z^2_i - 1) (v^2_j - 1) A^2_{ij} ~ \text{ and } ~  T_2 =  \sum_{i \neq l, j \neq m} z_i z_l v_j v_m    A_{ij} A_{lm}.
\end{gather*}
The independence between $z$ and $v$ means that 
\begin{align*}
\E[ T^2_1  ]  & =  \sum_{i_1, j_1,i_2, j_2} \E[(z^2_{i_1} - 1) (v^2_{j_1} - 1)  (z^2_{i_2} - 1) (v^2_{j_2} - 1) ] A^2_{i_1 j_1}  A^2_{i_2 j_2} \\ 
& = \sum_{i, j} \E[ (z^2_{i} - 1)^2 ] \E[ (v^2_{j} - 1)^2] A^4_{i j} = O( \Trace^2(A^2) ) \\
\E[ T^2_2  ]  & = \sum_{i \neq l, j \neq m} \E[z^2_i z^2_l v^2_j v^2_m]    A^2_{ij} A^2_{lm} 
\end{align*}
Therefore
\[
\E[  (z\tr A v)^2 - \Trace(A^2) )^2] = O(\Trace^2(A^2))
\]

\end{proof}

\section{Preliminary random matrix results}
\label{sec:rmt}
This section lists several concepts, definitions, and standard theoretical results from random matrix theory.

\begin{Assumption}
	\label{asmp:rmt}
	\begin{enumerate}
		\item  Suppose $X= Z\Sigma^{1/2}$ where $Z_{ij}$ are i.i.d. random variables with mean $\E[Z_{ij}] = 0$, $\Var(Z_{ij}) = 1$ and $\E[Z^4_{ij}] < \infty$. 
		
		\item The sample size $n \to \infty$ while the dimensionality $p \to \infty$ as well, such that the aspect ratio $p/n \to \alpha > 0$.
		
		\item The spectral norm of $\Sigma$ is bounded in $p$ and its spectral distribution $F_{\Sigma}$ converges to a proper probability distribution $H$ supported on $[0, \infty)$, called the population spectral distribution (PSD).
	\end{enumerate}
\end{Assumption}


Given a symmetric positive semi-definite matrix $A \in \mathbb{R}^{d \times d}$, we define its spectral distribution as the empirical distribution of its eigenvalues. This is denoted as $F_{A}$, and writing $\lam_i(A), i = 1, \ldots, d$ for the eigenvalues of $A$, we have 
\[
F_{A}(t) = \frac{1}{d} \sum^d_{ i = 1} \1\{ \lam_i(A) \leq t \}. 
\] 
Its resolvent is given by 
\[
G_{A}(z) = ( A - z \I)^{-1}
\]
where $z$ is a complex variable defined away from all the real eigenvalues of $A$.
Another tool  to introduce is Stieltjes transform defined for the distribution $F$ supported on $[0, \infty)$
\[
m_F(z) = \int^{\infty}_{0} \frac{1}{\lambda - z} dF(\lambda), ~~ z \in \mathbb{C} \setminus \mathbb{R}^{+}.
\]
For the matrix $A$, one can check that $m_{F_{A}}(z) =  {d}^{-1} \Trace( G_{A}(z) ).$ 


\ignore{
	There is also  a close connection between the Stieltjes transform of a measure F and its moments, which we denote by
	\[
	\mu_{k}(G) = \int \lam^k d F(\lam), k = 0, 1, 2 \cdots.
	\]
	By Taylor expansion, 
	\[
	\frac{1}{ s - z} = -\frac{1}{z} \frac{1}{ 1 - s/z} = - \frac{1}{z} \sum^{\infty}_{k = 0} (s/z)^k
	\]
}

\ignore{
}

Denote SVD of $ X = U D \Gamma\tr$  and $\widehat \Sigma = n^{-1}\Gamma D^2 \Gamma\tr$,  where $\Gamma = (\Gamma_{ij}) = (\Gamma_{,1}, \ldots, \Gamma_{,p})$ is a unitary matrix consisting of the orthonormal eigenvectors of $\widehat \Sigma$. The ESD of the sample covariance matrix \(
\widehat \Sigma = n^{-1} X\tr X 
\)  is 
\[
F_{\widehat \Sigma}(t) = \frac{1}{p} \sum^p_{ i = 1} \1\{ \lam_i(\widehat \Sigma) \leq t \}. 
\]
\ignore{
	The following theorem proves that with probability $1$, $ F_{\widehat \Sigma}$ weakly converges to a nonrandom probability measure $F_{\alpha, H}$ where $p/n \to \alpha$ is the aspect ratio.
	\begin{Theorem}[Yao et al, 2018]
		Under Assumption RMT, then almost sure  $ F_{\widehat \Sigma}$ weakly converges to a nonrandom probability measure $F_{\alpha, H}$.  The Stieltjes transform of the spectral measure of $\widehat \Sigma$ satisfies 
		\[
		m_{\widehat \Sigma} (z) = \frac{1}{p} \Trace\left( (\widehat \Sigma - z \I_{p \times p})^{-1} \right) \text{  converges to $m(z)$ }
		\] 
		both almost surely and in expectation, for any $ z \in \mathbb{C}^{+}$. Here $m(z)$ is implicitly defined by the equation 
		\[
		m(z) = \int \frac{1}{ t(1 - \alpha - \alpha z m(z)  ) - z } dH(t), z \in \mathbb{C}^{+}.
		\] 
	\end{Theorem}
}
Let $u$ be an arbitrary nonrandom unit vector with $\| u \| = 1$ and that $v = (v_1, v_2, \ldots, v_n) \tr = \Gamma\tr u.$ 
Define the weighted empirical distribution that puts mass $v_i^2$ at the place as  $\lambda_i$.
\[
F_{u, \widehat \Sigma}(t) =  \sum^p_{ i = 1} v_i^2 \1\{ \lam_i(\widehat \Sigma) \leq t \}. 
\]
The Stieltjes transform of  the Gram matrix $S = XX\tr/n$ is given as
\beq
\label{m-v}
\frac{p}{n} \left( m_{\widehat \Sigma  }( z) + \frac{1}{z} \right) = m_{ S  }( z) + \frac{1}{z} ~~ z \in \mathbb{C} \setminus \mathbb{R}^{+}. 
\eeq
The next theorem shows when the weighted empirical distribution $F_{u, \widehat \Sigma}(t)$ is asymptotically equivalent to the $F_{\widehat \Sigma}(t).$

\begin{Theorem}[\cite{bai2007asymptotics}]
	\label{thm-s-eigen}
	Under Assumption \ref{asmp:rmt} and denote $H(t)$ as a proper limit population spectral distribution function of $\Sigma$. Then for a fixed unit vector $u \in \mathbb{R}^{p}$ with $u\tr (\Sigma - z \I)^{-1} u \to  m_{F_{H}}(z)$, where $m_{F_{H}}(z) $ denotes the Stieltjes transform of $H(t)$, it follows that for 
	\[
	F_{u, \widehat \Sigma}(t)  \to F_{\alpha,H} \quad \text{ a.s.} 
	\]
	Consequently, if $f(t)$ is a bounded function and $v = (v_1, v_2, \ldots, v_n) \tr = \Gamma\tr u$, one has 
	\[
	\sum^n_{j = 1} |v^2_j| f(\lambda_j) - \frac{1}{n} \sum^n_{j = 1} f(\lam_j) \to 0 , a.s.
	\]
\end{Theorem}

Next theorem specifies the size of the largest and smallest eigenvalues of $Z Z\tr/p$. Denote the eigenvalues of $Z Z\tr/p$ by $\lam_1 \geq \lam_2 \geq \cdots \leq \lam_n$. Write $\lam_{\max} = \lam_1$.  For $n \leq p$, $\lam_{\min} = \lam_n$; for $n > p$,  $\lam_{\min} = \lam_p$.

\begin{Theorem}[ Theorem 5.11 in \cite{bai2010spectral}]
	\label{thm-s-extrem-eigen}
	Suppose that Assumption \ref{asmp:rmt} holds and $p/ n \to  \alpha$. The ESD of  $Z Z\tr/p$ converge to the Marchenko–Pastur law and surely
	\begin{align}
		\lim_{n \to \infty} \lambda_{\min}(Z Z\tr/p) = (1 - \sqrt{1/\alpha})^2
	\end{align}
	and 
	\begin{align}
		\lim_{n \to \infty} \lambda_{\max}(Z Z\tr/p) = (1 +  \sqrt{1/\alpha})^2.
	\end{align}
\end{Theorem}

We call a random orthogonal matrix distributed uniformly over the $k \times k$ orthogonal matrices space ${\bbO}_k$ as a k-dimensional Haar matrix. Next lemma establishes  the basic  properties on the Haar matrices. The proof is given in  \cite{bai2010spectral}, Section 10. 

\begin{Lemma}
	\label{append:haar}
	For a $k \times k$ random haar matrix $H$ distributed uniformly over the orthogonal matrices space $\mathbb{O}_k$ 
	we have the following properties:
	\begin{enumerate}[label=(\alph*)]
		\item for any unit k-dimensional vector $u$, $ H u $ is uniformly distributed on the unit $k-$sphere. 
		
		\item $H\tr$  is also Haar distributed.
		
		\item if $Z$ is a $k \times k$ matrix with entries i.i.d.  $N(0,1)$, then $Z(ZZ\tr)^{-1/2}$ and $(ZZ\tr)^{-1/2}Z$ are Haar distributed. 
		
		\item consider a $k$-dimensional  unit vector $u$,  $ \beta = (\beta_1, \cdots, \beta_k) = H u$ and $f$ is a bounded continuous function. Denote $\phi(x)$ as  the density of $N(0,1)$. Then, as $k \to \infty$, 
		\[
		\frac{1}{ k } \sum^k_{j = 1} f(\sqrt{k} \beta_j) \to \int f(x) \phi(x) dx, \text{ a.s. }, 
		\]
		which suggests  the empirical distribution function of the entries of $\sqrt{k} \beta$ converge weakly to the distribution function of $N(0,1)$.  
		
		\item if $Z$ is a $k \times k$ matrix with entries iid $N(0,1)$, then $Z(Z\tr Z)^{-1/2}$ and $(Z Z\tr)^{-1/2}Z$ are Haar distributed.

		
	\end{enumerate}
\end{Lemma}

\section{ Derivation of statistical procedures}

By Equation~\eqref{eq:eqv} in the main text, the
SMILE equations $\Psi_n=0$ are equivalently to the
Gaussian score-basis equation $\Psi_n^{\rm G}=0$ based on 
\begin{align}
\label{eq:psi-G-transform-supp}
\Psi_n(\theta^2,\sigma^2_{\eps})
=
\begin{pmatrix}
\theta^2 & \sigma^2_{\eps}\\
0 & 1
\end{pmatrix}
\Psi_n^{\rm G}(\theta^2,\sigma^2_{\eps}),
\end{align}
Differentiating \eqref{eq:psi-G-transform-supp} gives
\[
\frac{\partial \Psi_n}{\partial(\theta^2, \sigma^2_{\eps})\tr}
=
\begin{pmatrix}
\theta^2 & \sigma^2_{\eps}\\
0 & 1
\end{pmatrix}
\frac{\partial \Psi_n^{\rm G}}{\partial(\theta^2, \sigma^2_{\eps})\tr}
+
\begin{pmatrix}
\psi^{\rm G}_{\theta^2} &
\psi^{\rm G}_{\sigma^2_{\eps}}\\
0 & 0
\end{pmatrix}.
\]
Specifically,
\begin{align*}
& \frac{\partial \Psi_n}{\partial (\theta^2, \sigma^2_{\eps})\tr} =
\begin{pmatrix}
    -y\tr V^{-1} S V^{-1} y &  - y V^{-2} y \\
    \Trace( V^{-1} S V^{-1} ) -  2\left\{ y\tr V^{-1} S V^{-2}  y \right\} &  \Trace( V^{-2} ) - 2\left\{ y\tr V^{-3}  y \right\} \\
\end{pmatrix} \\
\end{align*} 
Under the similarity modeling, we have the moment equations $\E[ \Psi^G  ] = 0$, we have 
\[
\E\left[ \frac{\partial \Psi_n}{\partial(\theta^2, \sigma^2_{\eps})\tr} \right]
=
\begin{pmatrix}
\theta^2 & \sigma^2_{\eps}\\
0 & 1
\end{pmatrix}
\E\left[ \frac{\partial \Psi_n^{\rm G} }{ \partial(\theta^2, \sigma^2_{\eps})\tr } \right]. 
\]
This proves the invariance of the sandwich covariance under the
nonsingular transformation from $\Psi_n^{\rm G}$ to $\Psi_n$, i.e., 
\beq
\Var\begin{pmatrix}
	\hat \theta^2 \\
	\hat \sigma^2_{\eps}
\end{pmatrix} = \E[\dot{\Psi} ]^{-1} \E[\Psi \Psi\tr] \E[ \dot{\Psi} ]^{-1} = \E[\dot{\Psi}^G ]^{-1} \E[\Psi^G {\Psi^G} \tr] \E[ \dot{\Psi}^G ]^{-1}
	\eeq
Therefore, the covariance calculation can be carried out using $\Psi_n^{\rm G}$. Throughout this subsection, all quantities denoted by
$\Psi$, $\psi_{\theta^2}$, $\psi_{\sigma^2_{\eps}}$, and $\dot\Psi$
refer to the Gaussian score-basis estimating function
$\Psi_n^{\rm G}$. 

One can also check that fisher-scoring updates for $\Psi$ or $\Psi^G$ are equivalent.

\subsection{Derivation of  variance formula under the working Gaussian model }
\label{append:likelihood}
Furthermore, note $V = \sigma^2_e \I + \Qbeta S$ and 
\[
\frac{\partial V^{-1}}{ \partial \sigma^2_{\eps} } = -V^{-2}, \quad \frac{\partial V^{-1}}{ \partial \Qbeta } = -V^{-1}S V^{-1}.
\]
Therefore given that 
\begin{align*}
		\label{eq:esteq}
 & \Psi  =  \begin{cases} 
			\Trace( V^{-1}  S V^{-1}  y y\tr) - \Trace( V^{-1} S ) 
			\\
			\Trace(  V^{-2} y y\tr)  - \Trace(V^{-1})
		\end{cases}, 
	\end{align*}
and
\begin{align*}
\frac{\partial \Psi_n}{\partial (\theta^2, \sigma^2_{\eps})\tr} =
\begin{pmatrix}
   \Trace( V^{-1} S V^{-1} S ) - 2\left\{ y\tr V^{-1} S V^{-1} S V^{-1}  y \right\} &  \Trace( V^{-1} S V^{-1} ) -  2\left\{ y\tr V^{-1} S V^{-2}  y \right\}  \\
    \Trace( V^{-1} S V^{-1} ) -  2\left\{ y\tr V^{-1} S V^{-2}  y \right\} &  \Trace( V^{-2} ) - 2\left\{ y\tr V^{-3}  y \right\} \\
\end{pmatrix},
\end{align*}
with $\E[yy\tr |S] = V$, we have
\[
\E[\dot \Psi] =   -\begin{pmatrix}
	\Trace( V^{-1} S V^{-1} S ) 
	&  \Trace( V^{-1} S V^{-1})   \\
	\Trace( V^{-1} S V^{-1})   & \Trace(V^{-2})
\end{pmatrix} .
\]
Note $V^{-1} S = (I - \sigma^2_{\eps} V^{-1})/\theta^2$ when $\theta^2 \neq 0$, then
\begin{align*}
	 \E[\dot \Psi] 
	& = \frac{1}{\Qbeta} \begin{pmatrix}
		{  \Trace\left(  (\sigma^2_e  V^{-1} - \I)^2 \right) } /{\Qbeta} 
		& { \Trace(V^{-1}) - \sigma^2_e \Trace(V^{-2}) }  \\
		{ \Trace(V^{-1}) - \sigma^2_e \Trace(V^{-2}) }  & \Trace(V^{-2}) \Qbeta
	\end{pmatrix}
\end{align*}
The $2 \times 2$ matrix inverse is obtained as 
\begin{align*}
(\E[\dot \Psi])^{-1} & =  \det( \E[\dot \Psi]  )^{-1}   \begin{pmatrix}
	\E[\dot \Psi]_{22} & - \E[\dot \Psi]_{12} \\
	-\E[\dot \Psi]_{12} & \E[\dot \Psi]_{11}
\end{pmatrix}	\\
\det( \E[\dot \Psi]  ) 
& = \frac{ n \Trace(V^{-2}) - \Trace^2(V^{-1})  }{\theta^4}. 
\end{align*}
The expression for $\E[\Psi \Psi\tr]$ is given below, which involves the fourth-order moment of $y$,
\[
\small
\E[\Psi \Psi\tr | S] = \begin{pmatrix}
	\E[ (y\tr V^{-1} S V^{-1} y)^2 |S] - \Trace^2(V^{-1} S) & \E[ y\tr V^{-1} S V^{-1} y y\tr V^{-2} y | S ] - \Trace(V^{-1} S) \Trace(V^{-1}) \\ 
	\E[ y\tr V^{-1} S V^{-1} y y\tr V^{-2} y | S ] - \Trace(V^{-1} S) \Trace(V^{-1}) & E[ (y\tr V^{-2} y)^2 | S ] - \Trace^2(V^{-1})
\end{pmatrix}
\]
Under the normality condition, the quadratic property of multivariate normal $y$  simplifies the expression as  $ \E[\Psi \Psi\tr] = 2\E[\dot \Psi ]$ and  
  \begin{align}
  	\label{vareq}
		& \var({\widehat \theta}^2) =  \frac{2 \theta^4 \Trace(V^{-2})}{ n \Trace( V^{-2})  - \Trace^2(V^{-1}) } 
		\\ 
		&  \var(\widehat \sigma^2_{\eps}) =  \frac{ 2 \sigma^4_{\eps}
			\Trace\left(  (\sigma^2_\eps  V^{-1} - \I)^2 \right)  }{n \Trace( V^{-2})  - \Trace^2(V^{-1}) } 
	\end{align}
    \ignore{
 Let  $\kappa_i = 1/(1 + \gamma \lambda_i )$ and $\bar \kappa = n^{-1} \sum^{n}_{i = 1} \kappa_i$,
 \begin{align*}
	& \var({\widehat \theta}^2) =  
	\frac{2 \theta^4 }{n} \left(1 + \frac{ \bar \kappa^2}{ n^{-1} \sum^{n}_{i = 1} (\kappa_i - \bar \kappa)^2  } \right)		
	\\ 
	&  \var(\widehat \sigma^2_{\eps}) =  
\frac{2 \sigma_{\eps}^4 }{n} \left(1 + \frac{  (\bar \kappa - 1) ^2 }{ n^{-1} \sum^{n}_{i = 1} (\kappa_i - \bar \kappa)^2  } \right)		
\end{align*}
}
\ignore{
Another way to derive the estimator is through the likelihood function based on the normal distribution, 
$y|S \sim N(0, V)$ with $V = \sigma^2_e \I + \Qbeta S$ as 
\[
\ell(\Qbeta, \sigma^2_e; y, U )  = -n/2 \log(2 \pi) - 1/2 \log|V| - 1/2 y\tr V^{-1} y.
\]
The score function $\ell'$ of the likelihood function with respect to $\theta^2$ and $\sigma^2_{e}$ is 
\begin{align*}
	\ell'_{\theta^2} = \frac{\partial \ell }{\partial \theta^2 }
	& = -\frac{1}{2} \Trace( V^{-1} S )  + 
	\frac{1}{ 2 } \left\{ y\tr V^{-1} S V^{-1}  y \right\} 
	\\
	\ell'_{\sigma^2_{e}} = \frac{\partial \ell }{\partial \sigma^2_e} & = 
	-\frac{1}{2} \Trace(V^{-1}) + 1/2 y\tr V^{-1} V^{-1} y.
\end{align*}
}

\ignore{
\[
\small
\E[\Psi \Psi\tr] = \begin{pmatrix}
	\E[ (y\tr V^{-1} S V^{-1} y)^2 ] - \Trace^2(V^{-1} S) & \E[ y\tr V^{-1} S V^{-1} y y\tr V^{-2} y ] - \Trace(V^{-1} S) \Trace(V^{-1}) \\ 
	\E[ y\tr V^{-1} S V^{-1} y y\tr V^{-2} y ] - \Trace(V^{-1} S) \Trace(V^{-1}) & E[ (y\tr V^{-2} y)^2] - \Trace^2(V^{-1})
\end{pmatrix}
\]
}


\subsection{Computational procedure of fisher-scoring updates  with  $(y, b_y, \tf, \ts )$ }
\label{appd-compute}

Given that $V^{-1} = \sigma^{-2}_{\eps} (\I + \gamma S)^{-1} = \sigma^{-2}_{\eps} H^{-1}$, we have
\begin{gather*}
V^{-1} S = \sigma^{-2}_{\eps} S (\I + \gamma  S )^{-1} = \sigma^{-2}_{\eps}  \gamma^{-1} (\I + \gamma  S - \I ) (\I + \gamma  S )^{-1} = \sigma^{-2}_{\eps}  \gamma^{-1} (\I -  H^{-1})	\\
V^{-1} S V^{-1} = \sigma^{-4}_{\eps}  \gamma^{-1} H^{-1}  (\I -  H^{-1}) \\
\Trace( V^{-1} S V^{-1} S ) = \sigma^{-4}_{\eps}  \gamma^{-2} \Trace( (\I -  H^{-1})^2 ).
\end{gather*}
We express each term of the score function and the Fisher information matrix  by $(y, b_y, \tf, \ts )$ in \eqref{eq:simp}
\begin{gather*}
y\tr V^{-1} S V^{-1} y =  
\frac{1}{ \sigma^{4}_{\eps}  \gamma } (y\tr b_y - b_y \tr b_y), \quad y\tr V^{-2} y = \frac{1}{\sigma^{4}_{\eps}} b_y\tr b_y  \\
\Trace( V^{-1} S ) = \frac{1}{ \sigma^{2}_{\eps}  \gamma } (n - \tf), \quad
\Trace(V^{-1}) = \frac{1}{ \sigma^{2}_{\eps} } \tf, \quad \Trace(V^{-2}) = \frac{1}{ \sigma^{4}_{\eps} } \ts \\
\Trace( V^{-1} S V^{-1} S ) = \sigma^{-4}_{\eps}  \gamma^{-2} ( n - 2 \tf + \ts). 
\end{gather*}
Therefore for the fisher-scoring updates 
\begin{align*}
	\Psi = \begin{pmatrix}
	     -\frac{1}{ \gamma \sigma^2_{\eps}} (n - t^{(1)} ) +  \frac{1}{ \sigma^4_{\eps} \gamma } \left\{ y\tr b_y - b_y \tr b_y  \right\} \\
     -\frac{1}{\sigma^2_{\eps} } t^{(1)} + \frac{1}{ \sigma^4_{\eps} } b_y \tr b_y,    
	\end{pmatrix} 
\end{align*}
and
\[
\E[ \dot{\Psi}]  = -\begin{pmatrix}
	(\ts - 2\tf + n)/\theta^4   &  ( t^{(1)} - t^{(2)} ) / \theta^2 \sigma^2_\eps  \\
	{ ( t^{(1)} - t^{(2)} ) }/ \theta^2 \sigma^2_\eps  & t^{(2)}/\sigma^4_{\eps}  
\end{pmatrix}.
\]
Finally, according to \eqref{vareq}, we have
\beq
\label{eq:vareq-simple}
\Var(\hat \sigma_{\eps}^2) = \frac{ 2 \sigma^4_{\eps}  (\ts - 2 \tf + n) }{ n \ts - {\tf}^2 }, \quad  \Var(\hat \theta^2) = \frac{ 2 \theta^4 \ts }{ n \ts - {\tf}^2 }.
\eeq
\ignore{
Besides, we may express $\dot{\Psi}$ as
\[
\dot{\Psi} = \begin{pmatrix}
	\frac{(\ts - 2\tf + n)}{ \theta^4 } - 2 \frac{  y H^{-1} (\I - H^{-1})^2 y  }{\gamma^2 \sigma^6_{\eps} }    &  \frac{ ( t^{(1)} - t^{(2)} ) }{ \theta^2 \sigma^2_\eps} -2 \frac{ b_y (\I - H^{-1}) b_y }{ \sigma^6_{\eps} \gamma }  \\
	\frac{ ( t^{(1)} - t^{(2)} ) }{ \theta^2 \sigma^2_\eps} -2 \frac{ b_y (\I - H^{-1}) b_y }{ \sigma^6_{\eps} \gamma }  & \frac{t^{(2)}}{\sigma^4_{\eps}}  -  2\frac{ b_y\tr H^{-1} b_y }{ \sigma^6_{\eps} }
\end{pmatrix}.
\]
Denote $m_y = H^{-2}y$ then we have 
\[
\dot{\Psi} = \begin{pmatrix}
	\frac{(\ts - 2\tf + n)}{ \theta^4 } - 2 \frac{  (y - b_y) (b_y - m_y)  }{\gamma^2 \sigma^6_{\eps} }    &  \frac{ ( t^{(1)} - t^{(2)} ) }{ \theta^2 \sigma^2_\eps} -2 \frac{ b_y\tr b_y - b_y\tr m_y  }{ \sigma^6_{\eps} \gamma }  \\
	\frac{ ( t^{(1)} - t^{(2)} ) }{ \theta^2 \sigma^2_\eps} -2 \frac{ b_y\tr b_y - b_y\tr m_y  }{ \sigma^6_{\eps} \gamma }  & \frac{t^{(2)}}{\sigma^4_{\eps}}  -  2\frac{ b_y\tr m_y }{ \sigma^6_{\eps} }
\end{pmatrix}.
\] 
}

\ignore{
\begin{algorithm}[H]
	\DontPrintSemicolon
	\SetKwInOut{Input}{Input}
	\SetKwInOut{Output}{Output}
	\SetAlgoNoLine
	\KwIn{\( (y, X, W) \) and initial values $ (\theta^2(0), \sigma^2_{\eps}(0) ).$ }
	\KwOut{Estimated $(\hat \theta^2, \hat \sigma^2_{\eps})$ and their estimated covariance.}
	\caption{Estimation and inference of $\theta^2$ with large-scale data }
	
	
	\For{\(k = 0, 1, \ldots\)}{
		Calculate $b_y(k)$ using the conjugate gradient method.
		
		Calculate $\tf(k)$ and $\ts(k)$ with randomized trace estimators.
		
		
		Obtain $ \Psi(k) = ( \psi_{ \theta^2  }(k), \psi_{ \sigma_{\eps}^2 }(k) )$ and $\E[ \dot{\Psi} | S]$ by  
		\begin{align*}
			\small
			\psi_{ \theta^2  } & = -\frac{1}{ \gamma \sigma^2_{\eps}} (n - t^{(1)} ) +  \frac{1}{ \sigma^4_{\eps} \gamma } \left\{ y\tr b_y - b_y \tr b_y  \right\},
			\\
			\psi_{ \sigma_{\eps}^2 }  & = 
			-\frac{1}{\sigma^2_{\eps} } t^{(1)} + \frac{1}{ \sigma^4_{\eps} } b_y \tr b_y, \\
			\E[ \dot{\Psi} | S] & = -\begin{pmatrix}
				(\ts - 2\tf + n)/\theta^4   &  ( t^{(1)} - t^{(2)} ) / (\theta^2 \sigma^2_\eps)  \\
				{ ( t^{(1)} - t^{(2)} ) }/ (\theta^2 \sigma^2_\eps)  & t^{(2)}/\sigma^4_{\eps}  
			\end{pmatrix}.
		\end{align*}

		Update the estimators:
		\[\small
		\begin{pmatrix}
			\theta^2{(k+1)} \\
			\sigma^2_{\eps}(k+1)  
		\end{pmatrix} = \begin{pmatrix}
			\theta^2{(k)} \\
			\sigma^2_{\eps}(k)  
		\end{pmatrix} - \left(\E[ \dot{\Psi}(k)] \right)^{-1}  \Psi(k).   
		\]
	}

	\ignore{
		\begin{align*}
			\small
			\psi_{ \theta^2  } & = -\frac{1}{ \gamma \sigma^2_{\eps}} (n - t^{(1)} ) +  \frac{1}{ \sigma^4_{\eps} \gamma } \left\{ y\tr b_y - b_y \tr b_y  \right\},
			\\
			\psi_{ \sigma_{\eps}^2 }  & = 
			-\frac{1}{\sigma^2_{\eps} } t^{(1)} + \frac{1}{ \sigma^4_{\eps} } b_y \tr b_y, \\
			\E[ \dot{\Psi} | S] & = -\begin{pmatrix}
				(\ts - 2\tf + n)/\theta^4   &  ( t^{(1)} - t^{(2)} ) / (\theta^2 \sigma^2_\eps)  \\
				{ ( t^{(1)} - t^{(2)} ) }/ (\theta^2 \sigma^2_\eps)  & t^{(2)}/\sigma^4_{\eps}  
			\end{pmatrix}.
		\end{align*}
	}
	
	\label{alg1}
\end{algorithm}
}

\subsection{Analytical expression  with the idempotent similarity matrix}
\label{append:anova}
We give the derivation details for the estimators in \eqref{sp1} in the main text. For the idempotent matrix  $S = n/ r  U U\tr$ with $U\tr U = \I_r $ and $\tilde \gamma = n/r \gamma$, the Woodbury matrix identity suggests  
\begin{gather*}
	H^{-1} = \I - \tilde \gamma U( \I_{r \times r} + \tilde \gamma \I  )^{-1} U\tr  = \I_{n \times n} - \frac{n \gamma }{ r + n \gamma}  U_{n \times r} U_{n \times r} \tr. 
\end{gather*}
 When $r < n$,  $\kappa_i = (1 + \gamma n/r  )^{-1} $ for $1 \leq i \leq r$ and $\kappa_i = 1$ for others. According to    
\begin{gather*}
b_y = H^{-1} y  =  y - \frac{\tilde \gamma}{1 + \tilde \gamma}  U U\tr y, \\
\tf = \sum^n_{i = 1} \kappa_i  = r (1 + \tilde \gamma)^{-1} + (n - r), \\
\ts =  \sum^n_{i = 1} \kappa^2_i = r (1 + \tilde \gamma)^{-2} + (n - r).
\end{gather*}
Plugging them into the estimating equations
\begin{align*}
	\begin{cases} 
		y\tr y - y\tr U U\tr y + (1 + \tilde \gamma)^{-1} y\tr U U\tr y   =  n\sigma^2_e  \\
		y\tr U U\tr y =  r\sigma^2_e (1 + \tilde \gamma)
	\end{cases}.
\end{align*}
 The analytical expressions are then obtained as
\begin{gather*}
	\hat \sigma^2_{\eps}  = \frac{y\tr y - y\tr U U\tr y}{n-r}, \quad
	 \hQbeta =  \frac{  y\tr U U\tr y - \frac{r}{n} y\tr y }{n-r} \\
	\hat \gamma 
	= \frac{  y\tr U U\tr y - \frac{r}{n} y\tr y}{y\tr y -  y\tr U U\tr y} , \quad \hat h^2 = \frac{ y\tr U U\tr y - \frac{r}{n} y\tr y }{y\tr y}.
\end{gather*}

\section{Supplementary numerical results}

\subsection{Numerical comparison of SMILE estimators using different choices of \(W\).}
\begin{figure}[H]
	\centering
\resizebox{1\linewidth}{!}{
		\includegraphics[scale = 0.5]{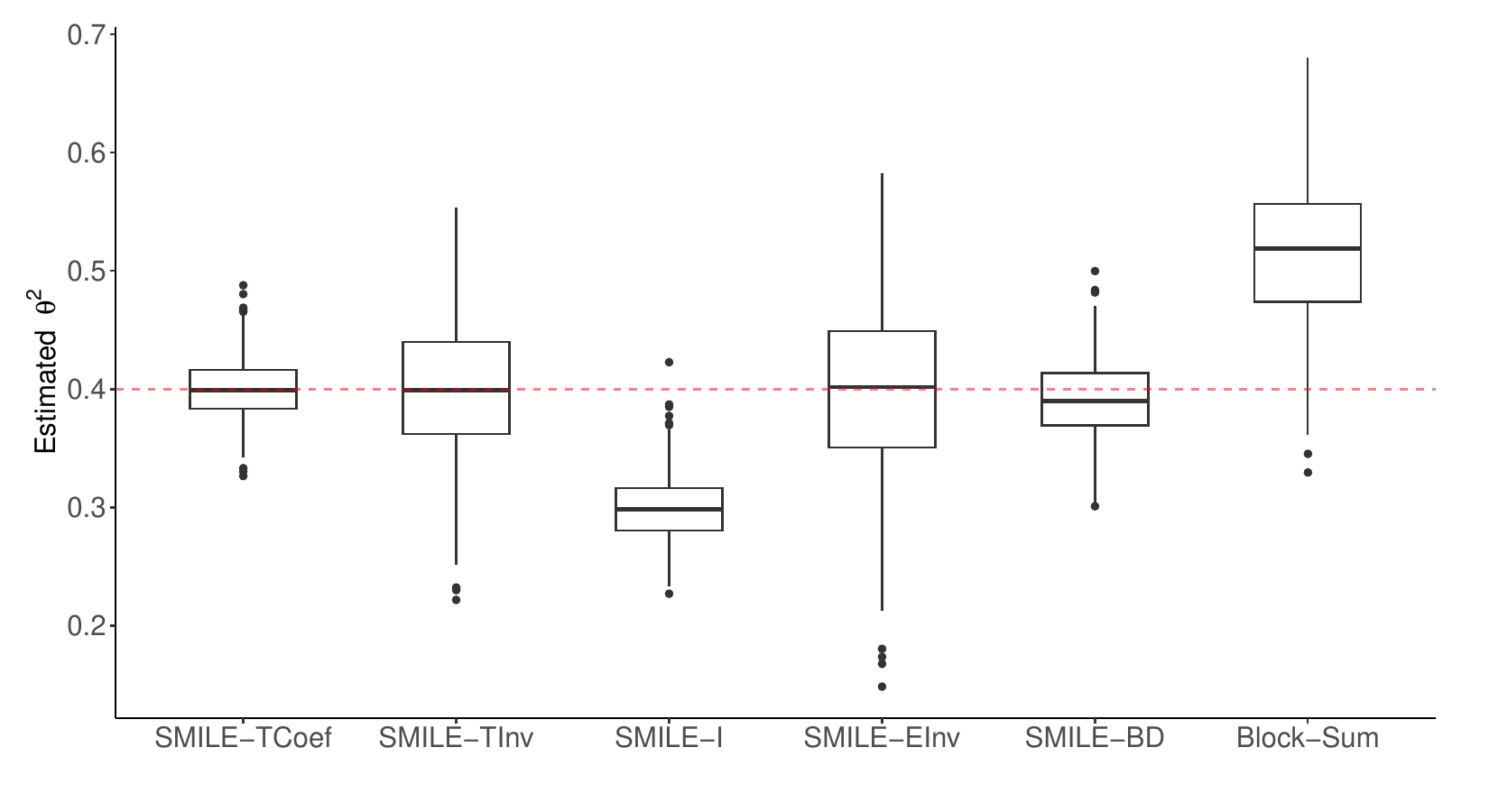}
	}
	\caption{  \it{ Boxplots of the estimation errors of the SMILE estimators $\widehat\theta^2$ 
using
different $W$.  
Here $(n, p) = (800, 1000)$ and the
    genotype covariance $\Sigma$ from chromosome~1 (4.8--5.5\,Mb) is used. 200 SNPs are randomly selected as  the support set $\cA$ of $\beta$ with the effect size $\beta_j = \omega_j \alpha_j$  for $j \in \cA$ with   $\omega_j = (\sum_{k} \Sigma^2_{jk} )^{1/2} $ and standard normal realizations \(\alpha_j \).  The first two estimators SMILE-TCoef and SMILE-TInv use the oracle weights $W = \beta\beta\tr/ \| \beta \|^2$ and the true inverse $\Sigma^{-1}$, confirming the accuracy of the SMILE estimator under the rank-1 or the full-rank similarity matrix. The other estimators use data-driven choices of $W$, which include the identity matrix $W = I$ (SMILE-I), the estimated inverse $\hat \Sigma^{-1}$ with $ \widehat \Sigma $ calculated from $n = 800$ in-sample observations combined with $1000$ supplementary $X$ (SMILE-EInv), and the block diagonal $W$ in \eqref{block} with 100 equal blocks (SMILE-BD). The corresponding block summation estimator $\hs$  has an upward bias when blocks are dependent. }
	}
	\label{fig:W}
\end{figure}

\subsection{Covariance approximation performance under  AR(1) covariance}

\begin{figure}[H]
	\centering
	\includegraphics[width=\linewidth]{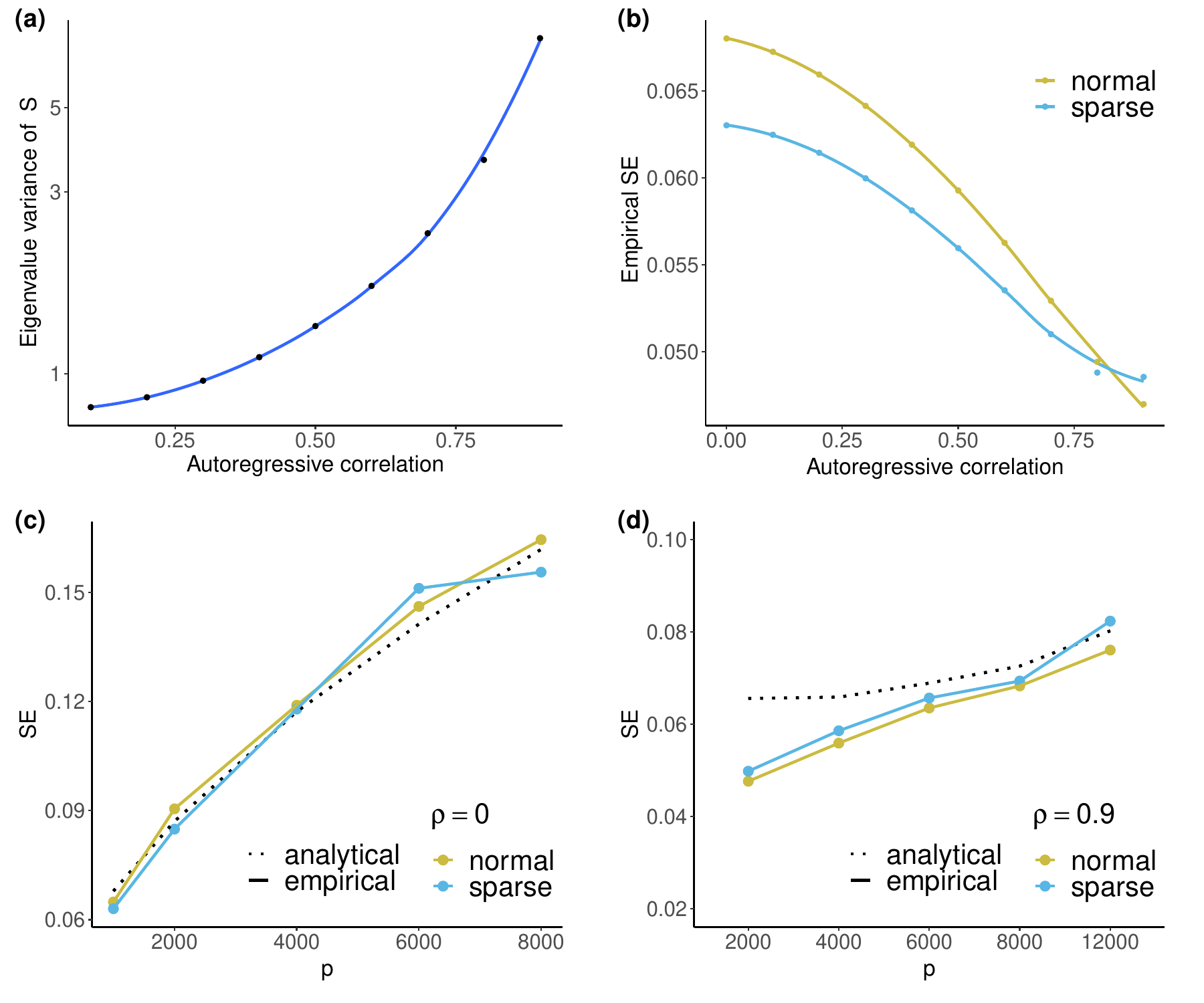}
	\caption{ We evaluate the covariance approximation (\ref{eq:mle}) in Section \ref{sec:implement} under the working Gaussian model and compare with the empirical variance calculated through the simulations. 
		The setting is the same as Figure \ref{fig1}.  The upper panel shows that as auto-correlation $\rho$ increases,  (a) the eigenvalue variation of $S$ increases, and (b) the empirical SE of $\hat \theta^2(XX\tr)$ decreases.
		The lower panel compares the analytical SE in \eqref{vareq} and empirical SE with the autoregressive correlation  (c) $\rho = 0$ and (d) $\rho = 0.9$. 
	}
	\label{fig:eigenvar}
\end{figure}

\subsection{Discussion of the block summation estimator}

\begin{figure}[H]
	\centering
		\begin{tabular}{cc}
			\includegraphics[scale = 0.5]{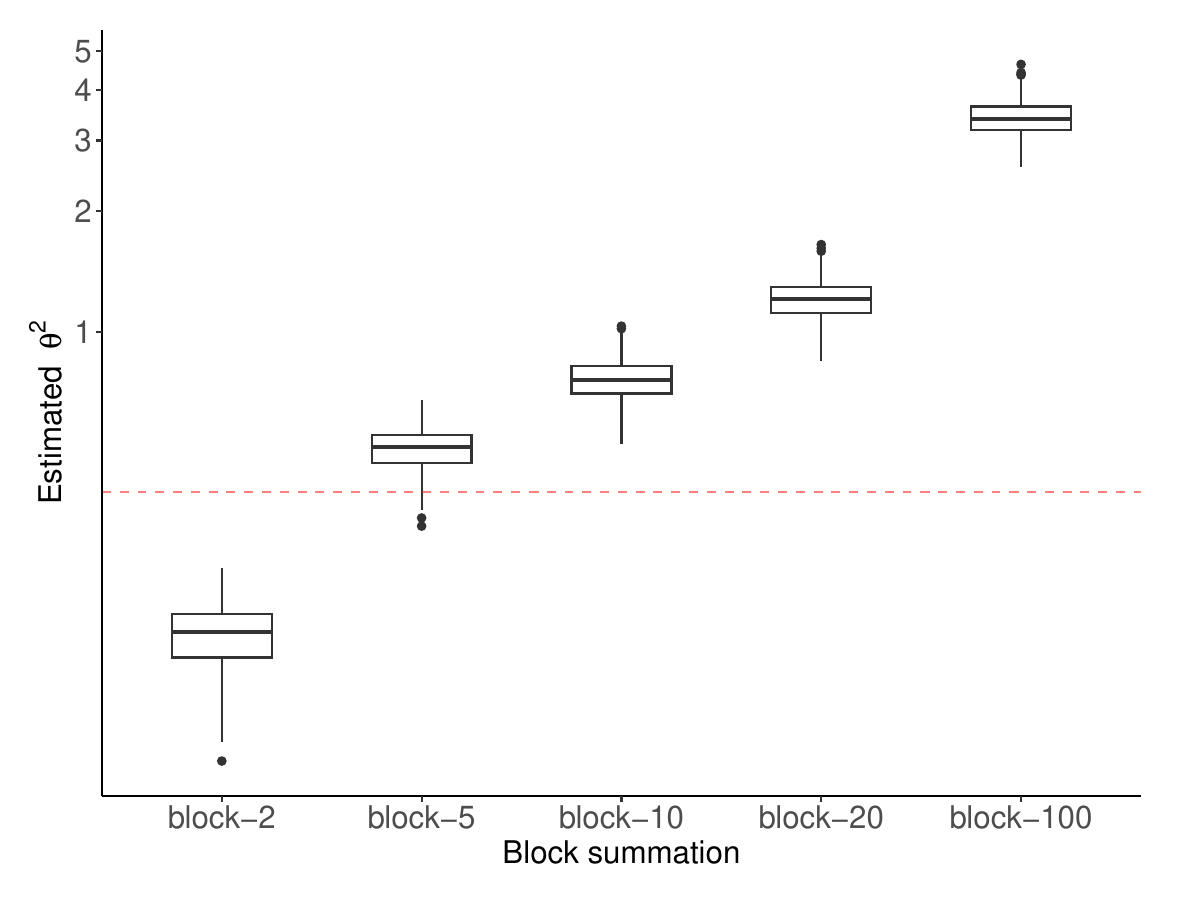}  \\
		\end{tabular}
	\caption{  
		 Block summation estimator $\hs$ with $2$, $5$, $10$ and $20$ equal blocks are also evaluated. The more blocks, the larger the bias is.  
	}
	\label{fig:bsum}
\end{figure}

As discussed in the main text,  the summation estimator $\hs$  is sensitive to block division and requires a larger sample size for accurate block heritability estimation. We illustrate this in Figure \ref{fig:bsum}, where    $\hs$ is evaluated with $5$, $10$, and $20$ equal blocks in the setting of  Figure \ref{fig:W}, showing the bias getting larger with the block number increasing.  However, for 2 blocks with $500$ SNPs,  a larger sample size for accurate block heritability estimation is needed.      

\begin{figure}[H]
	\centering
	\resizebox{\linewidth}{!}{
		\begin{tabular}{c}
			\includegraphics{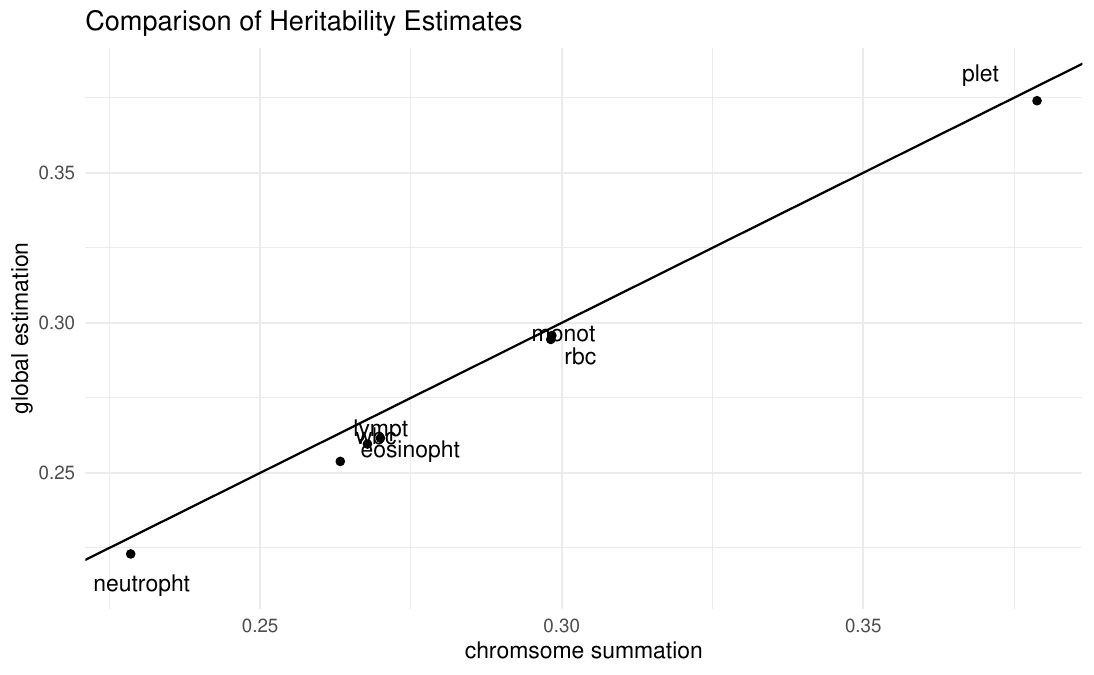}     
			\\
			\includegraphics{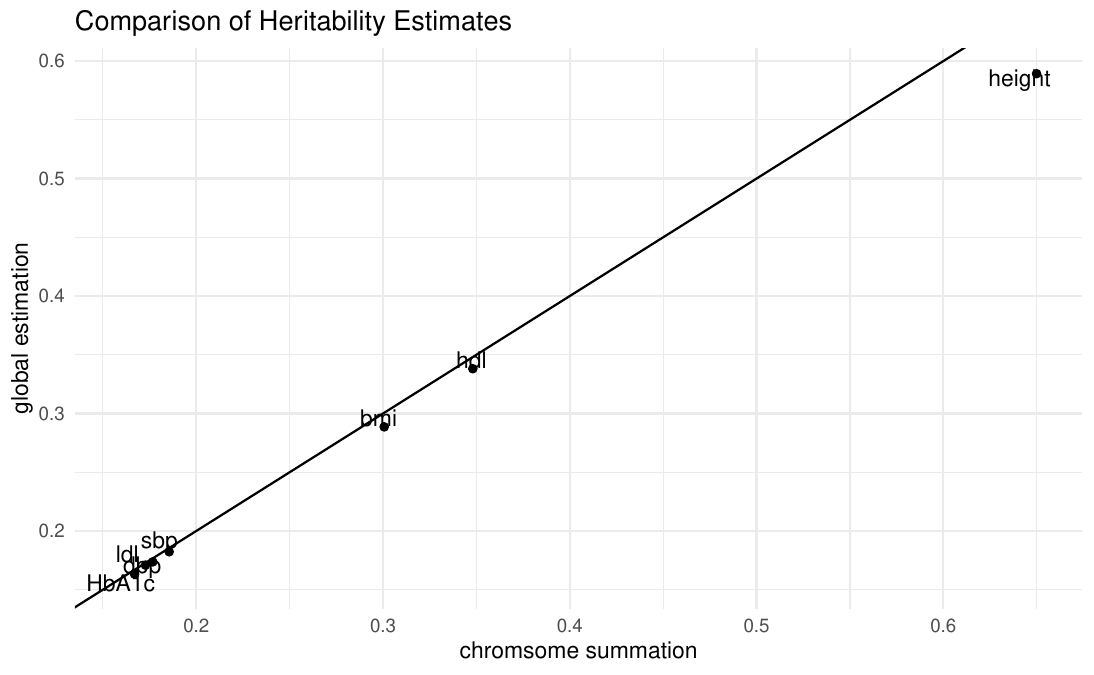}   
		\end{tabular}
	}
	\caption{ Comparison between the chromosome summation and global estimation. As expected, the chromosome summation gives higher estimates but for most outcomes with less extensive architecture, two estimates are similar. For the height with extensive genetic architecture and high heritability, the difference could be significant.  }
	\label{fig:chr-sum}
\end{figure}

To avoid bias caused by small dependent blocks,   for genetic application, one may  estimate  each chromosome heritability  through SMILE and  the total heritability is
\[
\widehat \theta_{ \text{sum-chr} }^2 = \sum^{22}_{i = 1}  \widehat \theta^2_{(i) }.
\]
Figure \ref{fig:chr-sum} evaluates the performance of $\widehat \theta_{ \text{sum-chr} }^2$  for complex traits, blood biomarkers, and cell counts.  As expected, the chromosome summation gives higher estimates, but for most outcomes with less extensive genetic architectures, the two estimates are similar. However, for the height with extensive genetic architecture and high heritability, the difference could be significant. This suggests the chromosome-level genotype data may not be totally independent of each other. 

\subsection{Simulation setup in Section \ref{sec:simulation}}
\label{supp_sec:setup}

In Section \ref{sec:simulation}, four different effect distributions are simulated as follows: 
\begin{itemize}	
	\item[(a)] {\bf Random location and random normal effects:} $\cA$ is selected randomly and  non-zero coefficients $\beta_j \sim N(0, 1)$ for $j \in \cA.$ The coefficients are then re-scaled so that $\Qbeta = 0.2.$ 
	
	\item[(b)] {\bf Random location and low-LD dependent effects:} Randomly selected $\cA$ and variants with low LD and low MAF tend to have large effects.    Specifically, for the $j$th variant,  we define the weight $\omega_j = \{f_j(1- f_j)\}^{-0.75}/\ell_j,$ where $f_j$ is the effect  allele frequency and $\ell_j$ is the corresponding LD score \citep{bulik2015ld}. Then $ \beta_j$ is a realization from the normal distribution $ N(0, \omega^2_j ) $  for $j \in \cA.$ 

	\item[(c)] {\bf Random location and high-LD dependent effects:}  Variants with high LD and high MAF tend to have large genetic effects, with the weight $\omega_j = \{f_j(1 - f_j)\}^{0.75} \times \ell_j.$ 
	
	\item[(d)] {\bf Region-dependent location and effects:}  $\cA$ is sampled from variants in MHC region (Chr6:$2.55$M - $3.35$M) with the sampling probability $\P(\beta_j \neq 0) = 1/[1 + \exp\{2 (\ell_j - \bar \ell ) \}]$ where $\bar \ell$ is the median of LD score. Once $\cA$ is determined, the elements of $\beta_\cA$ are assigned according to the entries of the $1000$th eigenvector of the LD matrix in the MHC region.

\end{itemize}

 To calculate the SMILE estimator,  we first standardize the genotype data and specify $W$ as a block diagonal matrix that is the inverse of the LD matrix, as discussed in Section \ref{sec:herit}.  Specifically, we modify LD blocks from  \cite{pickrell2016detection} by merging the LD blocks within the long-range LD regions specified in \cite{price2008long} to capture the main LD structure. Some LD blocks are weakly correlated with each other and are allowed by our method. Finally, we have 1683 blocks in total, and vary the block sizes, where the block of the MHC region on the chromosome 6   
 contains the largest number of  $7443$ SNPs.   
   In each block, we specify the generalized inverse of the LD matrix by inverting the eigenvalues that collectively explain more than $99.5\%$ of the total variance. 
   
   The calculation of  $W$ in \eqref{block} could be conducted using the genotypes of all $30K$ subjects, or the $12K$ in-sample individuals that match to the outcomes.  Their performances are similar, as we reported in the Supplementary Table \ref{tab:supp-cover-1} and \ref{tab:supp-cover-2}. In what follows, we report the results based on the $W$ using the genotypes of 30K subjects.

\subsection{Additional simulation results}

\begin{table}[H]
\centering
\caption{ 	
	 Comparison of coverage probability and confidence interval (CI) length in different settings. SMILE-I calculates   $W(X)$ with  $n = 12K$  samples and  SMILE-A calculates $W(X, X_E)$ incorporating all samples with $n + n_E = 30K$.  Two approaches have similar performances. 
}
\label{tab:supp-cover-1}
	\resizebox{\textwidth}{!}{
\begin{tabular}{l l r r r r r r r r}
	\toprule
	& & \multicolumn{4}{c}{Normal} & \multicolumn{4}{c}{L-LD} \\
	\midrule
	Method & Setting & Bias & Emp. SE & Est. SE & Cover & Bias & Emp. SE & Est. SE & Cover \\
 \midrule
SMILE-A & 10 & -0.014 & 0.078 & 0.076 & 0.950 & -0.007 & 0.078 & 0.077 & 0.943\\
 & $10^2$ & 0.017 & 0.078 & 0.079 & 0.953 & 0.008 & 0.077 & 0.075 & 0.967\\
 & $10^3$ & -0.007 & 0.078 & 0.078 & 0.960 & -0.003 & 0.078 & 0.078 & 0.950\\
 & $10^4$ & -0.020 & 0.077 & 0.076 & 0.953 & -0.011 & 0.078 & 0.076 & 0.950\\
 & $10^5$ & -0.011 & 0.077 & 0.080 & 0.940 & -0.007 & 0.078 & 0.081 & 0.957\\
\addlinespace
SMILE-I & 10 & -0.013 & 0.078 & 0.078 & 0.947 & -0.006 & 0.078 & 0.079 & 0.950\\
 & $10^2$ & 0.018 & 0.078 & 0.079 & 0.963 & 0.009 & 0.078 & 0.076 & 0.950\\
 & $10^3$ & -0.005 & 0.078 & 0.079 & 0.953 & -0.002 & 0.078 & 0.079 & 0.950\\
 & $10^4$ & -0.019 & 0.078 & 0.078 & 0.950 & -0.010 & 0.078 & 0.077 & 0.947\\
 & $10^5$ & -0.009 & 0.078 & 0.081 & 0.947 & -0.006 & 0.078 & 0.082 & 0.953\\
\addlinespace
GCTA & 10 & 0.032 & 0.043 & 0.046 & 0.867 & -0.038 & 0.043 & 0.046 & 0.847\\
 & $10^2$ & 0.033 & 0.043 & 0.042 & 0.903 & -0.094 & 0.042 & 0.042 & 0.387\\
 & $10^3$ & 0.038 & 0.042 & 0.042 & 0.837 & -0.107 & 0.041 & 0.043 & 0.283\\
 & $10^4$ & 0.049 & 0.042 & 0.043 & 0.797 & -0.098 & 0.042 & 0.044 & 0.383\\
 & $10^5$ & 0.040 & 0.042 & 0.043 & 0.833 & -0.100 & 0.041 & 0.046 & 0.337\\
\addlinespace
LDAK & 10 & -0.094 & 0.035 & 0.037 & 0.250 & -0.126 & 0.034 & 0.036 & 0.047\\
 & $10^2$ & 0.028 & 0.036 & 0.035 & 0.893 & -0.136 & 0.034 & 0.033 & 0.020\\
 & $10^3$ & 0.023 & 0.035 & 0.034 & 0.883 & -0.148 & 0.034 & 0.033 & 0.007\\
 & $10^4$ & 0.025 & 0.035 & 0.036 & 0.880 & -0.141 & 0.034 & 0.035 & 0.020\\
 & $10^5$ & 0.020 & 0.035 & 0.035 & 0.913 & -0.145 & 0.034 & 0.036 & 0.020\\
\addlinespace
LDMS & 10 & -0.021 & 0.052 & 0.053 & 0.927 & -0.035 & 0.052 & 0.054 & 0.873\\
 & $10^2$ & 0.013 & 0.052 & 0.049 & 0.940 & -0.030 & 0.052 & 0.050 & 0.895\\
 & $10^3$ & 0.005 & 0.052 & 0.053 & 0.933 & -0.023 & 0.052 & 0.053 & 0.910\\
 & $10^4$ & 0.008 & 0.052 & 0.052 & 0.967 & -0.014 & 0.052 & 0.049 & 0.945\\
 & $10^5$ & 0.008 & 0.052 & 0.054 & 0.937 & -0.013 & 0.052 & 0.056 & 0.917\\
\bottomrule
\end{tabular}
}
\end{table}

\begin{table}[H]
\caption{ 	
	 Comparison of coverage probability and confidence interval (CI) length in  H-LD and MHC-dependent settings. SMILE-I calculates   $W(X)$ with  $n = 12K$  samples and  SMILE-A calculates $W(X, X_E)$ incorporating all samples with $n + n_E = 30K$.  Two approaches have similar performances. 
}
\label{tab:supp-cover-2}
\resizebox{\textwidth}{!}{
\begin{tabular}{llrrrrrrrrr}
	\toprule
	& & \multicolumn{4}{c}{H-LD} & & \multicolumn{4}{c}{Region} \\
	\midrule
	Method & Setting & Bias & Emp. SE & Est. SE & Cover & Setting & Bias & Emp. SE & Est. SE & Cover \\
	\midrule
SMILE-A & 10 & -0.022 & 0.078 & 0.075 & 0.950 & 10 & 0.024 & 0.077 & 0.077 & 0.957\\
 & $10^2$ & 0.020 & 0.078 & 0.080 & 0.953 & 50 & 0.006 & 0.078 & 0.073 & 0.963\\
 & $10^3$ & -0.001 & 0.077 & 0.081 & 0.940 & 100 & -0.009 & 0.078 & 0.075 & 0.953\\
 & $10^4$ & 0.008 & 0.077 & 0.080 & 0.943 & 500 & -0.012 & 0.078 & 0.079 & 0.943\\
 & $10^5$ & -0.001 & 0.077 & 0.074 & 0.967 & 1000 & -0.002 & 0.078 & 0.076 & 0.950\\
\addlinespace
SMILE-I & 10 & -0.022 & 0.078 & 0.076 & 0.953 & 10 & 0.011 & 0.078 & 0.078 & 0.960\\
 & $10^2$ & 0.021 & 0.078 & 0.080 & 0.950 & 50 & -0.008 & 0.078 & 0.074 & 0.950\\
 & $10^3$ & -0.012 & 0.078 & 0.082 & 0.917 & 100 & -0.021 & 0.078 & 0.075 & 0.937\\
 & $10^4$ & -0.005 & 0.078 & 0.081 & 0.933 & 500 & -0.025 & 0.078 & 0.079 & 0.937\\
 & $10^5$ & -0.011 & 0.078 & 0.075 & 0.960 & 1000 & -0.016 & 0.078 & 0.077 & 0.937\\
\addlinespace
GCTA & 10 & 0.035 & 0.043 & 0.046 & 0.867 & 10 & 0.140 & 0.041 & 0.036 & 0.043\\
 & $10^2$ & 0.170 & 0.043 & 0.041 & 0.023 & 50 & 0.041 & 0.041 & 0.036 & 0.870\\
 & $10^3$ & 0.467 & 0.036 & 0.026 & 0.000 & 100 & 0.136 & 0.040 & 0.034 & 0.043\\
 & $10^4$ & 0.468 & 0.035 & 0.025 & 0.000 & 500 & 0.223 & 0.041 & 0.035 & 0.000\\
 & $10^5$ & 0.455 & 0.036 & 0.027 & 0.000 & 1000 & 0.302 & 0.039 & 0.031 & 0.000\\
\addlinespace
LDAK & 10 & -0.081 & 0.035 & 0.037 & 0.393 & 10 & 0.009 & 0.033 & 0.029 & 0.973\\
 & $10^2$ & 0.169 & 0.036 & 0.033 & 0.003 & 50 & -0.055 & 0.033 & 0.029 & 0.590\\
 & $10^3$ & 0.362 & 0.031 & 0.022 & 0.000 & 100 & 0.010 & 0.033 & 0.028 & 0.983\\
 & $10^4$ & 0.371 & 0.030 & 0.022 & 0.000 & 500 & 0.064 & 0.034 & 0.030 & 0.517\\
 & $10^5$ & 0.355 & 0.031 & 0.024 & 0.000 & 1000 & 0.160 & 0.033 & 0.027 & 0.000\\
\addlinespace
LDMS & 10 & -0.008 & 0.052 & 0.053 & 0.940 & 10 & -0.065 & 0.051 & 0.051 & 0.777\\
 & $10^2$ & 0.040 & 0.051 & 0.047 & 0.893 & 50 & -0.108 & 0.051 & 0.048 & 0.457\\
 & $10^3$ & 0.011 & 0.050 & 0.049 & 0.950 & 100 & -0.084 & 0.051 & 0.048 & 0.617\\
 & $10^4$ & -0.005 & 0.050 & 0.050 & 0.940 & 500 & -0.054 & 0.050 & 0.050 & 0.810\\
 & $10^5$ & 0.007 & 0.050 & 0.049 & 0.957 & 1000 & -0.038 & 0.050 & 0.048 & 0.880\\
\bottomrule
\end{tabular}
}
\end{table}

\newpage

\section{Stability analysis}
\label{sec:stability}
\subsection{Computational stability of  CG-SMILE} 
To handle large-scale genotype data, SMILE uses the conjugate gradient method to compute $(I + \gamma S)^{-1}y$.   The trace terms $(\tf, \ts)$ are approximated using a Monte Carlo trace estimator combined with the conjugate gradient method. We denote it as CG-SMILE  to distinguish it from the exact calculation. 

\begin{figure}[H]
	\centering   \includegraphics[width=0.8\linewidth]{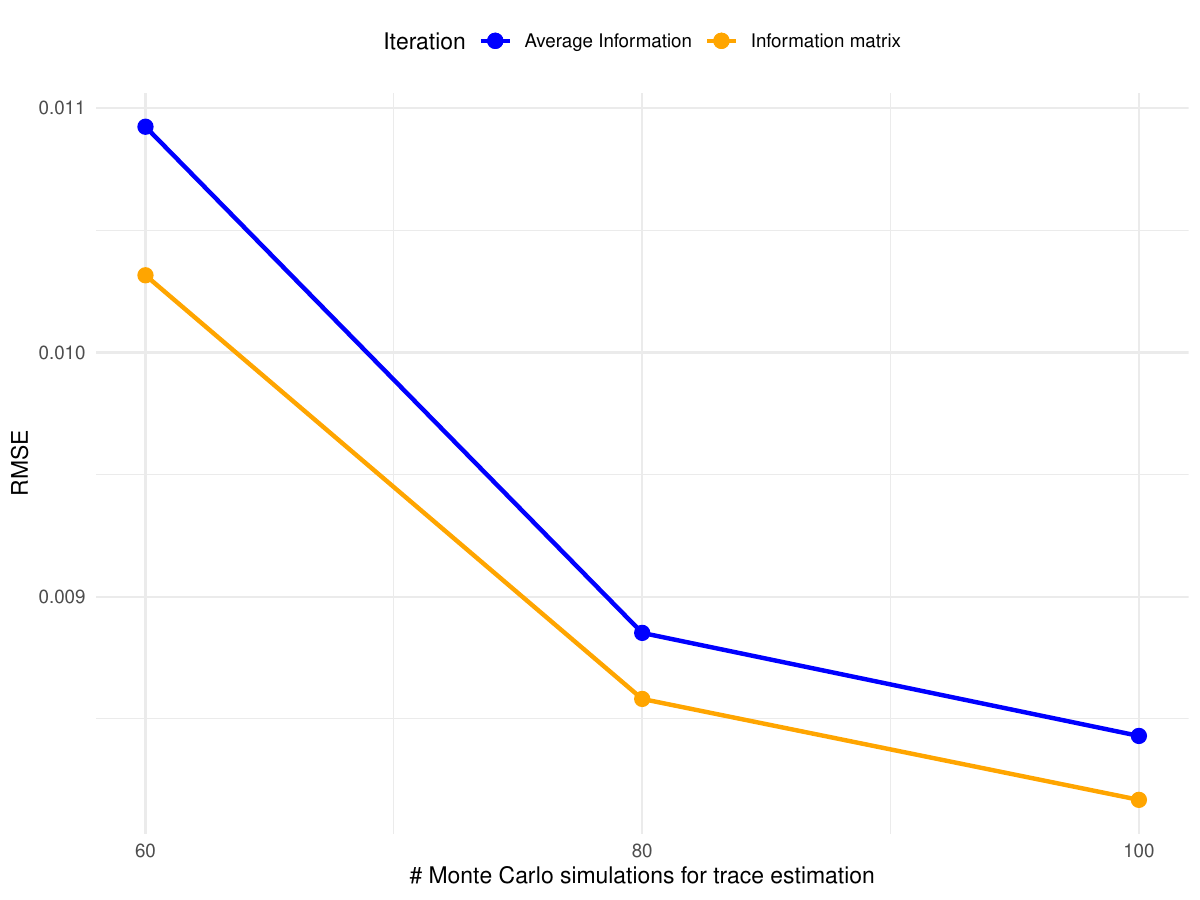}
	\caption{Robust performance of CG-SMILE  based on the Monte Carlo trace estimation, which samples 60, 80, and 100 random vectors.  RMSE is calculated between CG-SMILE and SMILE estimates across 200 repeats. We also evaluate an optimization using the average information method \citep{gilmour1995average} instead of the fisher information matrix.  The result shows CG-SMILE is close to SMILE and estimation gets more accurate with more random vectors drawn.  }
	\label{fig:cg}
\end{figure}

To evaluate the computational stability of CG-SMILE, we consider drawing varying numbers of random vectors for the trace calculation, and   RMSE is calculated between CG-SMILE and SMILE estimates across 200 repeats. Besides, we also evaluate a method that replaces $\E[\dot \psi ]$ (Fisher information up to a scaling)  by the average information  \citep{gilmour1995average},  which avoids the calculation of $\ts$. The setting is the same as Simulation setting (a) in the main text. The results are summarized in Figure \ref{fig:cg}

\begin{figure}[H]
    \centering
   			\begin{tabular}{c}
    \includegraphics[width=0.7\linewidth, height = 0.4\textheight]{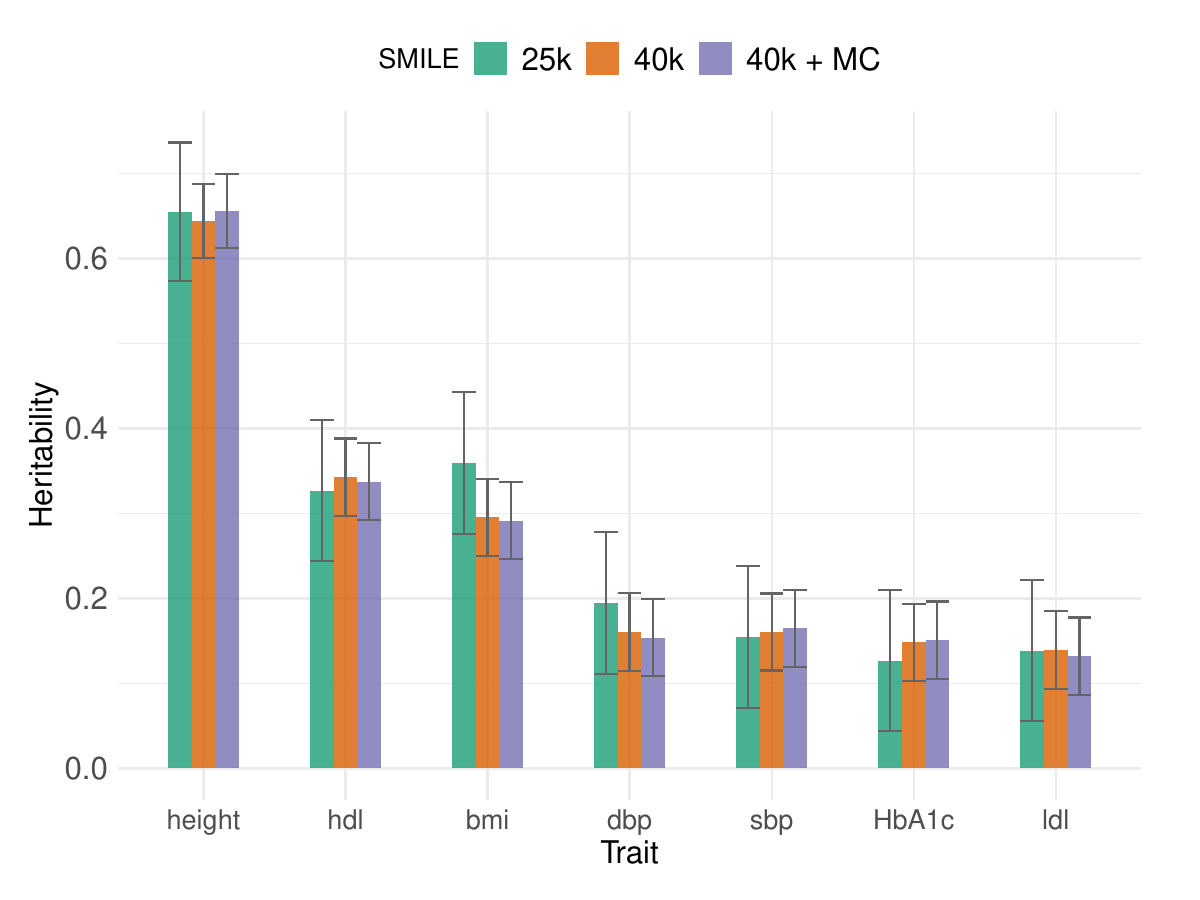} \\
    \includegraphics[width=0.7\linewidth, height = 0.4\textheight]{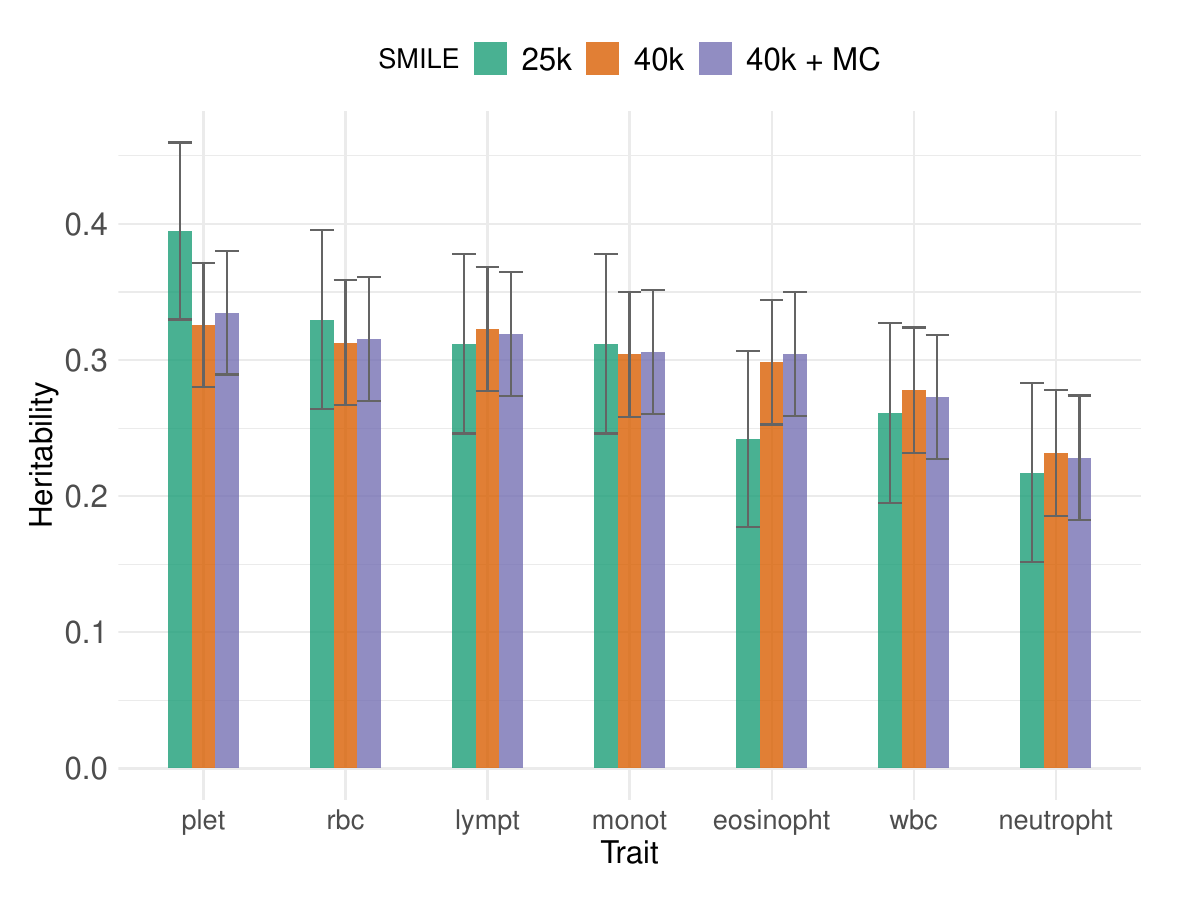}
    	\end{tabular}{c}
    \caption{Robust performance of SMILE under independent 25K or 40K samples or CG-SMILE based on the Monte-Carlo trace estimation}
\label{fig:sample_size}
\end{figure}

In addition, we also compare the estimates for real UK Biobank outcomes under different sample sizes in Figure \ref{fig:sample_size}.

\subsection{Stability analysis for outcome outliers and  outcome transformation}
\label{sec:outlier}
\begin{figure}[H]
    \centering
    \includegraphics[width=0.7\linewidth]{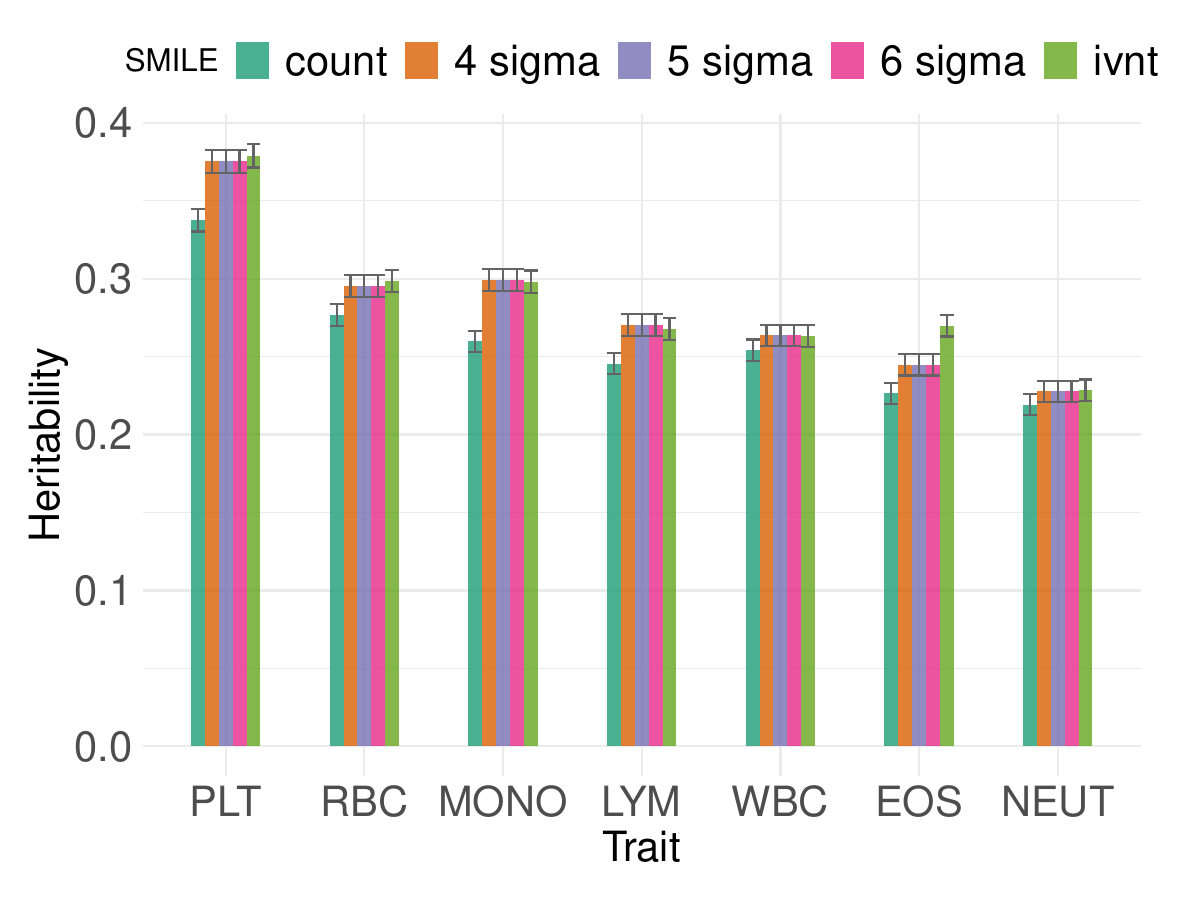}
    \caption{Heritability estimation when using the residual of the count outcome, remove the outliers outside the  4, 5, 6 standard deviation and inverse normal transformation of the residual. As expected, the estimated heritability for count residual is usually smaller, while the outcome transformation or outlier removal helps achieve similar heritability estimates for most outcomes.   Blood cell counts include white blood cells (WBC), red blood cells (RBC), platelets (PLT), lymphocytes (LYM), monocytes (MONO), neutrophils (NEUT), and eosinophils (EOS). }
    \label{fig:process}
\end{figure}

Heritability estimation is sensitive to the skewed outcome distribution and outliers, e.g., extreme values may significantly make $\Var(y_i)$ larger.  The blood cell count outcomes usually have extreme outliers even after log transformation, and we evaluate whether different outcome processing and transformation will alter the heritability estimation. Specifically, we conduct the log transformation for the outcome and take the residuals after adjusting the covariates.  We then estimate heritability using: (1) the raw count residuals, (2) residuals after removing outliers beyond 4, 5, or 6 standard errors, and (3) inverse-normal-transformed residuals. We compare the estimated heritability after different processing in Figure \ref{fig:process}.

As expected, the estimated heritability for count residual is usually smaller, while the outcome transformation or outlier removal helps achieve more stable and similar heritability estimates for most outcomes. It suggests researchers need to be careful about the skewed outcome distribution when estimating heritability.  

\newpage

\ignore{

\ignore{
	The variance of estimators can be obtained alternatively based on \eqref{vareq}. We have 	\(
	\tf = \sum^n_{i = 1} \kappa_i, \ts = \sum^n_{i = 1} \kappa^2_i
	\)  Specifically,  when $r < n$, we have $\kappa_i = (1 + \gamma n/r  )^{-1} $ for $1 \leq i \leq r$ and $\kappa_i = 1$ for others
	\[
	\Var(\hat \theta^2) = \frac{ 2 \theta^4 \ts }{ n \ts - {\tf}^2 }, \quad \Var(\hat \sigma_{\eps}^2) = \frac{ 2 \sigma^4_{\eps}  (\ts - 2 \tf + n) }{ n \ts - {\tf}^2 }
	\]
	Therefore
	\[
	n \ts - {\tf}^2 = n	\sum^n_{i = 1} (\kappa_i - \bar \kappa)^2, \quad \sum^n_{i = 1} (\kappa_i - \bar \kappa)^2 = r(1 - r/n) (1- (1 + \gamma \frac{n}{r}  )^{-1}  )^{2}, \quad \frac{\sum^n_{i = 1}(\kappa_i - 1)^2}{ \sum^n_{i = 1} (\kappa_i - \bar \kappa)^2  }  = \frac{n}{n-r}
	\]
	\[
	\ts = \sum^n_{i = 1} \kappa_i ^2 = r(1 + \gamma \frac{n}{r})^{-2} + (n-r), \quad  \frac{  \sum^n_{i = 1}\kappa_i^2 }{  \sum^n_{i = 1} (\kappa_i - \bar \kappa)^2 } = \frac{1}{\tilde \gamma^2}[ \frac{n}{n-r} + \frac{n}{r \kappa^2} ]
	\]
}

Next, we give some discussion of the small and independent block division. We have the following corollary from Theorem 1.3 in \cite{bryson2021marchenko}. 
\begin{Corollary}
Let $\Xw_{n \times p} = (U^{(1)}, \ldots, U^{(m)})$ with $p = \sum_{k} r_k$ be a random matrix following the block-independent model with the aspect ration $p/n$ converging to a number $\lambda \in (0, \infty)$ and $\max_{k}r_k = o(p)$ as $p \to \infty.$ Assume that all entries of the random matrix $\Xw$ have uniformly bounded fourth moments. Then with probability $1$, the empirical spectral distribution of the sample covariance matrix $W = m^{-1}X X\tr $ converges weakly in distribution to the Marchenko-Pastur distribution with parameter $\lambda.$	
\end{Corollary}
}

\ignore{
\begin{table}[]
    \centering
   \begin{tabular}{cccc}
\toprule
nsim & conjugate precision & rmse AI & rmse MC\\
\midrule
60 & 1e-05 & 0.0120162 & 0.0112560\\
60 & 1e-04 & 0.0115508 & 0.0118031\\
60 & 1e-03 & 0.0096141 & 0.0118830\\
80 & 1e-05 & 0.0098607 & 0.0099678\\
80 & 1e-04 & 0.0098538 & 0.0090671\\
\addlinespace
80 & 1e-03 & 0.0097533 & 0.0105159\\
100 & 1e-05 & 0.0088180 & 0.0092632\\
100 & 1e-04 & 0.0085948 & 0.0084288\\
100 & 1e-03 & 0.0097691 & 0.0098256\\
\bottomrule
\end{tabular}
    \caption{Stochastic trace estimation  for the final estimation approximation accuracy}
    \label{tab:my_label}
\end{table}
}

\bibliographystyle{apalike}
\bibliography{Heri}